\documentclass[aps,pre,twocolumn,unsortedaddress]{revtex4-2}

\usepackage{comment}

\usepackage{graphicx}
\usepackage{dcolumn}
\usepackage{amsfonts}
\usepackage{dsfont}
\usepackage{mathrsfs}
\usepackage{enumitem}
\usepackage{footnote}
\usepackage{booktabs}
\usepackage{braket}
\usepackage[overload]{empheq}
\usepackage{mathrsfs}
\usepackage{titlesec}
\usepackage{array}
\usepackage{upgreek}
\usepackage[usenames, dvipsnames]{xcolor}
\usepackage{siunitx}
\usepackage{tabularx}
\usepackage{cases}
\usepackage{epstopdf}
\usepackage{framed}
\usepackage{xspace}
\usepackage{pifont}
\usepackage{hhline}
\usepackage{multirow}
\usepackage{paracol}
\usepackage{mwe}
\usepackage[utf8]{inputenc}
\usepackage[T1]{fontenc}

\usepackage{graphicx}
\usepackage{dcolumn}
\usepackage{bm}
\usepackage{hyperref}

\usepackage{xcite}
\usepackage{xr}
\makeatletter
\newcommand*{\addFileDependency}[1]{
  \typeout{(#1)}
  \@addtofilelist{#1}
  \IfFileExists{#1}{}{\typeout{No file #1.}}
}
\makeatother

\newcommand*{\myexternaldocument}[1]{%
    \externaldocument{#1}%
    \addFileDependency{#1.tex}%
    \addFileDependency{#1.aux}%
}
\myexternaldocument{main}

\newcommand{\pard}[2]{\frac{\partial #1}{\partial #2}}
\newcommand{\pardd}[2]{\frac{\partial ^2 {#1}}{\partial {#2} ^2}}

\newcommand*\diff{\mathop{}\!\mathrm{d}}

\newcommand{\lB}{l_\mathrm{B}}
\newcommand{\kB}{k_\mathrm{B}}

\newcommand{\dens}{n}
\newcommand{\permz}{\varepsilon_0}
\newcommand{\permr}{\varepsilon_r}

\newcommand{\lD}{\lambda_\mathrm{D}}
\newcommand{\hrho}{\widehat{\rho}}

\newcommand\Du{\mbox{\textit{Du}}}
\newcommand{\munen}{\mu_{\mathrm{nen}}}

\newcommand{\dep}[2]{\ensuremath{\frac{\partial #1}{\partial #2}}}

\def\e{\mathrm{e}}

\def\O{\mathcal{O}}

\DeclareSIUnit[number-unit-product = {\,}] \cal{cal}

\graphicspath{{./Figures-TechnicalPaper/}}

\newcommand{\Eqref}[1]{Eq.~\eqref{#1}}
\newcommand{\Eqsref}[1]{Eqs.~\eqref{#1}}

\newcommand{\latin}[1]{{\it{#1}}}
\newcommand{\ie}{\latin{i.e.}}
\newcommand{\eg}{\latin{e.g.}}
\newcommand{\adhoc}{\latin{ad hoc}}
\newcommand{\caveat}{\latin{caveat}}
\newcommand{\interalia}{\latin{inter alia}}
\newcommand{\perse}{\latin{per se}}
\newcommand{\cf}{\latin{cf.}}

\newcommand{\german}[1]{{\it #1}}
\newcommand{\ansatz}{\german{ansatz}}

\usepackage{ulem}

\begin{document}

\title{Poisson-Nernst-Planck charging dynamics of an electric double layer capacitor: symmetric and asymmetric binary electrolytes}

\author{Ivan Palaia}
\affiliation{Department of Physics, King’s College London, WC2R 2LS, United Kingdom}
\affiliation{Institute of Science and Technology Austria, 3400 Klosterneuburg, Austria}
\author{Adelchi J. Asta}
\affiliation{Dessia, 92160 Antony, France}
\author{Megh Dutta}
\affiliation{Sorbonne Université, CNRS, Physico-chimie des Électrolytes et Nanosystèmes Interfaciaux, PHENIX, 75005 Paris, France}
\author{Patrick B. Warren}
\affiliation{Hartree Centre, Science and Technology Facilities Council (STFC), Sci-Tech Daresbury, Warrington WA4 4AD, United Kingdom}
\author{Benjamin Rotenberg}
\affiliation{Sorbonne Université, CNRS, Physico-chimie des Électrolytes et Nanosystèmes Interfaciaux, PHENIX, 75005 Paris, France}
\affiliation{Réseau sur le Stockage Electrochimique de l’Energie (RS2E), FR CNRS 3459, 80039 Amiens Cedex, France}
\author{Emmanuel Trizac}
\affiliation{Universit\'e Paris-Saclay, CNRS, LPTMS, 91405 Orsay, France}
\affiliation{ENS de Lyon, 69342 Lyon, France}

\date{\today}

\begin{abstract}
A parallel plate capacitor containing an electrolytic solution is the simplest model of a supercapacitor, or electric double layer capacitor. Using both analytical and numerical techniques, we solve the Poisson-Nernst-Planck equations for such a system, describing the mean-field charging dynamics of the capacitor, when a constant potential difference is abruptly applied to its plates. Working at constant total number of ions, we focus on the physical processes involved in the relaxation and, whenever possible, give its functional shape and exact time constants. We first review and study the case of a symmetric binary electrolyte, where we assume the two ionic species to have the same charges and diffusivities. We then relax these assumptions and present results for a generic strong (\ie~fully dissociated) binary electrolyte. At low electrolyte concentration, the relaxation is simple to understand, as the dynamics of positive and negative ions appear decoupled. At higher electrolyte concentration, we distinguish several regimes. In the linear regime (low voltages), relaxation is multi-exponential, it starts by the build-up of the equilibrium charge profile and continues with neutral mass diffusion, and the relevant time scales feature both the average and the Nernst-Hartley diffusion coefficients. In the purely nonlinear regime (intermediate voltages), the initial relaxation is slowed down exponentially due to increased capacitance, while bulk effects become more and more evident. In the fully nonlinear regime (high voltages), the dynamics of charge and mass are completely entangled and, asymptotically, the relaxation is linear in time. We finally discuss non-ideal behavior in real capacitors and provide conditions for which mean-field is expected to hold.
\end{abstract}

\maketitle

\section{\texorpdfstring{INTRODUCTION}{Introduction}}

In a wealth of physical systems, charged surfaces confine a liquid containing ions. 
If the surfaces are conductive a so-called electric double layer capacitor (EDLC) is formed, owing its name to the two layers of opposite charge that build up at the interface between each surface and the electrolytic solution. Compared to standard capacitors with insulating dielectrics, the ability of EDLCs to store charge and energy is enhanced by the local rearrangement of the confined electrolyte and very large capacities per unit weight can be reached when the conductive electrodes are made of porous or fibrous materials, as this hugely increases the surface area in contact with the electrolyte \cite{Simon2008,Chen2020, simon_perspectives_2020}. EDLCs of this kind can store more energy than common capacitors, can release it more rapidly than common batteries, and last longer than devices based on chemical reactions. These three characteristics, placing them precisely at the border between capacitors and batteries \cite{Winter2004}, explain their widespread usage as well as numerous potential applications: from consumer electronics and wearable devices \cite{Dubal2018}, to energy production \cite{Brogioli2009, Boon2011,Janssen2014} and means of transportation, where large power exchanges are needed to propel and halt electric vehicles, harvest braking energy, or promptly activate emergency devices \cite{Simon2008}.

The need to control the dynamics of charge and discharge of EDLCs has motivated substantial interest in fundamental sciences. Impedance spectroscopy experiments are routinely used to measure the linear response of the system to oscillating fields \cite{vivier_impedance_2022, Barbero2005a}.  Computer simulations have addressed paramount questions concerning \interalia\ the effects of electrode polarisability, ion and pore sizes, or electrostatic correlations \cite{Jeanmairet2022}. In parallel, analytical studies have proved fundamental to inform experiments and simulations, emphasizing the role played by the molecular structure of the double layer (not captured by usual mean-field approaches) in determining the EDLC capacitance \cite{Kornyshev2007, Bazant2011}. Nonetheless, recent works have highlighted the need to better understand the dynamics of simple model systems, such as a planar capacitor in mean-field, to clearly isolate the effect of geometry, hydrodynamics and non-idealities in real devices \cite{Wu2022,Janssen2019b,Asta2019JCP,Lian2020,Ma2022,Yang2022,Varela2021} or to develop strategies to speed up relaxation to equilibrium \cite{Breitsprecher2018,phd,Breitsprecher2020}. Namely, the relaxation dynamics of a planar EDLC subjected to a sudden change in the electric potential between plates has been described by the seminal works  \cite{Bazant2004,Janssen2018} for the simpler case of symmetric electrolytes and with particular attention for the linear (low-voltage) regime, of which \cite{Janssen2018} provided an exact solution.

\begin{figure}
\centering
\includegraphics[width=0.9\columnwidth]{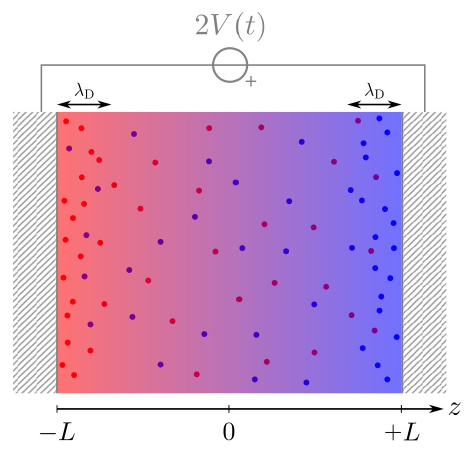}
\caption{Cartoon representing a charged, planar EDLC. Red cations and blue anions are treated at mean-field level, as suggested by the color of the solution, representing charge density. Within the mean-field approximation, for large enough $L$ and sufficiently low applied voltage, the thickness of the electric double layer at equilibrium is $\lD$.
}
\label{fig:EDLC}
\end{figure}

The charging dynamics of an ideal EDLC, such as the one presented in Fig.~\ref{fig:EDLC}, is an interesting problem \perse\ from a physical perspective. Several length scales are involved: the Bjerrum length $\lB$, 
setting the distance between two ions at which thermal energy becomes comparable with electrostatic repulsion; the Debye screening length $\lD$, setting the range of the electrostatic potential; the point along $z$ at which the electrostatic potential vanishes, coinciding with the geometrical center of the capacitor only for symmetric electrolytes; lastly, the size of the system $L$. In strongly polarised situations, a fifth length can in principle be relevant, the Gouy-Chapman length, a signature of the less efficient screening occurring when only counterions are in proximity of the plates. The relaxation to equilibrium observed when an external voltage is imposed can depend in principle on any these lengths, as well as on the diffusion coefficients of the ionic species, and on the applied voltage. These quantities can be combined in many ways to give plausible relaxation times: understanding which of those are relevant for the dynamics, for different electrolytes and at different times, is a challenge. Moreover, the relaxation is not simply exponential, as in many model physical systems, and can be hard to solve and describe in closed form.

In this paper, and in its companion Letter \cite{mainpaper}, we present a detailed study of the relaxation dynamics of an ideal, planar EDLC in mean-field. In Sec.~\ref{sec:PNPintro} we recapitulate the Poisson-Nernst-Planck formalism, within which we operate throughout the whole paper using the numerical method described in Appendix~\ref{sec:NumericalMethod}. Then, we present results concerning the case where the two ions species have the same valence and diffusion coefficients (fully symmetric case, Sec.~\ref{sec:sym}), the case where the two species have the same diffusion coefficient but different valences (partially asymmetric case, Sec.~\ref{sec:asymsym}) and the case where the two species have different valences and diffusion coefficients (fully asymmetric case, Sec.~\ref{sec:fullyasym}). For each of these cases we vary electrolyte concentration and applied voltage and we study the linear, purely nonlinear, and depleted nonlinear regimes. The relaxation involves in general more than one process, so that each regime features many relaxation time scales, each of which may become relevant at different times. We give a thorough characterisation of this complex phenomenon, focusing with more care on the cases for which a simple analytical interpretation is possible.

\section{\texorpdfstring{MODEL}{Model}}
\label{sec:PNPintro}

Our ideal EDLC (Fig.~\ref{fig:EDLC}) is treated within the Poisson-Nernst-Planck formalism \cite{Hunter}. We neglect hydrodynamic effects, and suppose purely Coulombic interactions between ions and Coulombic and hard-core interactions between ions and walls. No absorption phenomena are considered, so the compact part of the electric double layer, usually called Stern layer, is neglected. The two plates, distant $2L$ from each other, are connected to a time-dependent ideal voltage source, imposing a potential difference $2V(t)$ between them.

The salt is a strong binary electrolyte and is therefore completely dissociated in the solvent: densities are denoted $\dens_+$ and $\dens_-$, diffusion coefficients $D_+$ and $D_-$, and integer ionic valences $\bar{q}_+=+q_+$ and $\bar{q}_-=-q_-$ (for instance, for MgCl$_2$, $\bar{q}_+=+2$, $\bar{q}_-=-1$, and $q_+=2$, $q_-=1$). 
Electrodes are modelled as parallel, infinite, charged planes.
Densities $\dens_\pm$ and potential $\phi$ are functions of the sole spatial coordinate $z$ in the direction perpendicular to the planes and of time $t$.
Fig.~\ref{fig:EDLC} shows a sketch of the system, with size $2L$ and $z=0$ at the center of the capacitor. 

We work in the canonical ensemble, \ie~with a fixed number of ion pairs per unit surface $2\dens_0 L$, where $\dens_0$ is the uniform density of salt when the power source is off and the capacitor at equilibrium. The initial density for each species is $n_\pm^0$ and is such that $q_+ n_+^0 = q_- n_-^0 = q_+ q_- n_0$ by electroneutrality (the last equality holding when $q_+$ and $q_-$ are co-prime). 
Choosing to work in the canonical ensemble allows to avoid the question of where exactly ions are injected into the system from the reservoir during the dynamics, which in a real nanocapacitor might depend on the size and the topology of the pores.
In addition, the constant ion number ensemble is a good approximation for open systems where the electrodes have large lateral dimensions and it takes a long time to ensure the chemical equilibrium of the entire pore with the bath. In fact, studying relaxation in the canonical ensemble reveals precisely which time scales should be compared to chemical equilibration time scales to assess whether the system is effectively canonical or grand-canonical. Finally, there are regimes in which densities in the bulk solution (at the center of the capacitor) remain approximately constant, at least to first order, and the canonical ensemble is not quantitatively different from the grand-canonical ensemble. 

The Poisson-Nernst-Planck equation describes the change in time and space of the ionic densities \cite{Hunter}: it consists of a drift-diffusion model for ionic currents, complemented with a continuity equation. The average current density $j_\alpha(z,t)$ for ions of type $\alpha$ ($\in\{-,+\}$), moving along $z$ 
under the action of a potential $\phi(z,t)$, is  
\begin{equation}
    {j}_\alpha(z,t)=-\beta D_\alpha \,\dens_\alpha(z,t) \, \bar{q}_\alpha e \pard{\phi}{z}(z,t) - D_\alpha\pard{\dens_\alpha}{z}(z,t) \,,
    \label{eq:current}
\end{equation} 
where $\beta=1/(\kB T)$ is the inverse temperature and $e$ the elementary charge.
The first term represents a drift current, obtained as the product of a mobility $\beta D_\alpha$ (in agreement with Einstein's relation), a density and an electric force. The second term is a diffusion current, described by Fick's law. We assume a diagonal diffusivity tensor, meaning that the diffusion of a species is not influenced by the presence of other species. In general, the microscopic potential depends on the discrete positions of the ions. Within the Poisson-Nernst-Plank approximation, the potential is assumed to be a coarse-grained average of the microscopic potential, hence the mean-field nature of \Eqref{eq:current}.  In the static case, currents vanish and the Poisson-Boltzmann distribution is retrieved, with the average density proportional to the exponential of an average potential \cite{Hunter, Andelman2010}.

For currents $j_\alpha(z,t)$, the following exact continuity equation must hold, ensuring local ionic mass and charge conservation:
\begin{equation}
    \dep{\dens_\alpha}{t}(z,t)=-\dep{j_\alpha}{z}(z,t)\,.
\label{eq:continuityeq}
\end{equation}
Substituting \Eqref{eq:current} in \Eqref{eq:continuityeq} yields
\begin{equation}
    \dep{\dens_\alpha}{t}(z,t)=D_\alpha \dep{}{z} \left(\, \dens_\alpha(z,t)\, \beta \bar{q}_\alpha e\dep{\phi}{z}(z,t) + \dep{\dens_\alpha}{z} (z,t)\right).
\label{eq:PNP}
\end{equation}

The potential is related to the density by the Poisson equation, also exact:
\begin{equation}
    - \pardd{\phi}{z}(z,t) = \frac{\rho(z,t)}{\varepsilon_0\varepsilon_r}\,,
    \label{eq:poisson}
\end{equation}
where $\rho=q_+ e n_+ - q_- e n_-$ is the charge density, $\varepsilon_0$ the permittivity of vacuum, and $\varepsilon_r$ the relative permittivity.
\Eqsref{eq:PNP} and~\eqref{eq:poisson} constitute the Poisson-Nernst-Planck theory.

The fact that ions are confined within the two slabs of the capacitor imposes the boundary condition of vanishing currents at $+L$ and $-L$. From the definition in \Eqref{eq:current}:
\begin{align} \label{eq:bcflux}
\begin{split}
- \beta n_\alpha(-L, t) \bar{q}_\alpha e \dep{\phi}{z}(-L, t) - \dep{n_\alpha}{z}(-L, t) =0\,,
\\
- \beta n_\alpha(+L, t)  \bar{q}_\alpha e \dep{\phi}{z}(+L, t) - \dep{n_\alpha}{z}(+L, t) =0 \,.
\end{split}
\end{align}

We will focus on the situation where the EDLC is at equilibrium, with zero applied potential for times $t<0$, and is subject to a potential difference $2V_0$ for times $t>0$. The electric potential must be continuous between slab and solution, so that
\begin{equation} \label{eq:bcpotential}
\phi(\pm L, t)=\pm V(t)\,, \text{\quad with } V(t)=
\begin{dcases}
0 & \text{for } t<0\\
V_0 & \text{for } t\ge0
\end{dcases}.
\end{equation} 
The zero of the potential is arbitrary.

In the following, we solve the Poisson-Nernst-Planck equation analytically, when possible, and numerically, with a flux-conservative method whose details are given in the Appendix. The results of our numerical scheme were successfully compared to those of constant-potential Lattice-Boltzmann electrokinetics \cite{Asta2019JCP}, where Lattice-Boltzmann is coupled with an iterative resolution of the Poisson equation.


\section{\texorpdfstring{FULLY SYMMETRIC CASE: $q_+=q_-$, $D_+=D_-$}{Fully symmetric case: q+ = q-, D+ = D-}}
\label{sec:sym}

We focus first on the case of a symmetric binary electrolyte, whose species have valences $q_-=q_+=q=1$, and diffusivities $D_+=D_-=D$. The problem of studying relaxation in this case has been previously addressed in \cite{Bazant2004} and \cite{Janssen2018}, with particular focus on the linear regime. For the sake of completeness, we re-obtain here some of the results from Ref.~\cite{Janssen2018} following a simpler approach.
In the linear regime, the relaxation is multi-exponential. We then seek a clear characterisation of the nonlinear regime, building on Ref.~\cite{Bazant2004}.

\subsection{Linear regime}
\label{sec:DynPlanSymSymLinear}

Linearizing \Eqsref{eq:PNP} around the initial densities $n_+^0=n_-^0=n_0$ and taking the difference of the two equations for the two species, yields, together with \Eqref{eq:poisson}, the Debye-Falkenhagen equation for the charge density $\rho$:
\begin{equation} 
\label{eq:DebyeFalkenhagen}
\pard{\rho}{t}=D \left( \pardd{\rho}{z} - \lD^{-2} \rho \right)\,.
\end{equation}
Here we defined the Debye length $\lD$ as a constant in terms of the initial density $n_0$ by $\lD^{-2} = 2 n_0 \beta e^2/(\varepsilon_r \varepsilon_0) = 8\pi\lB n_0$, where $\lB=\beta e^2/(4\pi\varepsilon_r \varepsilon_0)$ is the Bjerrum length.

\begin{table}
\begin{center}
{\def\arraystretch{1.9}\tabcolsep=10pt
\begin{tabular}{ l | c | c }
Observable & Symbol & Unit \\ \hline
\rule{-4pt}{5ex}
Time & $t,\, \tau_i$ & \(\displaystyle \frac{L \lD}{ D} \) \\
Inverse time & $s,\, s_i$ & $\displaystyle \frac{ D}{L \lD}$ \\
Distance & $z$ & $L$ \\
Volumic ion density & $\dens_\pm$ & $n_0$ \\
Volumic charge density & $\rho$ & $2en_0$ \\
Electric potential & $\phi,\, V$ & $\displaystyle \frac{1}{\beta e}$ \\
Electric field & $E$ & $\displaystyle\frac{1}{\beta e L}$\\
Surface density & $\sigma$ & $\displaystyle \frac{\permz\permr}{\beta e^2 L} = \frac{1}{4\pi\lB L}$
\end{tabular}
}
\end{center}
\caption{List of units of the nondimensional quantities used. Note that they do not form a coherent system of units: for instance, the units of distance, volumic density and surface density are not related by simple powers.
}
\label{tab:units}
\end{table}

We proceed by making the equations nondimensional, according to the mapping described in Table \ref{tab:units}.
The system is completely described by the two dimensionless parameters 
\begin{equation}
\epsilon=\frac{\lD}{L} \quad\quad \text{and}\quad\quad v=\beta e V_0\,, 
\end{equation}
that we will keep using throughout the rest of the paper. Note that $\epsilon$ can be, but need not be, a small quantity, while, as long as we deal with the linear regime, $v$ has to be much smaller than unity. 

In these units, \Eqref{eq:DebyeFalkenhagen} is readily rewritten as
\begin{equation} \label{eq:DF}
\epsilon \pard{\rho}{t} = \epsilon^2 \pardd{\rho}{z} - \rho
\end{equation}
and the Poisson equation~\eqref{eq:poisson} reads
\begin{equation} \label{eq:P}
-\epsilon^2 \pardd{\phi}{z}=\rho \,.
\end{equation}
The boundary condition in \Eqref{eq:bcflux} can be written in terms of the charge density upon linearization:
\begin{equation} \label{eq:bcFlux}
-\pard{\rho}{z}(\pm 1,t) - \pard{\phi}{z}(\pm 1,t) =0 \,.
\end{equation}
Finally, the boundary condition in \Eqref{eq:bcpotential} reads
\begin{equation} \label{eq:bcPotential}
\phi(\pm 1,t)=\pm V(t) \,.
\end{equation}

We make the \ansatz\ that the potential $\phi(z,t)$ and the electric charge density $\rho(z,t)$ relax to equilibrium as 
\begin{equation} \label{eq:ansatzphi}
\phi(z,t)=v\frac{\sinh\left(\frac{z}{\epsilon}\right)}{\sinh\left(\frac{1}{\epsilon}\right)}+v\sum_{i=0}^{\infty} b_i(z)\operatorname{e}^{s_i t} 
\end{equation}
\begin{equation} \label{eq:ansatzrho}
\rho(z,t)=-v\frac{\sinh\left(\frac{z}{\epsilon}\right)}{\sinh\left(\frac{1}{\epsilon}\right)}+v\sum_{i=0}^{\infty} B_i(z)\operatorname{e}^{s_i t} \,,
\end{equation}
for $t>0$. The time-independent terms correspond to the only solutions of the steady-state Debye-Falkenhagen equation allowed by symmetry: indeed, $\phi$ and $\rho$ must be odd with respect to $z$ and $v$. Additionally, by \Eqref{eq:P}, we have $B_i(z)=-\epsilon^2 b_i''(z)$. We suppose $s_i<0$, so that the characteristic relaxation times are $1/|s_i|$.
Substituting \Eqref{eq:ansatzrho} in \Eqref{eq:DF} and enforcing an odd charge density gives
\begin{equation} \label{eq:Bnofx}
B_i(z)=c_i \sinh\left( \frac{\sqrt{1+\epsilon s_i}}{\epsilon} z \right)\,.
\end{equation}
This expression can be integrated to get the corresponding $b_i(z)$. Fixing the gauge $\phi(0)=0$ and imposing boundary conditions \Eqref{eq:bcFlux}, one finds
\begin{equation}
b_i(z)=-c_i \left[ \frac{\sinh\left(\frac{\sqrt{1+\epsilon s_i}}{\epsilon}z\right)}{1+\epsilon s_i} + z \frac{s_i \cosh\left(\frac{\sqrt{1+\epsilon s_i}}{\epsilon}\right)}{\sqrt{1+\epsilon s_i}} \right]\,.
\label{eq:bnofx}
\end{equation}
At this point, \Eqref{eq:bcPotential} gives, for any $i$, either $c_i=0$ or
\begin{equation} \label{eq:poleseq}
1+s_i \sqrt{1+\epsilon s_i} \coth\left(\frac{\sqrt{1+\epsilon s_i}}{\epsilon}\right) = 0\,.
\end{equation}
This equation is exactly equivalent to the one solved in \cite{Janssen2018} to obtain relaxation times. An alternative solution, consistent with our notation, was independently derived in \cite{phd}. The modes characterizing the linear response can also be considered in the frequency-domain via the impedance of the cell (see \textit{e.g.} \cite{Barbero2005b,Barbero2008} for symmetric electrolytes).

\begin{figure*}

\includegraphics[width=0.96\textwidth]{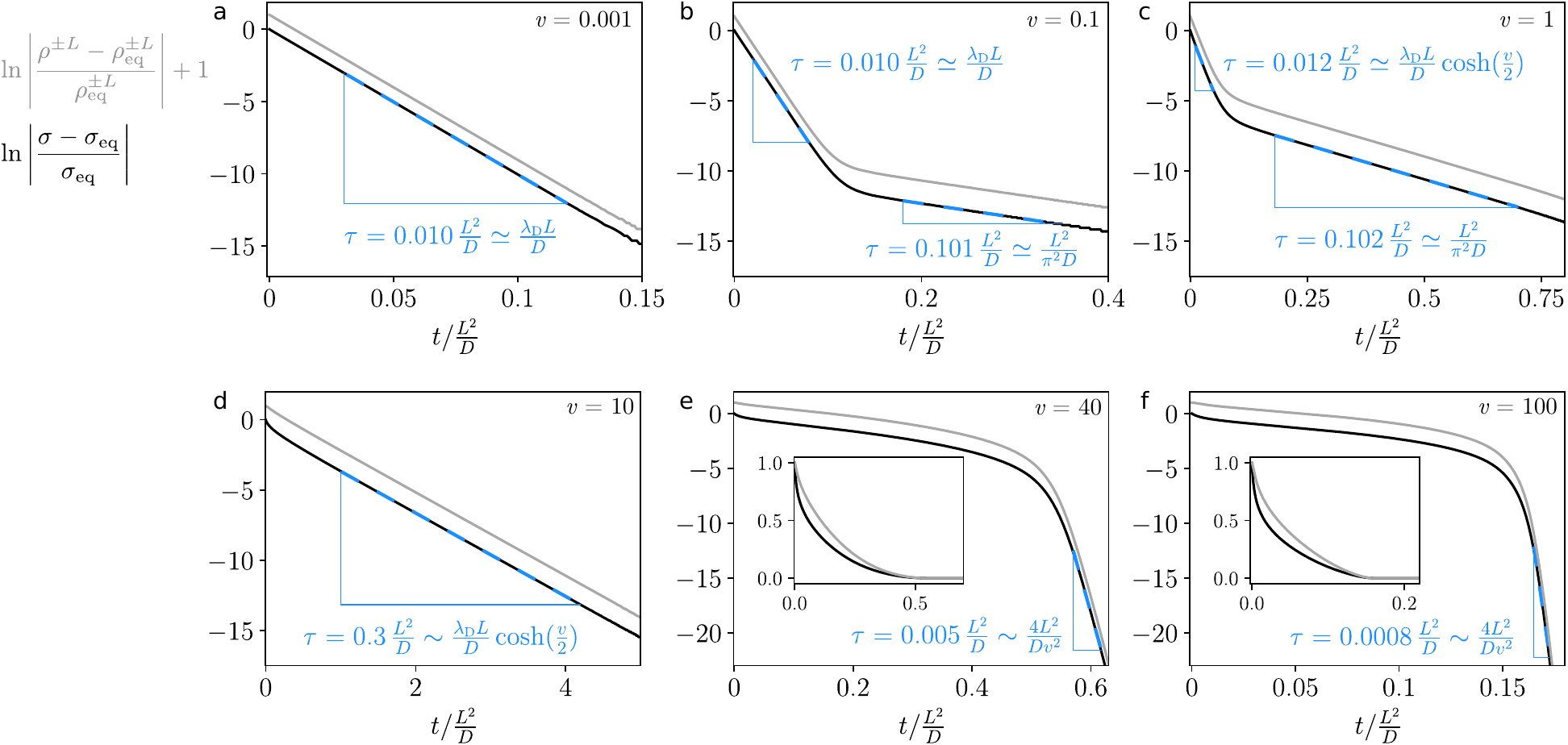}
\caption{Logarithmic plots of the quantities defined at the top left, where $\rho^{\pm L} = \rho(-L,t) = -\rho(L,t)$ is the charge density in the solution at contact with the electrode, $\sigma(t)$ is the surface charge density of the electrode, and $\rho^{\pm L}_{\mathrm{eq}}$ and $\sigma_{\mathrm{eq}}$ are their respective values at equilibrium. Here, $\epsilon=0.01$ and $v$ increases progressively, as indicated at the top right of each panel.
When the curves shown are linear, their slopes correspond to exponential relaxation rates. 
$s_0$ corresponds to a relaxation time $\lD L/D = 0.01\, L^2/D$ and $s'$ of $L^2/(\pi^2D) \simeq 0.10\, L^2/D$. 
At $v=0.001$ (a), only the double layer formation process is visible with its rate $s_0$ (unless too close to $t=0$, where other rates $s_i$ can play a role). At $v=0.1$ and $1$ (b-c), the first slope corresponds to the build-up of the double layer, while the second slope reflects its reorganization as the bulk is depleted from ions.
The system is in the purely nonlinear regime at $v=1$ and the double layer builds up in a time $\tau_\mathrm{PNL}\propto \cosh{(v/2)}$, as defined in \Eqref{eq:tauNLcosh} in Sec.~\ref{sec:DynPlanSymPurelyNL}. 
At $v=10$ (d) the system is partially depleted at equilibrium, as discussed in Sec.~\ref{sec:DynPlanSymFullyDepleted} (see Fig.~\ref{fig:DeplDuStat}c), but continues to relax with a time $\tau_\mathrm{PNL}$.
At $v=40$ and $100$ (e-f), the system is fully depleted (Sec.~\ref{sec:DynPlanSymFullyDepleted}): the early-time curves represent ion migration (the relaxation is not yet linear in time, but not perfectly exponential either); the late-time part represents relaxation of the counterionic double layers, in a time $\sim{\munen^2}/{D}$. Insets show $\vert (\rho^{\pm L}-\rho^{\pm L}_\mathrm{eq})/\rho^{\pm L}_\mathrm{eq} \vert$ and $\vert (\sigma-\sigma_\mathrm{eq})/\sigma_\mathrm{eq}\vert$ in linear scale.}
\label{fig:plotfit}
\end{figure*}

For $\epsilon = \lD/L \gg 1$, relaxation modes turn out to be of order $-\epsilon \pi^2 (i+1/2)^2$; the dominant (slowest) mode is $s_0\sim -\epsilon \pi^2/4$, corresponding to a dimensional relaxation time $\tau_0\sim 4L^2/(\pi^2 D)$. For $\epsilon = \lD/L \ll 1$, the dominant relaxation rate results exactly from the solution of the equation
\begin{equation} \label{eq:tanhZ}
    \tanh\left(\frac{Z}{\epsilon}\right)=-\frac{Z}{\epsilon}(Z^2-1),
\end{equation}
with $Z=\sqrt{1+\epsilon s_0}$. In physical units, this corresponds to a time 
\begin{equation}
\tau_0\simeq \frac{L\lD}{D} -\frac{\lD^2}{2D}\,.
\label{eq:tauBazant}
\end{equation}
The scaling $L\lD/D$ was pointed out in \cite{Bazant2004} and previously reported in \cite{MaCdonald1970,Kornyshev1977,Kornyshev1981}, while the second term represents the exact finite-double-layer correction as commented in \cite{Janssen2018,phd,Asta2019JCP}. The time scale from \Eqref{eq:tauBazant} is visible in Fig.~\ref{fig:plotfit}a, which shows numerical solutions of the ion density at contact with the electrode $\rho(-1,t)$ and of the electrode surface charge density $\sigma(t)$.

Note that assuming a monoexponential relaxation in \Eqsref{eq:ansatzphi} and~\eqref{eq:ansatzrho}, for $\epsilon\ll 1$, still leads to a surprisingly good approximation for the dominant relaxation rate \cite{phd}. 

The exact relaxation profile can be obtained by an analysis of \Eqsref{eq:DF}--\eqref{eq:bcPotential} in the Laplace domain \cite{Janssen2018,phd}, allowing to identify $c_i$ and write
\begin{equation} \label{eq:Resrho}
    \rho(z,t)=-v \frac{\sinh{\frac{z}{\epsilon}}}{\sinh{\frac{1}{\epsilon}}} + v \sum_{n=0}^{\infty}\frac{2 (1+\epsilon s_i)}{3-s_i^2} \frac{\sinh{\frac{z \sqrt{1+\epsilon s_i}}{\epsilon}}}{\sinh{\frac{\sqrt{1+\epsilon s_i}}{\epsilon}}} \operatorname{e}^{s_i t} \,.
    \end{equation}
A cumbersome computation shows that this equation, that we report for its perhaps more readable shape, is equivalent to Eq.~(40) in \cite{Janssen2018}.

As proposed in \cite{Bazant2004}, when $\epsilon=\lD/L \ll 1$, it is possible to establish an analogy between the EDLC and a simple RC circuit of the first order. We summarize this analogy here because it will prove useful in the following. Each double layer can be identified with a planar capacitor, opposing the charge of the electrode with an equal and opposite charge in the solution, distributed over a distance $\sim\lD$. Its capacitance per unit surface reads
\begin{equation}
C=\frac{\permz\permr}{\lD}\,.
\label{eq:Cplan}
\end{equation} 
Along the same spirit, the bulk, of length $2L-2\lD$, carries an electric resistance according to Drude's model of conduction. For some applied potential difference $\Delta V$, the current density can be expressed as
\begin{equation}
j = 2n_0 e^2 \beta D \frac{\Delta V}{2L-2\lD}\,.
\end{equation} 
The electric resistance of the bulk times unit surface (\ie~the inverse conductance per unit surface) is then
\begin{equation}
R=\frac{\Delta V}{j}= \frac{(L-\lD)}{n_0 e^2 \beta D}\,.
\label{eq:RNL}
\end{equation}
The circuit results as a series of a capacitance $C$, a resistance $R$ and a second capacitance $C$. Being the equivalent capacitance $C/2$, the characteristic time of the circuit turns out to be
\begin{equation}
\tau=R\frac{C}{2}= \frac{L-\lD}{2 n_0 \frac{\beta e^2}{\permz\permr}\lD D} = \frac{L \lD}{D} - \frac{\lD^2}{D}\,.
\label{eq:tauplan} 
\end{equation} 

This simple circuit analogy reproduces the dominant relaxation rate from \Eqref{eq:tauBazant} -- the correction of order $\lD^2/D$ is not exact, but has the right scaling.

\subsection{Depletion}
\label{sec:DynPlanSymDepletion}

In the nonlinear regime, it is impossible to write a Debye-Falkenhagen equation for the charge density and it is harder to make analytical predictions on the dynamics. The first clearly visible nonlinear effect is depletion.
We define depletion as the decrease in the bulk population from the initial value $n_0$ to some smaller final value. As $v$ increases, more ions are attracted to the oppositely charged electrode than are repelled by the like-charged one. Since the total number of ions (the integral of the ion densities) must be conserved, this calls for a decrease in the bulk density. Indeed, in the symmetric electrolyte case, \Eqref{eq:PNP} 
is not invariant under the transformation ($n_\pm-n_0 \to n_0-n_\pm,\ z\to-z,\ \bar{q}_\pm\to\bar{q}_\mp$) -- only valid for asymptotically small voltages -- meaning that exchanging the two ion species cannot be reduced to a simple flip of their excess densities about their initial values. 
More intuitively, if the applied potential is sufficiently high, all positive ions condense on the negative electrode and all negative ions condense on the positive electrode: in this extreme case the bulk of the capacitor is basically empty  and we have full depletion, as described in Sec.~\ref{sec:DynPlanSymFullyDepleted}. 
In this section, we focus on depletion in the linear regime $v\ll 1$, as a weakly nonlinear effect. The phenomenon has a clear signature on the relaxation dynamics and it introduces a purely diffusive time scale, that, at least in the $\epsilon\ll 1$ range, is slower than the dominant relaxation time $\tau_0$ from \Eqref{eq:tauBazant}, related to the electric double layer formation. 

To quantify depletion, we introduce a quantity, inspired by the Dukhin number \cite{BocquetRev2010, Lyklema1995}, that quantifies the fraction of ions that remains in the bulk after relaxation:
\begin{equation}
\Du_n=\frac{n_+(0,\infty)}{n_0}=\frac{n_-(0,\infty)}{n_0}\,.
\end{equation} 
This quantity is 1 when there is no depletion and 0 when ions are completely absent at the center of the capacitor. 
We say that the system is weakly depleted when $0.9<\Du_n<1$, depleted when $\Du_n<0.9$ and fully depleted when $\Du_n\simeq0$. 

\begin{figure}
\includegraphics[width=0.9\columnwidth]{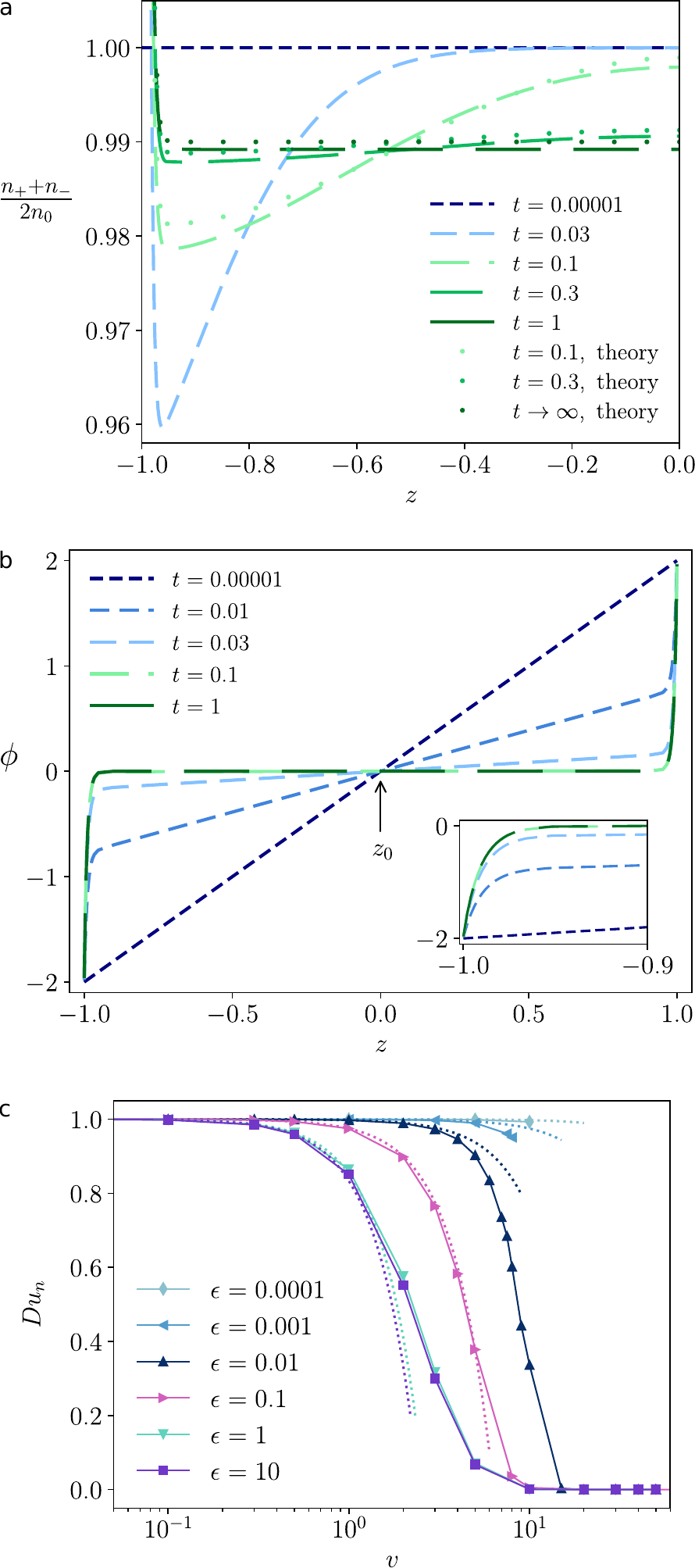}
\caption{(a) Mass as a function of $z$ (in units of Table \ref{tab:units}), at different times (here in units of ${L^2}/{D}$), for $\epsilon=0.01$ and $v=2$. In these units, the electric double layer formation occurs on a time scale $\sim 0.02$. At much shorter times (dark blue) the system has not moved yet. At later times (lighter greens) the double layer has formed already and the mass diffusion process manifests itself, with a relaxation time scale $\tau'\simeq0.1$: the sinusoid of \Eqref{eq:n2ndorder} is visible. At times much larger than $\tau'$ (dark green) the system is at equilibrium at a new value of bulk density, predicted by \Eqref{eq:n2ndorderbulk}. Dotted lines indicate the analytical predictions for later times as per \Eqref{eq:n2ndorder}, where the parameter $A$ was set to 0.006 for all curves.
(b)~Potential as a function of $z$, at different times (here in units of ${L^2}/{D}$), for $\epsilon=0.01$ and $v=2$. The inset is a zoom close to the left electrode, while the arrow indicates the point of null potential (see Sec.~\ref{sec:partiallyasym:pnl}), which in the present case is $z_0=0$.
(c)~Points show $\Du_n$ as a function of $v$, for different values of $\epsilon$, as extracted from numerical solutions of the Poisson-Nernst-Planck equation. Smaller $\Du_n$ represent stronger depletion. Solid lines are a guide to the eye. The curves at $\epsilon=100$ and $1000$, not shown here, coincide with the $\epsilon=10$ curve. In dotted lines, the prediction from \Eqref{eq:n2ndorderbulk}.
}
\label{fig:DeplDuStat}
\end{figure}
 
We perform a weakly nonlinear analysis of the Poisson-Nernst-Planck equations. Indeed, in the linear regime, the magnitude of the corrections to $\rho$ and $\phi$ are orders of magnitude below the dominant process (the formation of the two double layers), but a new depletion-related time scale appears, slower than the previously mentioned ones. We expand ionic densities, charge density and potential -- that we already know up to linear order in $v$ -- and consider small terms of order $v^2$ and $v^3$. Since $\rho$ and $\phi$ must be odd functions of $v$, we write
\begin{align}
n_\pm(z,t)&=n_\pm^{(0)}+ v n_\pm^{(1)}(z,t) + v^2 n_\pm^{(2)}(z,t) + \nonumber \\ & \quad \quad + v^3 n_\pm^{(3)}(z,t)+ O(v^4)
\\ 
\rho(z,t)&= v \rho^{(1)}(z,t) + v^3 \rho^{(3)}(z,t) + O(v^5)
\\
\phi(z,t)&= v \phi^{(1)}(z,t) + v^3 \phi^{(3)}(z,t) + O(v^5)\,,
\end{align} 
where all quantities are dimensionless and $\rho=\frac{n_+ - n_-}{2}$ according to the units in Table \ref{tab:units}.

Introducing the above expansions in the Poisson-Nernst-Planck equation~\eqref{eq:PNP} and collecting $\mathcal{O}(v^0)$ terms, we find that at order 0, $n_\pm^{(0)}=n_0$. From $\mathcal{O}(v^1)$ terms, we find that at linear order, $\phi^{(1)}(z,t)$ is given by \Eqref{eq:ansatzphi} and $\rho^{(1)}(z,t) =\pm n_\pm^{(1)}(z,t)$ by \Eqref{eq:ansatzrho}. Then, introducing these results into the equation for $\mathcal{O}(v^2)$ terms, we obtain
\begin{multline} \label{eq:PNP2ndorder}
\pard{n_\pm^{(2)}}{t}=\epsilon \pardd{n_\pm^{(2)}}{z} - \frac{1}{\epsilon}\frac{\cosh\frac{2z}{\epsilon}}{\sinh^2\frac{1}{\epsilon}} + \\
\sum_{i=0}^\infty \operatorname{e}^{s_i t} f_i(z) + \sum_{i=0}^\infty \sum_{j=0}^\infty \operatorname{e}^{(s_i+s_j) t} g_{ij}(z) \, 
\end{multline}
where the exact form of the functions $f_i$ and $g_{ij}$, which can be expressed in terms of the $B_i$ and $b_i$ given in \Eqsref{eq:Bnofx} and \eqref{eq:bnofx}, are in fact not relevant for the present analysis. Indeed, we look for relaxation modes slower than the purely linear one ($s_0$, as determined by \Eqref{eq:poleseq}) for large $t$, so that the last two terms can be neglected. Imposing mass conservation in the form $\int_{-1}^1 n_\pm^{(2)}(z,t)\diff z = 0$ and requiring $n_\pm^{(2)}$ to be even in $z$ to respect symmetry, one obtains
\begin{multline} \label{eq:n2ndorder}
n_\pm^{(2)}(z,t)= \\ -\frac{\epsilon}{4}\coth\frac{1}{\epsilon} + \frac{\cosh\frac{2z}{\epsilon}}{4 \sinh^2 \frac{1}{\epsilon}}+ A \cos(\pi z) \operatorname{e}^{s' t} +\ o(\operatorname{e}^{s' t})\,,
\end{multline} 
with $A$ constant and $s'=-\pi^2\epsilon$. It can be checked that the first two terms in the r.h.s. correspond to a particular solution of \Eqref{eq:PNP2ndorder}, while the last one is the standard solution for the corresponding homogeneous diffusion equation satisfying the symmetry and boundary conditions. In physical units, this corresponds to a diffusive time scale 
\begin{equation}
\tau'=\frac{L^2}{\pi^2 D}\,.
\label{eq:taudepl}
\end{equation}
This second order correction to the ionic density represents depletion in the $\epsilon\ll 1$ regime (where $s'$ is indeed slower than $s_0$ and~\eqref{eq:n2ndorder} makes sense). In this case, the equilibrium density in the bulk (\ie~at $|z|\ll 1-\epsilon$), which in our units is nothing but $\Du_n$, reads to second order
\begin{equation}
n_\pm(z,\infty) \simeq n_\pm(0,\infty) = Du_n = 1 - v^2 \frac{\epsilon}{4}\coth\frac{1}{\epsilon}\,.
\label{eq:n2ndorderbulk}
\end{equation}

Fig.~\ref{fig:DeplDuStat}a compares numerical results with our analytical approximation, showing that \Eqref{eq:n2ndorder} well predicts late-time density profiles as a function of $z$ and $t$. In this figure, similar for this symmetric case to Figs.~8a, 9 and 10 (top right) or Ref.~\citenum{Bazant2004}, the time-damped cosine shape is visible outside the double layer (light green curves). The depletion phenomenon emerges as a (neutral) mass diffusion of both species from the center of the EDLC toward the boundaries of the bulk region: ions to constitute the double layer are initially recruited from the regions close to the electrode, leaving a non-uniform mass distribution in the bulk (light blue curves in Fig.~\ref{fig:DeplDuStat}a). The corresponding potential across the capacitor is shown as a function of time in Fig.~\ref{fig:DeplDuStat}b.

The mass imbalance then evens out with a purely diffusive process on a length scale $L$ rather than $2L$ (ions move from the center to the borders and not from one border to another). This explains the absence of the diffusive mode $4L^2/(\pi^2 D)$ for symmetric electrolytes, otherwise legitimate. As shown in Fig.~\ref{fig:DeplDuStat}c, \Eqref{eq:n2ndorderbulk} predicts values of $\Du_n$ in the weakly nonlinear regime. 

Finally, the rate $s'$ can be proven to emerge naturally also in $\rho^{(3)}(z,t)$ and $\phi^{(3)}(z,t)$. This is why the relaxation rate $s'$ is visible in late-time profiles of $\rho(-1,t)$ and $\sigma(t)$, plotted in Fig.~\ref{fig:plotfit}b-c, for $\epsilon=0.01$.

To summarize, the linear regime features two processes: double layer formation and, at higher order, depletion. The double layer forms at a dominant rate $s_0$. During this process the bulk stays electroneutral, but becomes inhomogeneous as for mass density: the bulk region closer to the electrodes becomes less populated than the central region. A slower process, at least for $\epsilon<\pi^{-2}\simeq0.1$, then onsets with rate $s'$: this is a mere diffusion of neutral excess mass within the bulk, with positive and negative ions moving together from the center toward the double layer boundaries. The diffusive depletion process is asymptotically absent at $v\to0$, but its relative importance compared to the double layer formation process grows with $v$ (see the upwards shift of late-time  lines in Fig.~\ref{fig:plotfit}b and c). 

A last word of caution concerns the quantity $\lD$ defined after \Eqref{eq:DebyeFalkenhagen}, and its dimensionless equivalent $\epsilon=\lD/L$.  This `reservoir' Debye length is defined in terms of the \textit{initial} concentration $n_0$, but it is important to note that the physically relevant Debye length in the capacitor, defined for instance in terms of the depleted mid-point salt concentration, can be much larger than $\lD$.

\subsection{Purely nonlinear regime}
\label{sec:DynPlanSymPurelyNL}

Fig.~\ref{fig:DeplDuStat}c shows that for $\epsilon\gg1$ depletion occurs as soon as $v\gtrsim 1$, so it is impossible to tell apart the purely nonlinear effects, the ones that would appear even in a grand-canonical formulation of the problem, 
from the strictly canonical effects of depletion. In this case, we say that no purely nonlinear regime exists: upon increasing $v$, the system goes directly from the linear regime to a depleted nonlinear regime, that we will describe in Sec.~\ref{sec:DynPlanSymFullyDepleted}.
On the contrary, for $\epsilon\ll1$, more ions are available and depletion is only observed at voltages significantly higher than 1: this makes purely nonlinear effects visible at intermediate voltages, starting from $v\simeq1$. These consist in a clear asymmetry, for a given species, between left and right double layer at equilibrium; however the total number of ions is much larger than the number of ions involved in the double layers, so that the bulk population stays almost unaffected (no depletion). 

In this purely nonlinear regime, the linear analysis of the relaxation times from Sec.~\ref{sec:DynPlanSymSymLinear} is not valid and the double layer formation does not happen anymore at a rate $s_0$. However, it is possible to understand the change in the rate of formation of the electric double layer by using the Grahame equation \cite{Hunter,Andelman2010}. In the units of Table \ref{tab:units}, this equation reads
\begin{equation}
|\sigma(\infty)|=\frac{2}{\epsilon} \sinh\left(\frac{v}{2}\right)\,
\label{eq:Grahame}
\end{equation} 
and is exact in the $\epsilon\to 0$ limit, even for $v>1$. 
The following differential capacitance emerges: 
\begin{equation}
\dep{|\sigma(\infty)|}{v}=\frac{1}{\epsilon}\cosh\left( \frac{v}{2} \right) \,.
\end{equation}
If this capacitance is used in the circuit model, replacing the one of \Eqref{eq:Cplan}, the relaxation time in \Eqref{eq:tauplan} becomes
\begin{equation}
\tau_{\mathrm{PNL}}=\frac{\lD L}{D} \cosh\left( \frac{v}{2} \right)\,,
\label{eq:tauNLcosh}
\end{equation}
where PNL stands for purely nonlinear. 
The same scaling appears already in \cite{Macdonald1954,Bazant2004} and is in quantitative agreement with our numerical calculations (see Fig.~\ref{fig:plotfit}d and Fig.~\ref{figm:tausym} of \cite{mainpaper}, where time scales extracted from numerical simulations are summarized).

The same Grahame equation~\eqref{eq:Grahame} allows to estimate for what values of $v$ and $\epsilon$ depletion starts to become relevant, determining a boundary between the purely nonlinear and the depleted regimes for small $\epsilon$.
We introduce the Dukhin number 
\begin{equation}
\Du=\frac{|\sigma(t=\infty)|}{2 n_0 L}\,,
\label{eq:Dudef}
\end{equation} 
giving information on the maximum surface charge neutralisable by the ions \cite{BocquetRev2010,Lyklema1995}. We define a system as depleted when $\Du\simeq 0.1$, in a manner conceptually equivalent to the previously employed $Du_n<0.9$.
Using \Eqref{eq:Grahame}, this condition reads
\begin{equation}
\Du=2\epsilon \sinh\left(\frac{v}{2}\right)\simeq 0.1 \,.
\label{eq:DuConditionNL}
\end{equation} 
This relation can be used to identify the limit between purely nonlinear and fully depleted nonlinear regime in a ($v$, $\epsilon$) diagram, as we do in Fig.~\ref{figm:phasesym} of \cite{mainpaper}, showing that it is consistent with numerical results.

\subsection{Fully depleted nonlinear regimes}
\label{sec:DynPlanSymFullyDepleted}

Upon increasing $v$ at $\epsilon\ll 1$, depletion becomes more and more important. The Dukhin number $Du_n$ correspondingly decreases (Fig.~\ref{fig:DeplDuStat}c) and the late-time line representing depletion in Fig.~\ref{fig:plotfit}b-c gradually shifts upwards. At the same time, still for $\epsilon\ll 1$, the time scale related to charging increases as per \Eqref{eq:tauNLcosh}. In terms of Fig.~\ref{fig:plotfit}, the slope of the early-time curve gradually decreases as $v$ increases. Eventually the two processes (double layer charging and depletion) become indistinguishable and the system gradually enters the fully-depleted nonlinear regime. A similar thing happens at $\epsilon\gg 1$, where depletion coincides with nonlinearity, as discussed above. The transition to this new fully depleted regime is not abrupt and defines a band on the $(\epsilon,v)$ diagram where depletion is only partial (say, $0.1<\Du_n<0.9$ or $\Du\sim1$). Such band is relatively narrow: Fig.~\ref{fig:DeplDuStat}c shows that the passage from $\Du_n=1$ to $\Du_n=0$ happens in no more than a decade of $v$, for all probed values of $\epsilon$ (see also Fig.~3 of \cite{mainpaper}). 

For parameters at which depletion is fully achieved, the `bulk' ionic concentration at equilibrium $n_\pm(0,\infty)$ can be orders of magnitude lower than the initial one $n_0$. In this case, all positive ions are concentrated in proximity of the negative electrodes, and vice versa. At long times, the double layer contains only counterions, so that the equilibrium distribution is governed by a sort of Gouy-Chapman length for either subsystem made of one electrode and its counterions. Such subsystem is in general not electroneutral (it cannot be so if $Du>1$): the Poisson equation shows that its Gouy-Chapman length reads, in physical units, $\munen=(2\pi q \lB \sigma_{\mathrm{res}})^{-1}$, where $\sigma_{\mathrm{res}} = |\sigma(\infty)| - 2n_0 L$ is the unscreened residual part of the electrodes' surface charge (where `nen' stands for non-electroneutral and `res' for residual). After the counterionic double layers have formed, it is reasonable to think that the last dynamic phenomenon to happen is a rearrangement of each double layer on the smallest length scale available, $\munen$; this would correspond to a relaxation time ${\munen^2}/{D}$. 
In the following we show how to predict $\sigma(\infty)$ and thus $\munen$, and what dynamics leads to the formation of the counterionic double layers.

\begin{figure}[b]
\centering
\includegraphics[width=0.8\columnwidth]{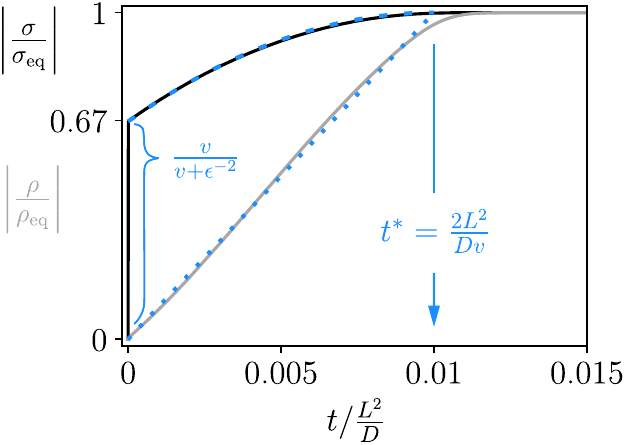}
\caption{Ratios of $|\sigma(t)|$ (black) and $|\rho(\pm L,t)|$ (gray) versus their equilibrium values, as a function of time. Here, $\epsilon=0.1$ and $v=200$. 
The blue dashed lines represent the ideal linear evolution of $\rho$ and the theoretical prediction for $\sigma$ from \Eqsref{eq:tstar}--\eqref{eq:parabolicsigma}. 
}
\label{fig:NLunscreened}
\end{figure}

We take a look at the large voltage asymptotics to simplify the problem. We suppose $v$ to be so high that even when equilibrium is reached, ions have negligible impact on the potential profile between the electrodes: the electric field, at any time, is approximately $V_0/L$ everywhere, except possibly within a very small distance from the electrodes. We will verify later when this assumption is correct. As soon as the power source is switched on, ions, of valence $q=1$, move by electric drag with a constant velocity $\nu =\beta D e V_0/L = vD/L$, directed toward the oppositely charged electrode. Once the ions reach the electrode, we suppose the latter to be so highly charged that a very thin double layer is formed, of negligible size compared to $L$. If the number of ions reaching the double layer is somehow proportional to the density at contact, we expect $\dens_\pm(\mp L,t)$, $\rho(\pm L,t)$ and, more rigorously, the double layer charge $\int_\mathrm{EDL} \rho(z,t)\,\diff z$ to grow linearly with time, approaching their final value in a time $t^*$. Such time is easily computed, in physical units, as the time needed for the furthermost ion to reach the oppositely charged electrode, a distance $2L$ away: 
\begin{equation}
t^*=\frac{2L}{\nu}=\frac{2L^2}{ D v}\,.
\label{eq:tstar}
\end{equation}
Fig.~\ref{fig:NLunscreened} confirms that $\rho(-L,t)$ is linear in time from $t=0$ until $t^*$.

This allows to compute the time evolution of $\sigma$. By definition, integrating the electric field to get the electric potential across a half-capacitor (from $-L$ to 0) for $0<t<t^*$, one must obtain the potential difference $V_0$. In physical units:
\begin{widetext}
\begin{equation}
V_0 = \frac{e|\sigma(t)|}{\permz\permr}L - \frac{e}{\permz\permr}\int_{-L}^0\diff z \int_{-L}^z \diff z' \underbrace{\left( 2 n_0 L \frac{t}{t^*}\delta(z'+L) + n_0 h(z',t) \right)}_{\rho(z',t)}\,
\label{eq:parabolicsigma0}
\end{equation}
\end{widetext}
with
\begin{equation}
h(z',t)= 
\begin{dcases}
I_{(-L,\, -L+\nu t)}(z')\quad &\text{if}\quad 0<t<\frac{t^*}{2} \\
I_{(-L,\, L - \nu t)}(z')\quad &\text{if}\quad \frac{t^*}{2}<t<t^*
\end{dcases}\,,
\label{eq:hoft}
\end{equation}
where $I_{(z_1,\,z_2)}(z')$ is the gate function, equal to 1 if $z'\in (z_1,z_2)$ and to $0$ otherwise. In \Eqref{eq:parabolicsigma0}, the inner integral represents the electric field in $z$ due to the ions between $-L$ and $z$. The first term in parenthesis represents positive ions that have adhered to the wall in $z=-L$ at time $t$; we make the drastic choice of a $\delta$-function distribution, but this is not crucial, as long as ions stay confined within a length $\ll L$. The second term in parenthesis is the charge density at point $z'$ in the rest of the solution, whose functional shape \Eqref{eq:hoft} is determined by the following observation. For $t<t^*/2$, the nonzero charge density is due to negative ions leaving the region $(-L,0)$ at velocity $\nu$ from left to right, while the concentration of positive ions stays constant in this half of the capacitor (though not at $-L$); for $t>t^*/2$ all negative ions have left the region and the last positive ions approach the negative electrode with velocity $\nu$ from right to left. Solving \Eqref{eq:parabolicsigma0} for $0<t<t^*$ gives, in dimensionless units,
\begin{equation}
|\sigma(t)|=v+ \frac{1}{\epsilon^2}\left( \frac{2t}{t^*} - \frac{t^2}{{t^*}^2} \right)\,.
\label{eq:parabolicsigma}
\end{equation}  
The electrodes' charge when the double layer formation is concluded ($t=t^*$) is then $|\sigma(t^*)|=v+\epsilon^{-2}$. This is of straightforward interpretation, since, in our units of surface charge density, $2n_0 L$ reads $\epsilon^{-2}$: the surface charge developed by the electrodes to maintain a field $v$ across the EDLC is indeed $v+\epsilon^{-2}$, because exactly $\epsilon^{-2}$ of its charge is screened by the counterions stuck at the electrode. Agreement of \Eqref{eq:parabolicsigma} with numerical data is shown in Fig.~\ref{fig:NLunscreened}. This parabolic time-dependence of the surface charge density is also consistent with the linear current derived in Ref.~\cite{Beunis2008} for the fully symmetric case considered in this section.

For times $t>t^*$, we expect to observe the fast relaxation mentioned at the beginning of this section, on a time scale ${\munen^2}/{D}$. Now, the residual surface charge to be used in $\munen$ is nothing but $\sigma_{\mathrm{res}} \simeq |\sigma({t^*})|-\epsilon^{-2} = v$.
This corresponds to a relaxation time 
\begin{equation}
\tau=\frac{4L^2}{D v^2}\,,
\label{eq:tauFDNL}
\end{equation}
which is indeed observed at sufficiently high voltages, for all values of $\epsilon$ (see Fig.~\ref{figm:tausym} of \cite{mainpaper}). This exponential relaxation and corresponding time scale were not reported in Ref.~\cite{Beunis2008}. We note that in the latter reference, the authors argued for a power-law current with exponent $-3/4$, which seem supported by the experimental results of Ref.~\cite{Beunis2007}. Here, we did not attempt to give scaling laws: the system transitions to the unscreened regime in a complex fashion, even more complex for asymmetric electrolytes (see Secs.~\ref{sec:asymsym} and~\ref{sec:fullyasym}), and, at least within the range of parameters that we can probe, it is hard to find scalings that work for the whole regime or to neatly separate sub-regimes.

We made so far the crucial assumption that ions do not affect the linear potential profile through almost all the capacitor. A necessary condition for this to happen is that the electrodes' surface charge be larger than the integrated density of ions, meaning ${|\sigma(t)|}/(2n_0 L)>1$. Imposing this to \Eqref{eq:parabolicsigma}, we obtain
\begin{equation}
v>\frac{1}{\epsilon^2}\,.
\label{eq:veps2}
\end{equation}
This defines the unscreened fully depleted nonlinear regime and was used to delimit such a region in Fig.~3 of \cite{mainpaper}.

There is a transition region from the purely nonlinear regime to the (unscreened) fully depleted regime we just described: the transition region is the area comprised approximately between the curve \eqref{eq:DuConditionNL} and the curve $v=1/\epsilon^{2}$ (see Fig.~\ref{figm:phasesym} of \cite{mainpaper}). We call this the partially screened fully depleted nonlinear regime. In such region, the relaxation process gradually changes from exponential, with a well defined relaxation time given by \Eqref{eq:tauNLcosh}, to linear in time, as in \Eqref{eq:tstar}, due to depletion. At equilibrium, the electric field generated by the electrodes is reduced (partially screened) to a fraction of $v$ in the fully depleted zone at the center of the capacitor: double layers are indeed sufficiently populated to screen a non negligible fraction of it. In addition, the electric field, as a function of $z$, changes with time, as more and more ions reach the electrodes. As a consequence, analytical examination is hard and the functional time dependence of relaxation processes had to be retrieved numerically. Numerical results (Fig.~\ref{fig:plotfit}e-f) show anyway the presence of a late-time relaxation on a fast scale $\sim{\munen^2}/{D}$, that, as $v$ increases, converges to the one \eqref{eq:tauFDNL} predicted for the unscreened fully-depleted nonlinear regime (see Fig.~2 of \cite{mainpaper}).

Results from this section are summarized in Fig.~\ref{figm:phasesym} of \cite{mainpaper}.


\section{\texorpdfstring{PARTIALLY ASYMMETRIC CASE: $q_+\neq q_-$, $D_+=D_-$}{Partially asymmetric case: q+ != q-, D+ != D-}}
\label{sec:asymsym}

\begin{figure*}
\centering
\includegraphics[width=0.99\textwidth]{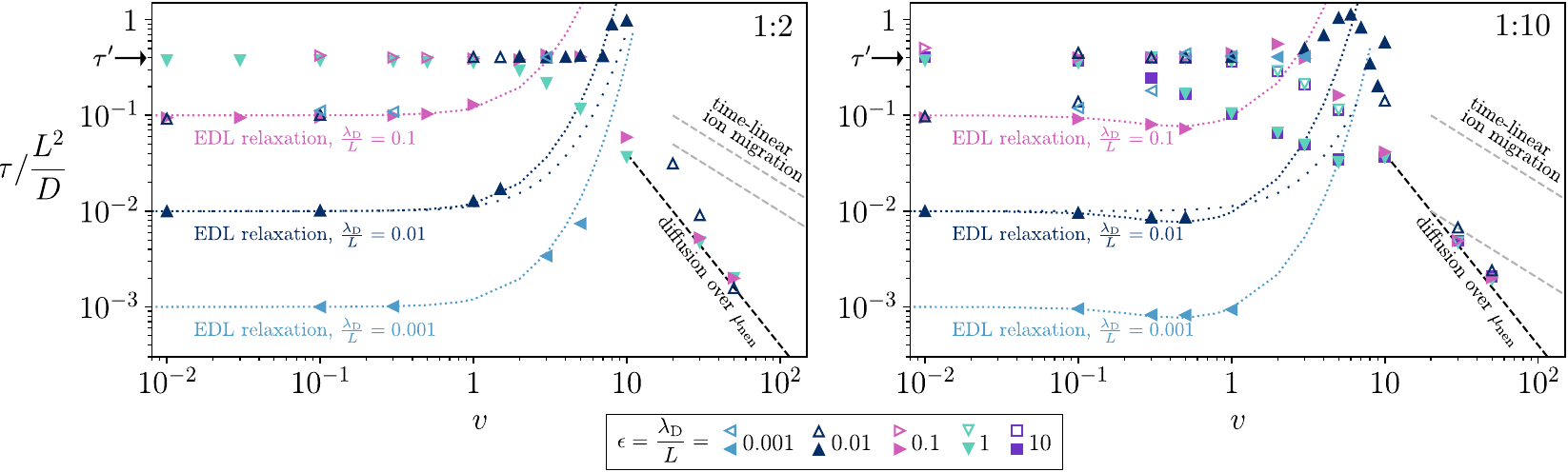}
\caption{Relaxation times $\tau$ as a function of $v$, for the $1:2$ case (left) and the $1:10$ case (right); different colors correspond to different values of $\epsilon = \lD/L$. Times are extracted from linear fits in logarithmic scale of $\sigma$, similar to what shown in Fig.~\ref{fig:plotfit} for the symmetric case. When two exponential relaxation times are visible from numerical data, one at short times and one at long times, they are both represented here (if the same $v$ and $\epsilon$ correspond to two symbols of the same color, the solid one represents early-time and the empty one late-time relaxation). The dense dotted lines represent the purely nonlinear time obtained from \Eqsref{eq:Ctotqq} and~\eqref{eq:RNLqq} and preceding ones; for comparison, we also plot the time given by \Eqref{eq:tauNLcosh}, relevant for the symmetric case and shown with sparser dots for $\epsilon=0.01$. The gray and black lines represent the times given by \Eqsref{eq:tstarqq} and~\eqref{eq:tauFDNLqq}, respectively.
}
\label{fig:NLtimescales_qq}
\end{figure*}

\subsection{Linear regime and depletion}

We now tackle the case of asymmetric valences: $q_+\neq q_-$. This requires a redefinition of our proxy for ion concentration $\lD$, so that in the linear regime this equals the Debye length \cite{Hunter}:
\begin{equation}
    \lD^{-2} = 4\pi\lB(q_+^2 n_+^0 + q_-^2 n_-^0)\,,
\end{equation}
where $n_+^0$ and $n_-^0$ are the initial concentrations of positive and negative ions. 

The linear regime exhibits the same dynamics as the symmetric valence case, as shown by a linearization of the Poisson-Nernst-Plank equation~\eqref{eq:PNP}. For $v>0$, depletion occurs analogously to Sec.~\ref{sec:DynPlanSymDepletion}, with one difference: with asymmetric valences, there is no reason to expect $n_\pm^{(2)}(z,t)$ to be even functions of $z$, nor to have any particular symmetry. This invalidates the reason why depletion-related diffusion took place on a length $L$ rather than $2L$ in a system with equal valences (\Eqref{eq:taudepl}). In an asymmetric-valence system, the relaxation mode corresponding to a time 
\begin{equation}
    \tau'= \frac{(2L)^2}{\pi^2 D} 
\end{equation} 
is then permitted. This characteristic time for depletion is observed numerically (see Fig.~\ref{fig:NLtimescales_qq}) at long times; for some parameter values (\eg~1:10 case, $\lD/L=0.01, v=0.1$), we observe however that after the end of the double layer relaxation and before the onset of this slower depletion mode, the faster depletion mode ${L^2}/({\pi^2D})$ is still visible.

\subsection{Purely nonlinear regime} 
\label{sec:partiallyasym:pnl}

The nonlinear regime also deviates from the symmetric-valence case: this is shown again in Fig.~\ref{fig:NLtimescales_qq} (with numerical results shown as symbols), that has to be compared to Fig.~2 in \cite{mainpaper}. The exponential increase of the relaxation time with the applied voltage at small $\epsilon$, which in the symmetric-valence case is explained by \Eqref{eq:tauNLcosh}, is not valid anymore in the general $q_+:q_-$ case. This is particularly evident in the 1:10 case (Fig.~\ref{fig:NLtimescales_qq}, right panel), where the relaxation time appears to be non-monotonic with $v$: an initial slight decrease, absent in the 1:1 case, is followed by an increase, steeper than in the 1:1 case. An analytical estimate for this curve can be obtained by a procedure analogous to the one leading to \Eqref{eq:tauNLcosh}, making use  of the simple RC circuit analogy. In the asymmetric valence situation, the Grahame equation~\eqref{eq:Grahame}, used to compute the capacitance, can be rewritten as per Ref.~\cite{Grahame1953}. The equilibrium (infinite time) charge density of the negative electrode, in the limit where the two double layers are completely separated ($\epsilon\to0$) reads:
\begin{equation}
|\sigma(\infty)|=\sqrt{\frac{n_0}{2\pi\lB}}\left( q_- \e^{-q_+\psi_-}+q_+ \e^{q_-\psi_-} - (q_- +q_+)\right)^{1/2}\,,
\label{eq:GrahameMinus}
\end{equation} 
where $\psi_- =\phi(-L)-\phi(0)<0$ is the potential of the negative electrode compared to the neutral bulk, in units of $(\beta e)^{-1}$, and $n_0=n_+^0/q_-=n_-^0/q_+$ is the salt concentration. If the potential on the right electrode is $\psi_+=\psi_- + 2v>0$, the surface charge on the positive electrode reads:
\begin{multline}
\hspace{-0.5cm} |\sigma(\infty)|=\sqrt{\frac{n_0}{2\pi\lB}} \\
\left( q_+ \e^{q_-(\psi_- +2v)} + q_- \e^{-q_+(\psi_- +2v)} - (q_+ +q_-)\, \right)^{1/2}\,.
\label{eq:GrahamePlus}
\end{multline} 
This corresponds to \Eqref{eq:GrahameMinus}, with the changes $q_\pm\to q_\mp$ and 
$\psi_-\to -\psi_+$.
The charge densities on the two planes must be equal in absolute value, because of global neutrality. This allows to equate \Eqsref{eq:GrahameMinus} and~\eqref{eq:GrahamePlus}, to obtain $\psi_-$. Assuming until the end of this section that  $q_+\le q_-$ without loss of generality, one can verify that the following is a solution:
\begin{equation}
\psi_-=\frac{1}{q_++q_-}\left(-2v q_- + \ln\frac{q_- \sum_{m=0}^{q_+-1} \e^{-2v m}}{q_+ \sum_{m=0}^{q_--1} \e^{-2v m}}   \right).
\label{eq:psiMinus}
\end{equation} 
Note that $\psi_-=-v+\O(v^2)$ as $v\to0$, as it should, and $\psi_-=-\frac{q_-}{q_++q_-}2v+\O(1)$ as $v\to\infty$. 

\begin{figure}
\centering
\includegraphics[width=1.05\columnwidth]{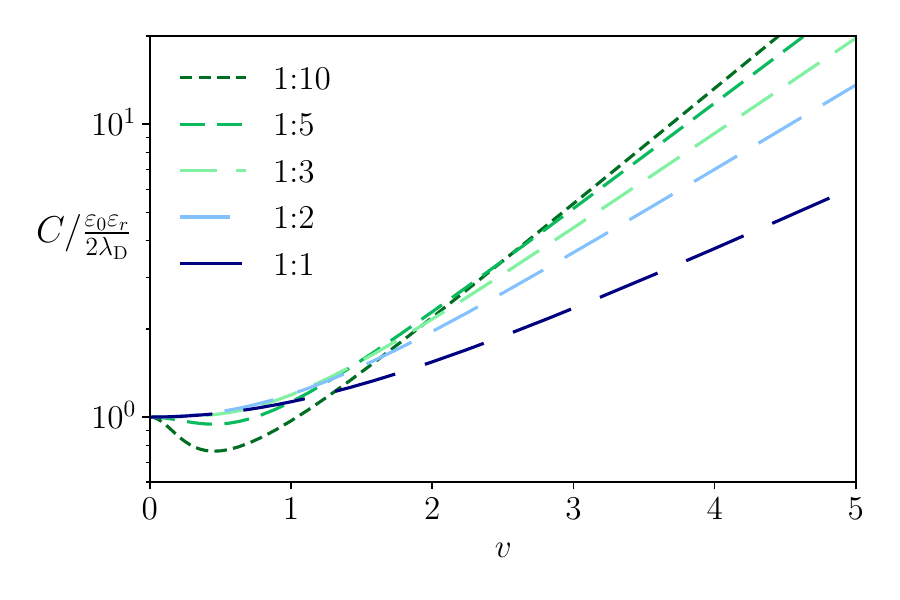}
\caption{Total capacitance per unit surface $C$, as a function of the applied voltage $v$, for $q_+=1$ and $q_-=1, 2,3,5, 10$. $C$ is computed from
  \Eqsref{eq:psiMinus}--\eqref{eq:Ctotqq}.
  The vertical scale is logarithmic and the slope of the obliquous asymptotes, for $v\to\infty$, is $q_+q_-/(q_++q_-)$.
  }
\label{fig:Cqq}
\end{figure}

The differential capacitance per unit surface of the negative electrode is the derivative of \Eqref{eq:GrahameMinus} with respect to $|\psi_-|$, computed at the point determined by \Eqref{eq:psiMinus}: 
\begin{equation}
C_-= \frac{\permz\permr}{\lD} \sqrt{\frac{q_+ q_-}{2(q_++q_-)}} \frac{-\e^{q_-\psi_-}+\e^{-q_+\psi_-}}{\sqrt{ q_- \e^{-q_+\psi_-}+q_+ \e^{q_-\psi_-} - (q_++q_-) }}\,.
\label{eq:Cminusqq}
\end{equation}
Analogously, for the positive electrode we have:
\begin{widetext}
\begin{equation}
C_+= \frac{\permz\permr}{\lD} \sqrt{\frac{q_+ q_-}{2(q_++q_-)}} \frac{\e^{q_-(\psi_- +2v)}-\e^{-q_+(\psi_- +2v)}}{\sqrt{ q_- \e^{-q_+(\psi_- +2v)}+q_+ \e^{q_-(\psi_- +2v)} - (q_++q_-) }}\,.
\label{eq:Cplusqq}
\end{equation}
\end{widetext}
Interestingly, for $q_-/q_+\gtrsim 3.18$, $C_-$ exhibits a non-monotonic behavior, reflected in the total capacitance per unit surface
\begin{equation}
C=\Bigl({\frac{1}{C_-}+\frac{1}{C_+}}\Bigr)^{-1}\,,
\label{eq:Ctotqq}
\end{equation}
which is represented in Fig.~\ref{fig:Cqq}.

The resistance times unit surface of the equivalent $RC$ circuit for the general $q_+:q_-$ case, in analogy with \Eqref{eq:RNL}, reads
\begin{equation}
R=\frac{2L-2\lD}{\beta e^2 n_0 q_+q_-(q_++q_-)D}\,,
\label{eq:RNLqq}
\end{equation} 
which can be rewritten in terms of $\lD$ using the fact that $q_+n^0_+ = q_-n^0_- = n_0q_+q_- $. Multiplying this resistance by the capacitance from \Eqsref{eq:Ctotqq} and preceding ones, one obtains an equivalent-RC-circuit time scale that is of order $L\lD/D$ at small $v$ and diverges exponentially with $v$. This time is represented by the densely dotted curves in Fig.~\ref{fig:NLtimescales_qq} and well captures the numerical results, both in the non-monotonic behavior and in the steeper ascent compared to the 1:1 case.

\begin{figure}
\centering
\includegraphics[width=0.99\columnwidth]{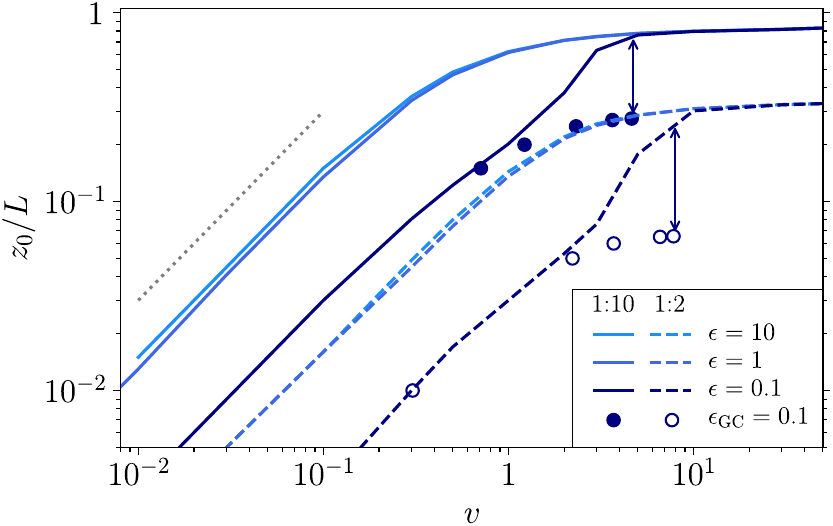}
\caption{Point of null potential $z_0$, as a function of applied voltage, in logarithmic scale. $z_0$ is retrieved numerically as the point where the charge density is zero. For both the 1:2 and the 1:10 case, different values of $\epsilon$ (10, 1, 0.1) are shown.
Circles 
represent results from a numerical solution of the nonlinear Poisson-Boltzmann equation, for a grand-canonical system with Debye length $\epsilon_\mathrm{GC} L=0.1L$; the arrows highlight that depletion, absent in the grand-canonical ensemble, pushes $z_0$ toward the positive electrode.
The dotted gray line is a guide with slope 1. Numerical errors on $z_0/L$ are of order $10^{-3}$. }
\label{fig:ZeroPotqq}
\end{figure}

The loss of left/right symmetry in the composition of the electric double layers and in the potential profile comes with a shift in the zero of the potential, \ie~the point $z_0$ on the $z$ axis where the solution is neutral.
This point is at the center of the capacitor ($z_0=0$) when $v\to0$; upon increasing $v$ it moves toward the electrode producing the smaller voltage drop, \ie~the electrode of the same sign as the least charged species. This is shown in Fig.~\ref{fig:ZeroPotqq}. The shift appears linear in $v$ at small applied voltages. At high voltages (where depletion is present, though), it seems to saturate at $(q_- - q_+)/(q_- +q_+ )$,
thus dividing the cell into two parts, whose sizes are in the same ratio $q_-/q_+$ as the potential drops across the two electrodes.  

Finally, to determine the boundary between the purely nonlinear and the fully depleted regime, we extend our definition of Dukhin number in \Eqref{eq:Dudef} to the asymmetric valences case, where to a salt concentration $n_0$ corresponds a maximum charge of $q_+n_+^0 = q_-n_-^0 = q_+ q_- n_0$:
\begin{equation}
    \Du = \frac{|\sigma(\infty)|}{2Lq_+q_-n_0}\,.
\end{equation}
Using \Eqsref{eq:GrahameMinus} or~\eqref{eq:GrahamePlus}, the condition $\Du\simeq 0.1$ reads  
\begin{equation}
    \epsilon \sqrt{\frac{q_++q_-}{2q_+q_-}} \left( q_- \e^{-q_+\psi_-}+q_+ \e^{q_-\psi_-} - (q_++q_-)\right)^{1/2} \simeq 0.1,
    \label{eq:DuConditionNLqq}
\end{equation}
to be complemented with \Eqref{eq:psiMinus}. Note that \Eqref{eq:DuConditionNLqq} reduces to \Eqref{eq:DuConditionNL} when $q_+=q_-$, as it should.

\subsection{Fully depleted nonlinear regime} 
\label{sec:fullydeplNL_qq}

The unscreened fully depleted nonlinear regime, for asymptotically high applied voltages, is now characterized by the fact that the two ionic species have different velocities: a constant electric field drags the negative species $q_-/q_+$ times faster (or slower) than the positive one. The charge in the two Debye layers grows linearly in time, like in the symmetric case, but arrives at its final value at two different times. Compared to Fig.~\ref{fig:NLunscreened}, this means that the curve ${\rho(-L,t)}/{\rho(-L,\infty)}$ has a different slope than the curve ${\rho(+L,t)}/{\rho(+L,\infty)}$, which in turn translates to an abrupt change in curvature of the $\sigma(t)$ curve.

\begin{figure*}
    \centering
    \includegraphics[width=0.7\textwidth]{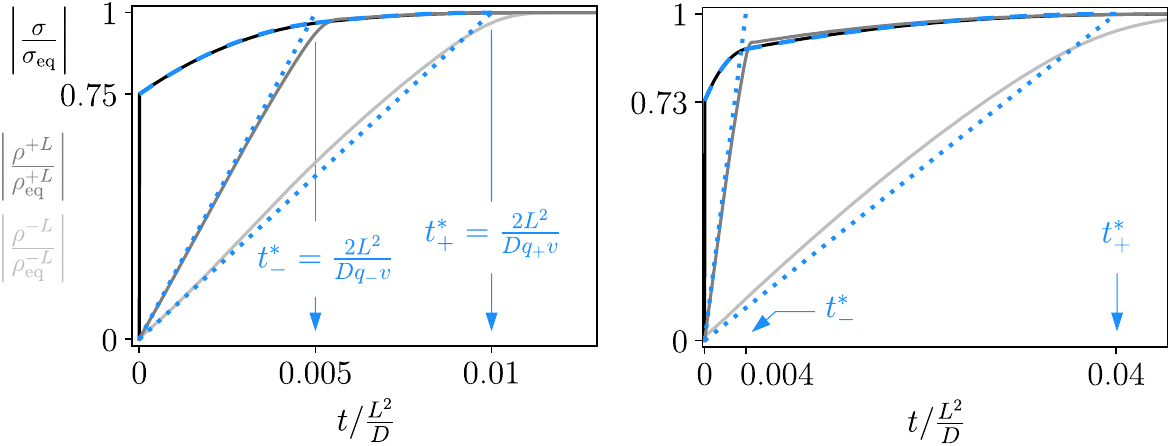}
    \caption{Ratios of $\sigma(t)$, $\rho^{+L}=\rho(+L,t)$ and $\rho^{-L}=\rho(-L,t)$ versus their equilibrium values, as a function of time. The electrode charge density is piecewise parabolic. The blue dotted lines represent the ideal linear evolution of $\rho(\pm L,t)$, while the blue dashed line represents \Eqref{eq:parabolicsigmaqq}. On the left: $v=200$, $\epsilon=0.1$ and $q_+\colon q_- = 1\colon 2$, so that $t_-^*=0.005$ and $t_+^*=0.01$; the curve for the electrode charge density starts at $v/[v+2\epsilon^{-2}(q_++q_-)^{-1}]=0.75$, as predicted by \Eqref{eq:parabolicsigmaqq}. On the right: $v=50$, $\epsilon=0.1$ and $q_+\colon q_- = 1\colon 10$, so that $t_-^*=0.004$ and $t_+^*=0.04$; the curve for the electrode charge density starts at $0.73$.}
    \label{fig:NLunscreenedqq}
\end{figure*}

In the quantitative analysis of this regime, we proceed as in the symmetric case (Sec.~\ref{sec:DynPlanSymFullyDepleted}). As said, the velocities of the two species are now different: $\nu_\pm= Dq_\pm e \beta V_0/L$, assuming again very large electric field and neglecting the effect of the electrolyte distribution on the velocities. If each species travels at its (constant) velocity, the times at which the furthermost ion of each species has reached the oppositely charged electrode are:
\begin{equation}
    t_\pm^*=\frac{2L}{\nu_\pm}=\frac{2L^2}{ D q_\pm  v}\,.
\label{eq:tstarqq}
\end{equation}
After a time $t_+^*$, the total number of positive charges in the system $2n_0'L$ (we define $n_0'=q_+n_+^0=q_-n_-^0=q_+ q_- n_0$) are adsorbed on the negative electrode, and after a time $t_-^*$, the same number of negative charges are adsorbed on the positive electrode. As in Sec.~\ref{sec:DynPlanSymFullyDepleted}, we assume that the electrolyte density at contact with the electrodes grows linearly in time and is localised in a region much smaller than the system size. 
For simplicity, we take $q_->q_+$, \ie~$t^*_-<t^*_+$. The analogous of \Eqref{eq:parabolicsigma0} for the asymmetric case, for $0<t<t_+^*$, is:
\begin{widetext}
\begin{equation}
\hspace{-4cm} 
2 V_0 = \frac{2 e|\sigma(t)|}{\permz\permr}L - \frac{e}{\permz\permr}\int_{-L}^L\diff z \int_{-L}^z \diff z' \underbrace{\left( 2 n_0' L \,\frac{t}{t_+^*}\delta(z'+L) - 2 n_0' L\, \mathrm{min}\left(\frac{t}{t_-^*},1\right)\delta(z'-L) + n_0' h(z',t) \right)}_{\rho(z',t)}\,,  
\label{eq:parabolicsigma0qq}
\end{equation}
with
\begin{equation}
h(z',t)= 
\begin{dcases}
I_{(-L,\, -L+\nu_- t)}(z')-I_{(L-\nu_+ t,L)}(z')\quad &\text{if}\quad 0<t<t_\mathrm{meet} \\
I_{(-L,\, L - \nu_+ t)}(z')-I_{(-L+\nu_- t,L)}(z')\quad &\text{if}\quad t_\mathrm{meet}<t<t^*_- \\
I_{(-L,\, L - \nu_+ t)}(z') \quad &\text{if}\quad t_-^*<t<t^*_+
\end{dcases}\,.
\label{eq:hoftqq}
\end{equation}
\end{widetext}
Again, $I_{(z_1,\,z_2)}(z')$ is the gate function. In \Eqref{eq:parabolicsigma0qq}, the first term in parenthesis represents positive ions that have adhered to the wall in $z=-L$ at time $t$; the second term in parenthesis represents negative ions that have adhered to the wall in $z=L$ at time $t$ (we write it for clarity, but this term does not contribute to the integral, as the presence of adsorbed ions is already encoded in the charge neutrality condition that sets the electrodes' field inside the capacitor); the third term in parenthesis is the charge density at point $z'$ in the rest of the system. 
It helps to think of two trains of ions rigidly moving toward the oppositely charged electrodes, where the head of each train continuously brings new ions to each corresponding double layer. Outside the double layer, on the $z<0$ side of the capacitor, the nonzero charge density is initially due to the negative ions leaving altogether toward positive $z$, while the concentration of positive ions stays constant. The same happens on the $z>0$ side, with reversed roles. In other words, as the tails of the two trains of ions travel at speeds $\nu_+$ and $\nu_-$ toward the centre of the electrode, they define three regions of positive charge (no negative ions present), zero charge (both species present) and negative charge (no positive ions present).
The two tails meet at $t=t_\mathrm{meet}$: at this time the neutral region has shrunk to a point. For $t_\mathrm{meet}<t<t_-^*$, the neutral region re-expands, this time because no ions are present in the central region anymore. The two trains continue to move until time $t=t_-^*$, when the fastest species (negative ions) has reached the oppositely charged electrode. After that, the right part of the bulk solution is neutral, whereas the left part is still populated by positive ions that have not reached the electrode yet. They do so at time $t=t_+^*$: after this time, all ions are adsorbed and the whole bulk is neutral.

Solving the integral in \Eqref{eq:parabolicsigma0qq} and nondimensionalising, one gets the following surface charge as a function of time:
\begin{widetext}
\begin{equation}
\hspace{-2cm} |\sigma(t)|=
\begin{dcases}
v+ \frac{2}{(q_++q_-)\epsilon^2}\left[ t \left(\frac{1}{t_-^*}+\frac{1}{t_+^*}\right) - \frac{1}{2}t^2\left(\frac{1}{{t_-^*}^2}+\frac{1}{{t_+^*}^2}\right) \right] \quad & \text{if}\quad 0<t<t_-^* \\
v+ \frac{2}{(q_++q_-)\epsilon^2}\left[\frac{1}{2} + \frac{t}{t_+^*} - \frac{t^2}{2{t_+^*}^2}\right] \quad & \text{if}\quad t_-^*<t<t_+^*
\end{dcases}\,.
\label{eq:parabolicsigmaqq}
\end{equation}  
\end{widetext}
At equilibrium, the electrodes' charge is $|\sigma(t^*_+)|=v+2/[(q_++q_-)\epsilon^{2}]$, as it should be, the second term representing the final amount of charge adsorbed on either electrode ($2n_0'L$ in physical units). The simple $t^2$ dependence observed in the symmetric case is here split into two parts, each relevant before or after the faster species has reached the electrode. The two parabolas in \Eqref{eq:parabolicsigmaqq} have different curvatures and are centered at different times.
\Eqref{eq:parabolicsigmaqq} agrees with numerical data, as shown in Fig.~\ref{fig:NLunscreenedqq}.

As for the symmetric case, for times $t>t^*_\pm$ we expect a fast relaxation over the Gouy-Chapman lengths of the two counterionic double layers. However, the relaxation of $\sigma$ is necessarily dominated by the relaxation of the slower double layer. Since the double layers are coupled through the electrodes' surface charge (equal and opposite on the two electrodes), the relaxation of the fast ions is limited by slow changes in $\sigma$ and therefore happens with the same time scale as the slow double layer. Hence the exponential relaxation for $t>t_+^*$ occurs in a time 
\begin{equation}
\tau=
\frac{4L^2}{D v^2 q_+^2}\,,
\label{eq:tauFDNLqq}
\end{equation}
which is observed at sufficiently high voltages, for all values of $\epsilon$ (see Fig.~\ref{fig:NLtimescales_qq}).

Again, to obtain \Eqref{eq:parabolicsigmaqq}, it was crucial to assume that ions do not affect the linear potential profile through almost all the capacitor. This happens if the change in electric potential (or field) caused by their motion is small with respect to the applied voltage. The analogue of condition \Eqref{eq:veps2}, determining the phase-space boundary of the regime we just described, is then 
\begin{equation}
v>\frac{2}{(q_++q_-)\epsilon^2}\,.
\label{eq:veps2qq}
\end{equation}
We also assumed that the distance an ion can travel by diffusion in a time $t_+^*$ is much smaller than $2L$, so that we could neglect diffusion currents and only consider drift: this amounts to $v>1/q_-$. In summary, the regime we just discussed, as for the symmetric case, is defined by \Eqref{eq:veps2qq} at small $\epsilon$ and coincides with the nonlinear regime for large $\epsilon$.

\begin{figure}
    \centering
    \includegraphics[width=\columnwidth]{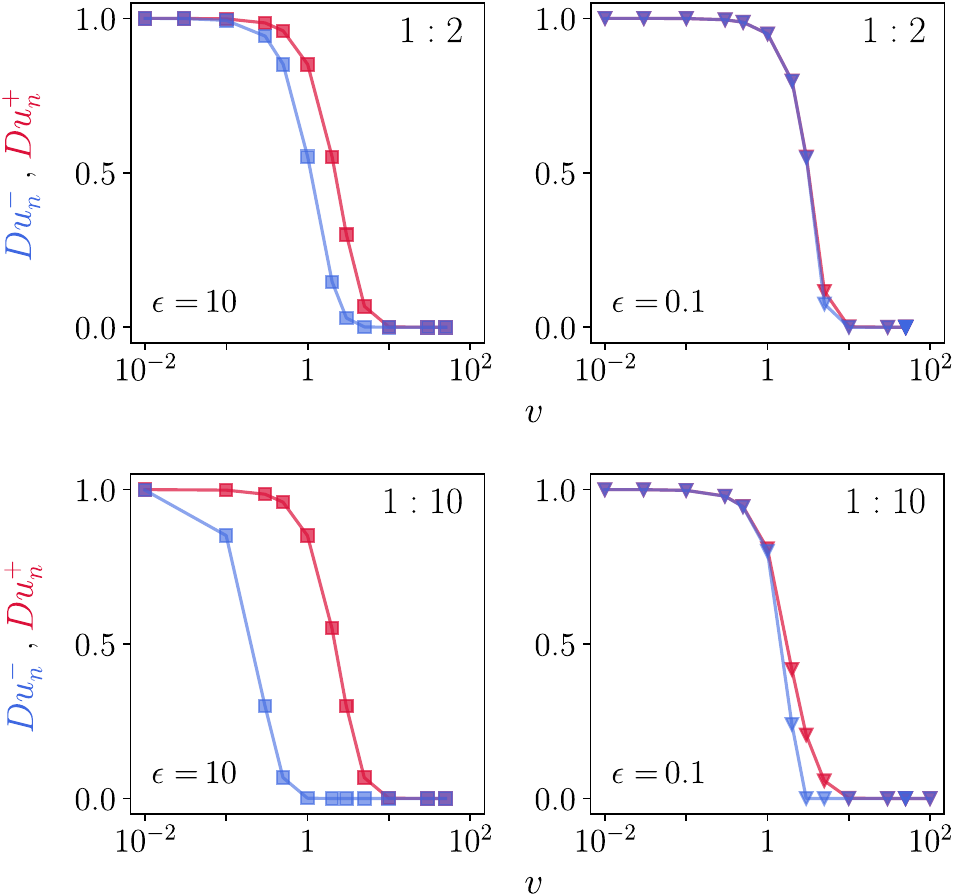}
    \caption{Dukhin number $\Du_n^\pm=n_\pm(0,t=\infty)/n_\pm(0,t=0)$, quantifying depletion, for $q_+:q_-=1:2$ (top row) and $1:10$ (bottom row). At large $\epsilon=\lambda_D/L$ (left column), depletion sets in at lower voltages for the large-valence negative ions (blue) than for the small-valence positive ones (red). In the $1:10$ case, there exists a range of voltages at which one species is fully depleted while the other is almost non-depleted. This does not happen in the symmetric valences case (Fig.~\ref{fig:DeplDuStat}c).}
    \label{fig:Depletionqq}
\end{figure}

In the latter large-$\epsilon$ case, it is worth mentioning that the asymmetric-valence phenomenology is richer than for symmetric valences: in the $\epsilon\gtrsim1$ regime, ions of different valence are depleted in different proportions (Fig.~\ref{fig:Depletionqq}). This can give rise to a hybrid behavior in terms of relaxation rates, that however covers a substantial range of voltages $v$ only for large, non realistic valences.  
For small $\epsilon$, between the purely nonlinear regime (non-depleted) and the unscreened nonlinear regime described by \Eqref{eq:parabolicsigmaqq} (fully depleted), we have what we called in Sec.~\ref{sec:DynPlanSymFullyDepleted} a partially screened fully depleted regime. At such intermediate $v$, the initial relaxation is neither exponential nor polynomial in time. Such relaxation can be divided in two parts, though, similarly to what happens in the unscreened regime: a first part before migration and depletion are completed, and a second part concerning relaxation inside the two counterion-only double layers. The latter process cannot but happen on the same time scale as in \Eqref{eq:tauFDNLqq}. This intermediate regime is in the asymmetric valence case even more complex, because at a given applied voltage the two species present two different levels of depletion.

\section{\texorpdfstring{FULLY ASYMMETRIC CASE: $q_+\neq q_-$, $D_+\neq D_-$}{Fully asymmetric case: q+ != q-, D+ != D-}}
\label{sec:fullyasym}

\subsection{Linear regime}
\label{sec:fullyasym_linear}

We finally consider the fully asymmetric case, with ions of different valences $q_+\neq q_-$ and diffusivities $D_+ \neq D_-$. In the linear regime, we solve the Poisson-Nernst-Planck equations analytically in the Laplace domain, as done for the fully symmetric case \cite{Bazant2004,Janssen2018,phd}. Since \Eqsref{eq:PNP} and \eqref{eq:poisson} are coupled PDEs, two successive nontrivial diagonalisations, one in the time domain and one in the Laplace domain are needed, leading to solvable uncoupled equations for linear combinations of the ion densities. In summary, the first transformation aims to diagonalize drift currents. After passing to the Laplace domain and diagonalising a second time, the diffusion equation in space can be solved. Once time and space derivatives do not appear anymore, a cumbersome calculation leads back to the original basis $\{\widehat{n}_+(z,s),\widehat{n}_-(z,s)\}$ (Laplace transforms of the ion densities) and boundary conditions can be imposed (potential at the electrodes, no flux through the electrodes, electroneutrality). 
As in the symmetric case, the modes characterizing the linear response can also be considered in the frequency-domain via the impedance of the cell (see \textit{e.g.}~\cite{Lelidis2005,Barbero2007,Chassagne2016,Antonova2020} for partially asymmetric electrolytes with opposite valencies but unequal diffusivities).

The non-zero poles of these functions represent the relaxation rates of the system and can be analyzed numerically. In the following, we take as reference 
$\hrho(-L,s)$, the Laplace transform of the charge density at contact $\rho(-L,t)$. This quantity depends on $\epsilon$, on the valences, and on the additional dimensionless parameter 
\begin{equation}
\delta=\frac{D_+}{D_-}\,.
\end{equation}
We assume here that the positive species is the slower one, so that $\delta\leq1$.
A numerical analysis of $\hrho$ reveals, for any $\delta<1$, the presence of new poles that did not exist in the case $\delta=1$. 

For large $\epsilon$, $\hrho$ has twice as many poles (and zeroes).
The number of poles (and zeroes) stays infinite, but each pole from the $\delta=1$ case splits into two poles and a zero as soon as $\delta<1$. The poles can be classified into two independent hierarchies of diffusive time scales: one, $\{\tau_{+,i}\}$, for the positive species and one, $\{\tau_{-,i}\}$, for the negative species. While their exact values can be retrieved numerically, these times are well described by the following equation
\begin{equation}
\tau_{\pm,i}\simeq\frac{4L^2}{(2i+1)^2\pi^2D_\pm}\,\quad\quad i=0,1,2...\,,
\label{eq:taupmn}
\end{equation}   
asymptotically exact for $\epsilon\to\infty$.
Still assuming that positive ions are slow compared to negative ones, the slowest time scale, \ie~the dominant one at large times, is $\tau_{+,0}=4L^2/(\pi^2 D_+)$. Note that the valences of the two species do not enter in the expression for the relaxation times $\tau_{\pm,i}$; they do play a role, though, in determining the importance of each mode (\eg~the weight of positive modes increases with $q_+$).

\begin{figure*}
\centering
\includegraphics[width=0.99\textwidth]{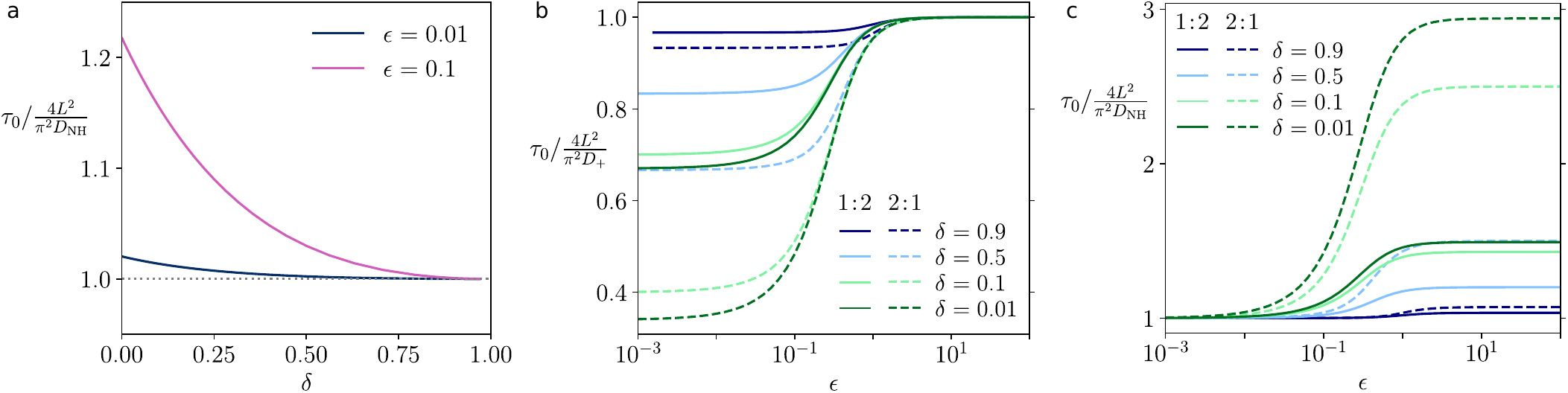}
\caption{a) Slowest relaxation time $\tau_0$ extracted numerically from the analytical $\hrho(-L,s)$ in units of its asymptotic value from \Eqref{eq:tauzeroqqDD}, as a function of $\delta$, for small values of $\epsilon$. The dotted horizontal line represents \Eqref{eq:tauzeroqqDD}. $q_+:q_-=1:1$. b-c) Same quantity, in units of its large-$\epsilon$ value $\tau_{+,0}$ from \Eqref{eq:taupmn} (b), and in units of its small-$\epsilon$ value from \Eqref{eq:tauzeroqqDD} (c). Continuous lines are for a 1:2 electrolyte, dashed lines are for a 2:1 electrolyte.}
\label{fig:tauzeroqqDDcombo}
\end{figure*}

For small $\epsilon$, the picture changes and the slowest time scale approaches
\begin{equation}
\tau_\mathrm{NH} = \frac{4L^2}{\pi^2 D_\mathrm{NH}}, \quad \text{with } D_\mathrm{NH}=\frac{(q_+ + q_-) D_+ D_-}{q_+ D_+ + q_- D_-}\,,
\label{eq:tauzeroqqDD}
\end{equation} 
featuring the so-called Nernst-Hartley diffusion coefficient $D_\mathrm{NH}$ \cite{Robinson1959}. Its presence reflects the fact that ions diffuse together under the action of an internal electric field that pulls the slowest species, while slowing down the fastest one \cite{Robinson1959}. The scaling of the relaxation time is shown in Fig.~\ref{fig:tauzeroqqDDcombo}a for the equal-valence case. Figs.~\ref{fig:tauzeroqqDDcombo}b-c show instead how the slowest relaxation time $\tau_0$ transitions from $\tau_{+,0}$ 
at large $\epsilon$ (\ref{fig:tauzeroqqDDcombo}b) to $\tau_\mathrm{NH}$ 
at small $\epsilon$ (\ref{fig:tauzeroqqDDcombo}c), for the 1:2 and 2:1 cases, at any $\delta$. 

The pole corresponding to \Eqref{eq:tauzeroqqDD}, the closest one to the origin, is however not the one with the largest residue. This is important because the residue of a pole is proportional to the weight of its corresponding mode in the time domain: a weight larger than the slowest mode's weight allows a fast mode to be visible, at least at short times. A fast, but large-weight mode represents a relaxation phenomenon that is dominant for a certain time and eventually fades out giving way to slower processes. Returning to the RC-circuit analogy, one can estimate what the time with the largest weight $\tau_\mathrm{w}$ might look like, assuming that it represents double layer charging. Recomputing the electric resistance times unit surface of \Eqsref{eq:RNL} and~\eqref{eq:RNLqq} for a bulk with asymmetric diffusivities gives
\begin{equation}
R=\frac{2L-2\lD}{\beta e^2 n_0 q_+q_-(q_+D_+ +q_-D_-)}\,.
\label{eq:RNLqqDD}
\end{equation}
Using a capacitance per unit surface $\permz\permr/\lD$ for each electrode, one finds the characteristic time
\begin{equation}
\hspace{-0.2cm}\tau_\mathrm{RC}=\frac{L\lD - \frac{\lD^2}{2}}{D_\mathrm{ave}}\,, \quad  \mathrm{with}\ D_\mathrm{ave}=\frac{q_+ D_+ + q_- D_-}{q_+ + q_-}\,,
\label{eq:tauRCqqDDlin_corr}
\end{equation}
where the factor 1/2 in front of the $\lD^2$ term was added \adhoc\ so as to match the fully symmetric case (\Eqref{eq:tauBazant}). 

\begin{figure*}
\centering
\includegraphics[width=0.97\textwidth, angle=0]{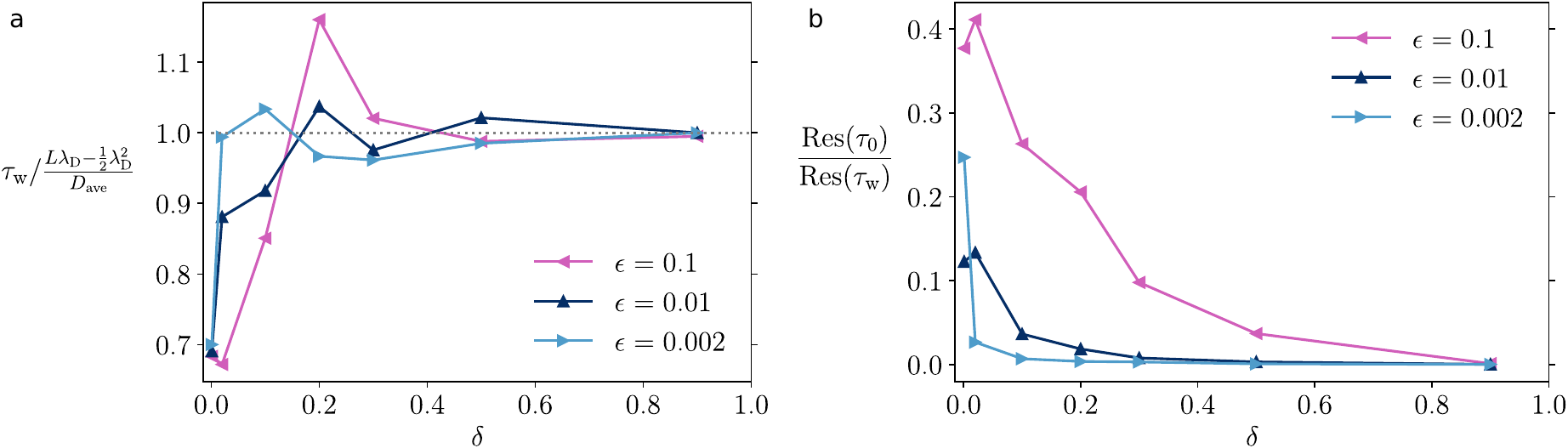}\\
\caption{a) Largest-weight relaxation time $\tau_\mathrm{w}$ extracted numerically from $\hrho(-L,s)$, in units of its theoretical value from \Eqref{eq:tauRCqqDDlin_corr}, as a function of $\delta=D_+/D_-$, for different values of $\epsilon=\lD/L$. 1:1 electrolyte. b) Ratio of the residues of $\hrho(-L,s)$ with respect to the pole corresponding to $\tau_0$ and to the pole corresponding to $\tau_\mathrm{w}$, as a function of $\delta$. 1:1 electrolyte.}
\label{fig:tau0_and_tauhw}
\end{figure*}

This expression contains an average of the diffusion coefficients weighted by ion valences, $D_\mathrm{ave}$, rather than the Nernst-Hartley diffusivity of \Eqref{eq:tauzeroqqDD}. It explains results both from the numerical analysis of $\hrho$ (Fig.~\ref{fig:tau0_and_tauhw}a) and from solutions of the Poisson-Nernst-Planck equations (Fig.~\ref{fig:NLtimescales_qqDD}, small $v$ and small $\lD/L$).
An analysis of the weights of the mode $\tau_\mathrm{w}\simeq\tau_\mathrm{RC}$ and of the slowest mode $\tau_0\simeq\tau_\mathrm{NH}$ is presented in Fig. \ref{fig:tau0_and_tauhw}b\,; $\tau_\mathrm{w}$ has a larger weight on the whole range of $\delta$. For $\delta\to 1$, the weight of $\tau_0$ goes to 0 – the time $\tau_\mathrm{NH}$ is indeed absent from the linear analysis of the symmetric valence case.

In summary, at large $\epsilon$ the relaxation is fully described by \Eqref{eq:taupmn}, with positive and negative ions decoupled as in the fully symmetric case. At small $\epsilon$, the double layer charging occurs at early times with a time scale $\tau_\mathrm{w}\simeq\tau_\mathrm{RC}$, described by \Eqref{eq:tauRCqqDDlin_corr}. Subsequently, the purely diffusive mode from \Eqref{eq:tauzeroqqDD} sets in, featuring the Nernst-Hartley coefficient and signaling that positive and negative ions diffuse together. An inspection of the curves for $n_\pm(z,t)$ highlights the origin of this two-step relaxation. Initially, the double layer builds up mostly thanks to the faster species: the equilibrium \textit{charge} density is reached through an immediate rearrangement of the fast species around the instantaneous local concentration of the slow species. Mass concentrations are therefore different from the equilibrium ones: there is a neutral excess of mass at the electrode with the same sign as the slow species (as the slow species has not had time to escape), and a defect of mass at the opposite side. This requires a mass relaxation process, which happens precisely with the diffusive time scale of \Eqref{eq:tauzeroqqDD}. Note that even though the diffusive scaling might recall depletion, \cf~\Eqref{eq:taudepl}, the phenomenon we just described is linear in all respects and occurs also for $v\to0$.

\subsection{Nonlinear regimes}
\label{sec:fullyasym_NL}

\begin{figure*}
\centering
\includegraphics[width=0.99\textwidth]{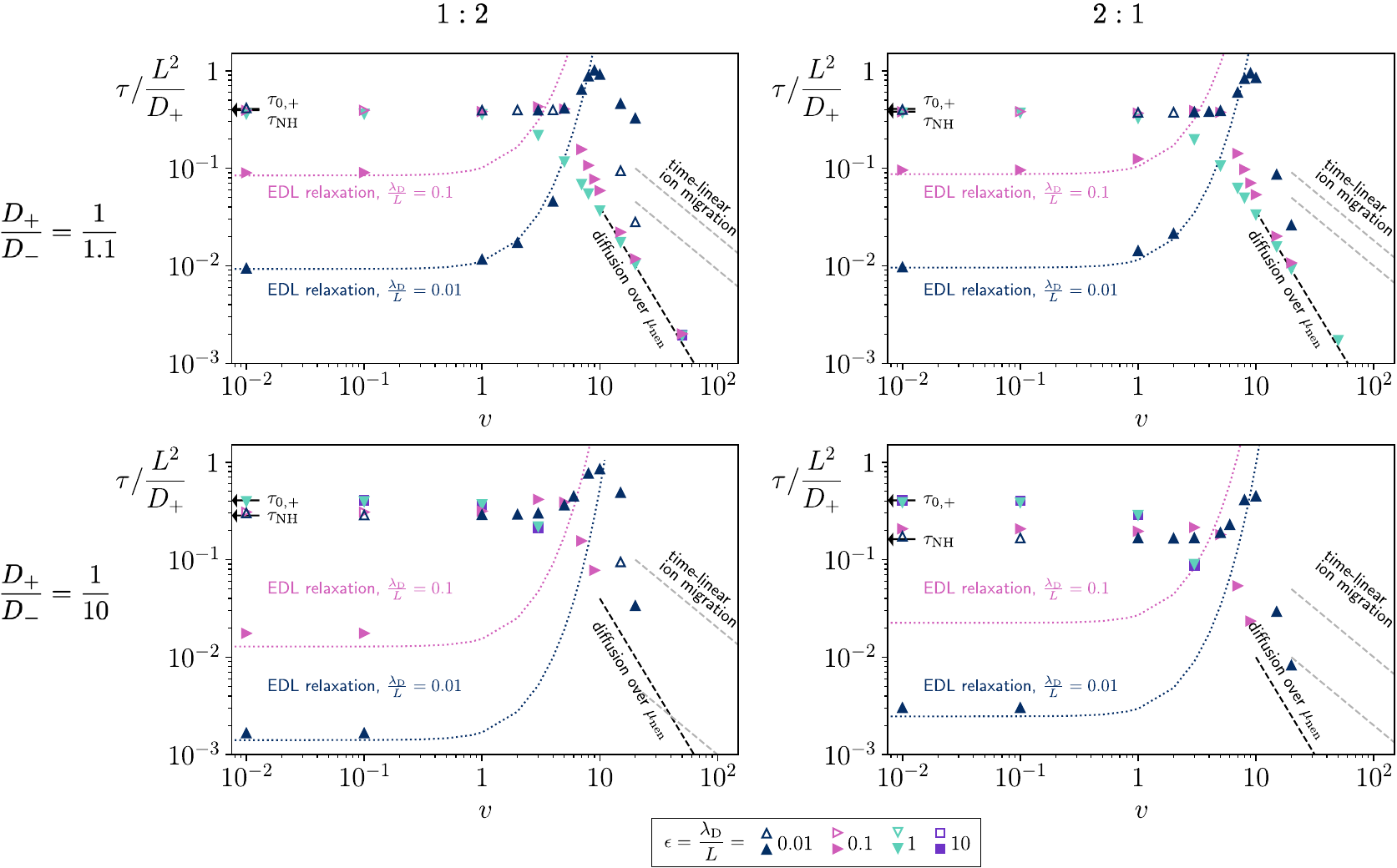}
\caption{Relaxation times $\tau$ as a function of $v$, for diffusivity ratios $\delta=D_+/D_-=1/1.1$ or 1/10 and valence ratios $q_+$:$q_-=1:2$ or $2:1$. Times are in units of ${L^2}/{D_+}$ and different colors correspond to different values of $\epsilon=\lD/L$. As in Fig.~\ref{fig:NLtimescales_qq}, times are extracted from exponential relaxation(s) of $\sigma$ and, when present, empty symbols distinguish late-time processes from earlier-time processes (full symbols). Two black arrows point at the values of $\tau_{0,+}$ and $\tau_\mathrm{NH}$, respectively from \Eqsref{eq:taupmn} and~\eqref{eq:tauRCqqDDlin_corr}. The colored dotted lines represent the purely nonlinear time obtained from \Eqsref{eq:RNLqqDD} and~\eqref{eq:Ctotqq} and preceding ones (see Fig.~\ref{fig:Cqq}): by construction, their small-$v$ value is $\tau_\mathrm{RC}$ from \Eqref{eq:tauRCqqDDlin_corr}. The dashed black line represents the time from \Eqref{eq:tauFDNL} with $D=D_+$. The dashed gray lines are the two times from \Eqref{eq:tstarqqDD} observed at short times, where the relaxation is linear in time.}
\label{fig:NLtimescales_qqDD}
\end{figure*}

An analysis of the relaxation times observed in the nonlinear regime is presented in Fig.~\ref{fig:NLtimescales_qqDD}, for $\delta=1/1.1$ and 1/10 and for valences 2:1 and 2:1.
Results obtained for the symmetric diffusivity cases seem to extend naturally to asymmetric diffusivities. In the purely nonlinear regimes, the capacitance computed in \Eqsref{eq:psiMinus}--\eqref{eq:Ctotqq} is still applicable and can be combined with the resistance from \Eqref{eq:RNLqqDD}. Qualitatively, at least for small valences, this result in a simple vertical shift of the exponential curve representing increased nonlinear capacitance in Fig.~\ref{fig:NLtimescales_qqDD}, such that at small $\epsilon$ the curve coincides with the linear regime time from \Eqref{eq:tauRCqqDDlin_corr}. The curve is most evident for the $\delta=1/1.1$, 1:2 case at $\epsilon = 0.01$ and is partially masked by high-weight diffusive time scales for other parameters. Its shape mostly depends on the valence ratio through the capacitance, as per Fig.~\ref{fig:Cqq}.

In the unscreened ($v\to\infty$) fully depleted regime, the discussion for the symmetric and asymmetric valence cases still applies, with the \caveat\ that asymmetry in the diffusivities will affect the initial linear drag phase. The times from \Eqref{eq:tstarqq} must then be replaced by
\begin{equation}
    t_\pm^*=\frac{2L}{\nu_\pm}=\frac{2L^2}{ D_\pm q_\pm  v}\,.
\label{eq:tstarqqDD}
\end{equation}
Once each ion train has reached the electrode, each counterionic double layer will start its internal exponential relaxation with relaxation time $4L^2/(D_\pm v^2 q_\pm^2)$. For the same reasons that led us to \Eqref{eq:tauFDNLqq} in the partially asymmetric case, the last relaxation should feature the slower of the two relaxation times of the two counterionic double layers:
\begin{equation}
    \tau = \frac{4L^2}{v^2\min(D_+ q_+^2, D_- q_-^2)}\,.
\end{equation}
This seems compatible with the numerical results presented in Fig.~\ref{fig:NLtimescales_qqDD}, that are however computationally more involved to obtain for significantly distinct values of valences and diffusivities.

Results from this section are summarized in Fig.~\ref{figm:phaseasym} of \cite{mainpaper}.

\section{\texorpdfstring{NON-IDEAL BEHAVIOR}{Non-Ideal Behavior}}\label{sec:nonideal}

In this section, we analyze in brief the applicability range of the Poisson-Nernst-Planck theory. Specifically, we discuss the impact of ionic correlations due to electrostatics and to ion packing by estimating
the electrostatic coupling parameter and the maximum ion density
in the device, across our regime diagram. Before doing so, we emphasize that our model is meant to provide a simple framework to benchmark further simulations, theories or experiments, allowing for a more profound assessment of the effect of non-idealities (\textit{e.g.}~electrostatic correlations, ion size, polarizability, electrode geometry).

\subsection{Deviation from mean-field}

Ionic correlations are an intrinsically discrete phenomenon, whose magnitude depends on local density and whose effect can be anisotropic and counterintuitive. A possible approach consists in introducing a constant effective correlation length scale within mean-field models. The effects of this effective correlation length on the small-voltage EDLC dynamics have been studied in \cite{Zhao2011,Fertig2025}, which reported small deviations from the mean-field dynamics for reasonable values of correlation length. Nonetheless, the mapping between the effective correlation length and physical parameters is somewhat elusive. We take here a different approach, relying only on the electrostatic coupling parameter. 

When only one kind of ions is present in the double layer, which is often the case in the nonlinear regimes, a useful tool to quantify the importance of ion-ion electrostatic correlations is the electrostatic coupling parameter \cite{Netz2000EPJE,Naji2013,Varenna,Palaia2022a}. It represents the squared ratio between the electrostatic energy between two ions of charge $eq$ at a distance $a$ and the thermal energy $\kB T$: 
\begin{equation}
    \Xi = \left(\frac{q^2 \lB }{a} \right)^2 .
    \label{eq:Xi}
\end{equation}
If ions adhere to a wall of charge density $e\Sigma$ and the system is electroneutral, the typical distance is fixed by the wall charge density ($a = \sqrt{q/\Sigma}$) and the coupling parameter assumes the familiar expression $\Xi = 2 \pi \lB^2 q^3 \Sigma$ \cite{Netz2000EPJE,Naji2013,Varenna,Palaia2022a}. Poisson-Boltzmann is an excellent approximation when $\Xi \lesssim 1$, while for $\Xi \gtrsim 10$ the ion density noticeably deviates from the mean-field one \cite{Moreira2000,Santangelo2006,Samaj2018a}. In addition, since usual ion sizes are smaller than $\lB = \SI{0.7}{nm}$ in water at ambient temperature, one can reasonably assume that absence of electrostatic correlations implies absence of ion packing effects. $\Xi \lesssim 1$ is therefore a good proxy for the absence of ionic correlations.

\subsection{Fully depleted nonlinear regime}

We analyze here the fully depleted nonlinear regime, described in Secs.~\ref{sec:DynPlanSymFullyDepleted}, \ref{sec:fullydeplNL_qq} and \ref{sec:fullyasym_NL}, and represented in the bottom-right regions of Figs. 3 and 4 in \cite{mainpaper}. In this regime, the vicinity of the electrodes is populated by counter-ions only. In addition, the system (electrode and counterions) is in general not electroneutral as the electrode charge $\Sigma$ can be larger than the amount of ions present in the system $2n_0 L$.
Looking, for simplicity, at the symmetric 1:1 case, at equilibrium all ions are in the vicinity of the electrode. In the worst case scenario, they are adsorbed on the plate and the typical distance between them is $a \simeq 1/ \sqrt{2n_0L}$, so that the coupling parameter \eqref{eq:Xi} takes the form
\begin{equation}
\label{eq:XiFD}
\Xi = \frac{\lB}{4\pi L \epsilon^2} = 2 \lB^2 L n_0 \,,
\end{equation}
which is independent of $v$.

For $v>1$ and $\epsilon>1$, we always have $\Xi<1$, as realistically $\lB \ll L$ for any aqueous-electrolyte EDLC (where $\lB=\SI{0.7}{nm}$), so that correlations are negligible. 
Referring to the regime diagram in Fig.~\ref{figm:phasesym} of \cite{mainpaper}, in the no-screening part of the bottom-right sector ($v>1$ and $v^{-1/2}<\epsilon<1$ as per Eq.~\eqref{eq:veps2}), we have $\Xi< v\lB/4\pi L$. For $v \lesssim 50$ for aqueous electrolytes, it is still $\Xi\lesssim 1$ for realistic EDLCs and correlations have little relevance. 
Finally, in the partially screened depleted sector ($v>1$ and $1/\sinh({v/2}) < \epsilon < v^{-1/2}$), $\Xi$ may exceed unity for some combinations of voltage, concentration, permittivity, pore size and temperature, and the validity of mean-field should be checked on a case-by-case basis computing $\Xi$ through Eq.~\eqref{eq:XiFD}. 
For example, a 5-mM 1:1 aqueous electrolyte in a 3-mm-wide EDLC (experimental parameters from \cite{Zhao2024}) would result in $\Xi$ exceeding 4000. In contrast, a 1-µm cell in the same conditions would experience a fairly weak coupling, with $\Xi$ close to 1.

\subsection{Purely nonlinear regime}

We now turn to the purely nonlinear regime, where depletion is absent. This is described in Secs.~\ref{sec:DynPlanSymPurelyNL}, \ref{sec:partiallyasym:pnl} and \ref{sec:fullyasym_NL}, represented in the central-bottom regions of Figs. 3 and 4 in \cite{mainpaper}, and identified, in the fully symmetric case, by the conditions $v>1$ and $\epsilon < (20\sinh(v/2))^{-1} $ as per Eq.~\eqref{eq:DuConditionNL}. In this regime, the system (electrode and double-layer) is electroneutral. While a theory of electrostatic coupling for two-species double layers does not exist, for large voltages counterions are expelled from the EDL, so that one can consider $\Xi = 2\pi \lB^2 q^3 \Sigma$. Once equilibrium is reached, Eq.~\eqref{eq:Grahame} for the electrode charge gives 
\begin{equation}
    \Xi = \frac{\lB}{L \epsilon}\sinh\left(\frac{v}{2}\right)
    \label{eq:XiPNL}
\end{equation}
for monovalent ions. By Eq.~\eqref{eq:DuConditionNL} again, in this sector $\Xi \leq \frac{\lB}{20 L \epsilon^2}$, with the equality holding only at the boundary with the depleted region. Replacing for $\epsilon$, this means 
\begin{equation}
    \Xi \leq \frac{4\pi}{10} \lB^2 L n_0 \simeq  \frac{LX}{\SI{3}{nm}} \,,
    \label{eq:XiPNLUpperBound}
\end{equation} where $X$ is the numeric value of the molarity in mol/L and we assumed an aqueous solution at ambient temperature. In practice, for nanopores with $L\approx\SI{100}{nm}$ and concentrations below $\SI{30}{mM}$ (\ie\ $X<0.03$), $\Xi$ stays below 1 in the whole sector. While Eq.~\eqref{eq:XiPNLUpperBound} sets an upper bound,  $\Xi$ decreases exponentially with decreasing $v$, rapidly moving away from this bound as per \Eqref{eq:XiPNL}. As a result, for $v\sim 1$, our results stay relevant even at substantially higher concentrations.
For the millimeter-scale EDLCs from \cite{Zhao2024}, at concentrations between 5 and \SI{400}{mM} and applied voltages between 0.05 and \SI{0.6}{V}, \Eqref{eq:XiPNL} yields a $\Xi$ ranging from 0.08 to 230, corresponding to a very weak to moderately strong coupling. 

\subsection{Early-time behavior ($t\rightarrow 0$)}
Even when mean-field is in principle bound to fail, it should still capture the early response, at least for initial concentrations low enough that correlations can be neglected at small times. At $t=0$ it is $a=n_0^{-1/3}$, which translates to
\begin{equation}
\Xi_\mathrm{bulk} = (\lB n_0^{1/3})^2 \lesssim 1 \,.
\end{equation}
This makes our analysis valid at least for short times for any electrolyte concentrations of order \SI{1}{M} or below. 
In cells from \cite{Zhao2024}, the early-time electrostatic coupling is completely negligible, with $\Xi$ between $0.01$ and $0.2$.

\subsection{Final comments}

Overall, while in many regions of parameter space the mean-field assumption strictly holds, it is in general not possible to determine regions in the $v$, $\epsilon$ diagram where mean-field surely fails, so that the value of the electrostatic coupling parameter $\Xi$ should be checked case by case. This is because electrostatic correlations can introduce new length scales that make it impossible to reduce the $\Xi \lesssim 1$ condition to a simple condition on $v$ and $\epsilon$, with $\Xi$ becoming in general a different function of the system size, ion concentration, voltage, solvent permittivity and temperature.
We also note that \Eqsref{eq:PNP} for the ion dynamics neglect the coupling with that of the solvent, not only in terms of electrokinetic effects (see \textit{e.g.} \cite{Lobaskin2016, Asta2019JCP}), but also Maxwell-Stefan fluxes which may contribute as the salt concentration increases (see \textit{e.g.} \cite{Balu2018}).

Finally, in the present work we have assumed that there are no electrochemical reactions (in particular involving the solvent) during the whole charging process. Considering the most used media, the electrochemical window of water is $v\simeq 50$, that of organic solvents $v\simeq 60$, while that of ionic liquids 
at most $v\simeq120$ \cite{Armand2009}. Our numerical results beyond these limits have the purpose of highlighting the dominant process, which is in general more evident far away from the transitions with other regimes.

\section{\texorpdfstring{CONCLUSION}{Conclusion}}

Despite its applications in electrochemistry, the relaxation to equilibrium of electric double-layer capacitors (EDLCs) has been somewhat elusive, in particular for the most common case of electrolytes with different valences and/or diffusivities. This work, together with \cite{mainpaper}, tries to fill this gap, by giving a new perspective on existing results for the linear fully symmetric case, and by subsequently addressing the asymmetric cases and the nonlinear regimes. 
We characterise the relaxation behavior focusing on the dominant time scales, that are usually the slowest ones. We define different regimes in the parameter space spanned by applied voltage and ion concentration, providing analytical boundaries between regimes whenever possible. For low ion concentrations, the behaviors of positive and negative ions can be mutually decoupled, with a relaxation that is diffusive (exponential in time) in the small-voltage regime and drag-dominated (linear in time) in the large-voltage one. Beyond this low concentration limit, the picture is different. At asymptotically small voltage, the relaxation is faster than diffusion, yet followed by a slower diffusion process of neutral mass in the case of asymmetric diffusivities. As voltage increases, the relaxation time increases exponentially due to nonlinearity, until the potential difference is strong enough to deplete ions from the bulk. Then relaxation gradually goes from exponential to linear in time, across a regime where the electric field across the bulk is highly varying. Lastly, if the electrode charge is much larger than the charge that the confined electrolyte can possibly neutralise, the picture from low salt concentration is recovered, where ions are dragged at constant velocity to the electrodes.

The results presented here rely on the validity of the mean-field approximation. We discussed in section VI how exceedingly large ion concentrations, caused either by too large initial concentrations or applied voltages, might bring about effects beyond mean-field that can at least partially invalidate our results, in particular in the partially screened fully depleted regime. The significance of electrostatic correlations \cite{Naji2013, Storey2012, Kanduc2010, Bonneau2023}, but also of ion pair formation \cite{Adar2017} and finite-size effects \cite{Borukhov1997,Ben-Yaakov2009, Jiang2014, Hartel2017, Gupta2020, Ma2022, Bultmann2022}, may depend on the specific kind of ions used, the permittivity of the medium \cite{Schlaich2019,Palaia2022a}, the pore size \cite{Hartel2017,Mendez-Morales2018}, the temperature and the nature of the electrode.

Our analysis contributes to the understanding of electrochemical devices and of confined charged materials, providing a clear mapping between any point in parameter space and mean-field relaxation times. This facilitates comparison among theories that include ion specificity, finite ion sizes, ionic correlations, hydrodynamics or geometric effects \cite{Werkhoven2018,Janssen2019,Lian2020,Gupta2020, Ma2022,Yang2022}. Finally, our work paves the way to the design of optimisation procedures for the charging of EDLCs \cite{Breitsprecher2018, Breitsprecher2020, phd}, with promising applications to energy production and recovery.

\section*{\texorpdfstring{ACKNOWLEDGEMENTS}{Acknowledgements}}
This work has received funding from the European Union's Horizon 2020 and Horizon Europe research and innovation programs under the Marie Sk\l{}odowska-Curie grant agreements Nos.~674979-NANOTRANS (I.P., P.B.W., B.R., E.T.), 101034413 (I.P.), and 101119598-FLUXIONIC (M.D., B.R., E.T.), as well as from the European Research Council under grant agreement No.~863473 (B.R.). B.R. acknowledges financial support from the French Agence Nationale de la Recherche (ANR) under Grant No.~ANR-21-CE29-0021-02 (DIADEM). I.P. thanks An{\dj}ela \v{S}ari\'{c} for further support at ISTA.

\appendix

\section{\texorpdfstring{APPENDIX: NUMERICAL METHODS}{Appendix: Numerical Methods}}
\label{sec:NumericalMethod}

The Poisson-Nernst-Planck equations are solved numerically using a flux-conservative finite-difference method. We discretize space into nodes, positioned at $z_{k+\frac{1}{2}}$ for $k=0,..., N-1$, and edges, located at $z_{k}$ for $k=0,..., N$. If the spacing between nodes and between edges is constant, we have $z_{k+\frac{1}{2}}=-L+(k+\frac{1}{2}) \Delta z$ and $z_k=-L+k \Delta z$, with $\Delta z= 2L/N$. The extension to a nonlinear spacing is straightforward and useful. For simplicity, we will describe the algorithm assuming constant spacing.

Ion densities $n_\pm$ and potential $\phi$ are defined on nodes $z_{k+\frac{1}{2}}$, while electric field and ionic currents are defined on edges $z_{k}$; this reduces the error associated with numerical derivation or integration. Densities are initially set to their $t=0$ value, $n_\pm(z,0)=n^0_\pm$, for all nodes. Then the density and potential profiles are evolved in time in steps of $\Delta t$. More precisely, at every step $i$:
\begin{enumerate}[leftmargin=*]
\setlength{\itemsep}{0.2em}
\item The ionic contribution to the electric field is computed via Gauss' theorem, by numerically integrating the following charge density:
$$\rho(z_{k+\frac{1}{2}},i\,\Delta t)=\bar{q}_-\dens_-(z_{k+\frac{1}{2}},i\,\Delta t) + \bar{q}_+\dens_+(z_{k+\frac{1}{2}},i\,\Delta t)\,.$$
\item The ionic contribution to the potential is computed by numerically integrating the ionic contribution of the electric field.
\item The electrodes' (constant) contribution to the electric field is computed by imposing that the overall potential difference across the capacitor (ions' and electrodes' contributions) be $2V(i\,\Delta t)=2V_0$ for $i>0$. This corresponds to computing the surface charge density $\sigma$ of the electrodes at time $i\, \Delta t$, which is proportional to the electrodes' electric field through the dielectric permittivity.
\item The overall electric field (ions' and electrodes' contributions) determines the ionic currents through the discrete analogous of \Eqref{eq:current}.
\item Ion densities at time $(i+1)\Delta t$ are computed through the discrete analogous of \Eqref{eq:continuityeq}, starting from the just determined currents at time $i\,\Delta t$. 
\end{enumerate}  
Iteration stops when the prescribed final time is reached.

The algorithm is flux-conservative \cite{NumericalRecipes} because it conserves the numerical integral of the densities for both ionic species, \ie~the total number of ions, up to machine precision. This is a consequence of the fact that density updates are computed by the finite-difference equivalent of \Eqref{eq:continuityeq}. At any time, the total number of ions is indeed
\begin{eqnarray}
& & \sum_k \dens_\pm(z_{k+\frac{1}{2}}, t+\Delta t)\, \Delta z = \nonumber \\
&= & \sum_k \left( \dens_\pm(z_{k+\frac{1}{2}},t) + \Delta t\, \frac{j_\pm(z_{k},t)-j_\pm(z_{k+1},t)}{\Delta z} \right) \Delta z \nonumber
\\ &=&\left( \sum_k \dens_\pm(z_{k+\frac{1}{2}},t) \Delta z \right) + \Delta t\,\left({j_\pm(z_{0},t)-j_\pm(z_{N},t)}\right) \nonumber
\\ &=& \sum_k \dens_\pm(z_{k+\frac{1}{2}},t) \Delta z\,.
\label{eq:fluxconservative}
\end{eqnarray}
The last equality follows from the zero-current condition on the electrodes, placed at $z_0=-L$ and $z_N=+L$. This property is exact for constant spacing $\Delta z$, as shown, but also for irregular spacings, provided that $\Delta z$ in \Eqref{eq:fluxconservative} is replaced everywhere by $(\Delta z)_{k+\frac{1}{2}}=z_{k+1}-z_k$. The usage of irregular spacings, with nodes and edges more dense close to the walls and less dense in the bulk, is essential to speed up the calculation when $L$ is orders of magnitude larger than the thickness of the double layer and it also favors stability in the nonlinear regime, as mentioned below.

A necessary condition for the stability of the algorithm is that the time step $\Delta t$ verify the Courant-Friedrichs-Lewy condition \cite{NumericalRecipes}, requiring that $\Delta t < \Delta z^2 /\max_\alpha\{D_\alpha\}$. Otherwise said, assuming purely diffusive motion, the typical distance traveled by the faster species in a time $\Delta t$ must be smaller than the lattice spacing. For the numerical calculations reported in the paper, the Courant factor $\Delta t / (\Delta z^2 /\max_\alpha\{D_\alpha\})$ is between 0.05 and 0.9.  

Lastly, it is necessary to have a sufficient number of nodes inside the double layer, for the results to be accurate and stable. This is particularly important when nonlinear effects set in, as densities and potential curves become much steeper in the double layer. For this reason, irregular node spacing is fundamental and was chosen in such a way that the distribution of nodes was linear in the bulk region and became exponentially dense closer to the electrodes. In addition, nonlinear effects favor electromigration with respect to diffusion and can elicit a rather violent response to the perturbation. This makes the Courant condition insufficient at strong voltages, calling for a further reduction of the time step. 

In contrast to canonical simulations, the results for grand-canonical systems (shown in Fig.~\ref{fig:ZeroPotqq}) are obtained by solving numerically the nonlinear Poisson-Boltzmann equation describing the final equilibrium state. In that case the Debye length sets the salt concentration in the reservoir instead of that in the system. In practice, we solve the Poisson-Boltzmann equation by an iterative procedure.

\newpage


\begin{thebibliography}{72}%
\makeatletter
\providecommand \@ifxundefined [1]{%
 \@ifx{#1\undefined}
}%
\providecommand \@ifnum [1]{%
 \ifnum #1\expandafter \@firstoftwo
 \else \expandafter \@secondoftwo
 \fi
}%
\providecommand \@ifx [1]{%
 \ifx #1\expandafter \@firstoftwo
 \else \expandafter \@secondoftwo
 \fi
}%
\providecommand \natexlab [1]{#1}%
\providecommand \enquote  [1]{``#1''}%
\providecommand \bibnamefont  [1]{#1}%
\providecommand \bibfnamefont [1]{#1}%
\providecommand \citenamefont [1]{#1}%
\providecommand \href@noop [0]{\@secondoftwo}%
\providecommand \href [0]{\begingroup \@sanitize@url \@href}%
\providecommand \@href[1]{\@@startlink{#1}\@@href}%
\providecommand \@@href[1]{\endgroup#1\@@endlink}%
\providecommand \@sanitize@url [0]{\catcode `\\12\catcode `\$12\catcode
  `\&12\catcode `\#12\catcode `\^12\catcode `\_12\catcode `\%12\relax}%
\providecommand \@@startlink[1]{}%
\providecommand \@@endlink[0]{}%
\providecommand \url  [0]{\begingroup\@sanitize@url \@url }%
\providecommand \@url [1]{\endgroup\@href {#1}{\urlprefix }}%
\providecommand \urlprefix  [0]{URL }%
\providecommand \Eprint [0]{\href }%
\providecommand \doibase [0]{https://doi.org/}%
\providecommand \selectlanguage [0]{\@gobble}%
\providecommand \bibinfo  [0]{\@secondoftwo}%
\providecommand \bibfield  [0]{\@secondoftwo}%
\providecommand \translation [1]{[#1]}%
\providecommand \BibitemOpen [0]{}%
\providecommand \bibitemStop [0]{}%
\providecommand \bibitemNoStop [0]{.\EOS\space}%
\providecommand \EOS [0]{\spacefactor3000\relax}%
\providecommand \BibitemShut  [1]{\csname bibitem#1\endcsname}%
\let\auto@bib@innerbib\@empty
\bibitem [{\citenamefont {Simon}\ and\ \citenamefont
  {Gogotsi}(2008)}]{Simon2008}%
  \BibitemOpen
  \bibfield  {author} {\bibinfo {author} {\bibfnamefont {P.}~\bibnamefont
  {Simon}}\ and\ \bibinfo {author} {\bibfnamefont {Y.}~\bibnamefont
  {Gogotsi}},\ }\bibfield  {title} {\bibinfo {title} {Materials for
  electrochemical capacitors},\ }\href {https://doi.org/10.1038/nmat2297}
  {\bibfield  {journal} {\bibinfo  {journal} {Nature Materials}\ }\textbf
  {\bibinfo {volume} {7}},\ \bibinfo {pages} {845} (\bibinfo {year}
  {2008})}\BibitemShut {NoStop}%
\bibitem [{\citenamefont {Chen}\ \emph {et~al.}(2020)\citenamefont {Chen},
  \citenamefont {Qiu},\ and\ \citenamefont {Cheng}}]{Chen2020}%
  \BibitemOpen
  \bibfield  {author} {\bibinfo {author} {\bibfnamefont {S.}~\bibnamefont
  {Chen}}, \bibinfo {author} {\bibfnamefont {L.}~\bibnamefont {Qiu}},\ and\
  \bibinfo {author} {\bibfnamefont {H.-M.}\ \bibnamefont {Cheng}},\ }\bibfield
  {title} {\bibinfo {title} {Carbon-{{Based Fibers}} for {{Advanced
  Electrochemical Energy Storage Devices}}},\ }\href
  {https://doi.org/10.1021/acs.chemrev.9b00466} {\bibfield  {journal} {\bibinfo
   {journal} {Chemical Reviews}\ }\textbf {\bibinfo {volume} {120}},\ \bibinfo
  {pages} {2811} (\bibinfo {year} {2020})}\BibitemShut {NoStop}%
\bibitem [{\citenamefont {Simon}\ and\ \citenamefont
  {Gogotsi}(2020)}]{simon_perspectives_2020}%
  \BibitemOpen
  \bibfield  {author} {\bibinfo {author} {\bibfnamefont {P.}~\bibnamefont
  {Simon}}\ and\ \bibinfo {author} {\bibfnamefont {Y.}~\bibnamefont
  {Gogotsi}},\ }\bibfield  {title} {\bibinfo {title} {Perspectives for
  electrochemical capacitors and related devices},\ }\href
  {https://doi.org/10.1038/s41563-020-0747-z} {\bibfield  {journal} {\bibinfo
  {journal} {Nature Materials}\ }\textbf {\bibinfo {volume} {19}},\ \bibinfo
  {pages} {1151} (\bibinfo {year} {2020})}\BibitemShut {NoStop}%
\bibitem [{\citenamefont {Winter}\ and\ \citenamefont
  {Brodd}(2004)}]{Winter2004}%
  \BibitemOpen
  \bibfield  {author} {\bibinfo {author} {\bibfnamefont {M.}~\bibnamefont
  {Winter}}\ and\ \bibinfo {author} {\bibfnamefont {R.~J.}\ \bibnamefont
  {Brodd}},\ }\bibfield  {title} {\bibinfo {title} {What {{Are Batteries}},
  {{Fuel Cells}}, and {{Supercapacitors}}?},\ }\href
  {https://doi.org/10.1021/cr020730k} {\bibfield  {journal} {\bibinfo
  {journal} {Chemical Reviews}\ }\textbf {\bibinfo {volume} {104}},\ \bibinfo
  {pages} {4245} (\bibinfo {year} {2004})}\BibitemShut {NoStop}%
\bibitem [{\citenamefont {Dubal}\ \emph {et~al.}(2018)\citenamefont {Dubal},
  \citenamefont {Chodankar}, \citenamefont {Kim},\ and\ \citenamefont
  {{Gomez-Romero}}}]{Dubal2018}%
  \BibitemOpen
  \bibfield  {author} {\bibinfo {author} {\bibfnamefont {D.~P.}\ \bibnamefont
  {Dubal}}, \bibinfo {author} {\bibfnamefont {N.~R.}\ \bibnamefont
  {Chodankar}}, \bibinfo {author} {\bibfnamefont {D.-H.}\ \bibnamefont {Kim}},\
  and\ \bibinfo {author} {\bibfnamefont {P.}~\bibnamefont {{Gomez-Romero}}},\
  }\bibfield  {title} {\bibinfo {title} {Towards flexible solid-state
  supercapacitors for smart and wearable electronics},\ }\href
  {https://doi.org/10.1039/C7CS00505A} {\bibfield  {journal} {\bibinfo
  {journal} {Chemical Society Reviews}\ }\textbf {\bibinfo {volume} {47}},\
  \bibinfo {pages} {2065} (\bibinfo {year} {2018})}\BibitemShut {NoStop}%
\bibitem [{\citenamefont {Brogioli}(2009)}]{Brogioli2009}%
  \BibitemOpen
  \bibfield  {author} {\bibinfo {author} {\bibfnamefont {D.}~\bibnamefont
  {Brogioli}},\ }\bibfield  {title} {\bibinfo {title} {{Extracting renewable
  energy from a salinity difference using a capacitor}},\ }\href
  {https://doi.org/10.1103/PhysRevLett.103.058501} {\bibfield  {journal}
  {\bibinfo  {journal} {Physical Review Letters}\ }\textbf {\bibinfo {volume}
  {103}},\ \bibinfo {pages} {058501} (\bibinfo {year} {2009})}\BibitemShut
  {NoStop}%
\bibitem [{\citenamefont {Boon}\ and\ \citenamefont {{van
  Roij}}(2011)}]{Boon2011}%
  \BibitemOpen
  \bibfield  {author} {\bibinfo {author} {\bibfnamefont {N.}~\bibnamefont
  {Boon}}\ and\ \bibinfo {author} {\bibfnamefont {R.}~\bibnamefont {{van
  Roij}}},\ }\bibfield  {title} {\bibinfo {title} {`{{Blue}} energy' from ion
  adsorption and electrode charging in sea and river water},\ }\href
  {https://doi.org/10.1080/00268976.2011.554334} {\bibfield  {journal}
  {\bibinfo  {journal} {Molecular Physics}\ }\textbf {\bibinfo {volume}
  {109}},\ \bibinfo {pages} {1229} (\bibinfo {year} {2011})}\BibitemShut
  {NoStop}%
\bibitem [{\citenamefont {Janssen}\ \emph {et~al.}(2014)\citenamefont
  {Janssen}, \citenamefont {H{\"{a}}rtel},\ and\ \citenamefont {van
  Roij}}]{Janssen2014}%
  \BibitemOpen
  \bibfield  {author} {\bibinfo {author} {\bibfnamefont {M.}~\bibnamefont
  {Janssen}}, \bibinfo {author} {\bibfnamefont {A.}~\bibnamefont
  {H{\"{a}}rtel}},\ and\ \bibinfo {author} {\bibfnamefont {R.}~\bibnamefont
  {van Roij}},\ }\bibfield  {title} {\bibinfo {title} {{Boosting Capacitive
  Blue-Energy and Desalination Devices with Waste Heat}},\ }\href
  {https://doi.org/10.1103/PhysRevLett.113.268501} {\bibfield  {journal}
  {\bibinfo  {journal} {Physical Review Letters}\ }\textbf {\bibinfo {volume}
  {113}},\ \bibinfo {pages} {268501} (\bibinfo {year} {2014})}\BibitemShut
  {NoStop}%
\bibitem [{\citenamefont {Vivier}\ and\ \citenamefont
  {Orazem}(2022)}]{vivier_impedance_2022}%
  \BibitemOpen
  \bibfield  {author} {\bibinfo {author} {\bibfnamefont {V.}~\bibnamefont
  {Vivier}}\ and\ \bibinfo {author} {\bibfnamefont {M.~E.}\ \bibnamefont
  {Orazem}},\ }\bibfield  {title} {\bibinfo {title} {Impedance {Analysis} of
  {Electrochemical} {Systems}},\ }\href
  {https://doi.org/10.1021/acs.chemrev.1c00876} {\bibfield  {journal} {\bibinfo
   {journal} {Chemical Reviews}\ }\textbf {\bibinfo {volume} {122}},\ \bibinfo
  {pages} {11131} (\bibinfo {year} {2022})}\BibitemShut {NoStop}%
\bibitem [{\citenamefont {Barbero}\ and\ \citenamefont
  {Evangelista}(2005)}]{Barbero2005a}%
  \BibitemOpen
  \bibfield  {author} {\bibinfo {author} {\bibfnamefont {G.}~\bibnamefont
  {Barbero}}\ and\ \bibinfo {author} {\bibfnamefont {L.~R.}\ \bibnamefont
  {Evangelista}},\ }\bibfield  {title} {\bibinfo {title} {Impedance
  {{Spectroscopy}} of a {{Cell}}: The {{Role}} of the {{Ions}}},\ }in\ \href
  {https://doi.org/10.1201/9781420037456.ch11} {\emph {\bibinfo {booktitle}
  {Adsorption {{Phenomena}} and {{Anchoring Energy}} in {{Nematic Liquid
  Crystals}}}}}\ (\bibinfo  {publisher} {{CRC Press}},\ \bibinfo {year}
  {2005})\ Chap.~\bibinfo {chapter} {11}\BibitemShut {NoStop}%
\bibitem [{\citenamefont {Jeanmairet}\ \emph {et~al.}(2022)\citenamefont
  {Jeanmairet}, \citenamefont {Rotenberg},\ and\ \citenamefont
  {Salanne}}]{Jeanmairet2022}%
  \BibitemOpen
  \bibfield  {author} {\bibinfo {author} {\bibfnamefont {G.}~\bibnamefont
  {Jeanmairet}}, \bibinfo {author} {\bibfnamefont {B.}~\bibnamefont
  {Rotenberg}},\ and\ \bibinfo {author} {\bibfnamefont {M.}~\bibnamefont
  {Salanne}},\ }\bibfield  {title} {\bibinfo {title} {Microscopic
  {{Simulations}} of {{Electrochemical Double-Layer Capacitors}}},\ }\href
  {https://doi.org/10.1021/acs.chemrev.1c00925} {\bibfield  {journal} {\bibinfo
   {journal} {Chemical Reviews}\ }\textbf {\bibinfo {volume} {122}},\ \bibinfo
  {pages} {10860} (\bibinfo {year} {2022})}\BibitemShut {NoStop}%
\bibitem [{\citenamefont {Kornyshev}(2007)}]{Kornyshev2007}%
  \BibitemOpen
  \bibfield  {author} {\bibinfo {author} {\bibfnamefont {A.~A.}\ \bibnamefont
  {Kornyshev}},\ }\bibfield  {title} {\bibinfo {title} {Double-{{Layer}} in
  {{Ionic Liquids}}:\, {{Paradigm Change}}?},\ }\href
  {https://doi.org/10.1021/jp067857o} {\bibfield  {journal} {\bibinfo
  {journal} {The Journal of Physical Chemistry B}\ }\textbf {\bibinfo {volume}
  {111}},\ \bibinfo {pages} {5545} (\bibinfo {year} {2007})}\BibitemShut
  {NoStop}%
\bibitem [{\citenamefont {Bazant}\ \emph {et~al.}(2011)\citenamefont {Bazant},
  \citenamefont {Storey},\ and\ \citenamefont {Kornyshev}}]{Bazant2011}%
  \BibitemOpen
  \bibfield  {author} {\bibinfo {author} {\bibfnamefont {M.~Z.}\ \bibnamefont
  {Bazant}}, \bibinfo {author} {\bibfnamefont {B.~D.}\ \bibnamefont {Storey}},\
  and\ \bibinfo {author} {\bibfnamefont {A.~A.}\ \bibnamefont {Kornyshev}},\
  }\bibfield  {title} {\bibinfo {title} {{Double layer in ionic liquids:
  Overscreening versus crowding}},\ }\href
  {https://doi.org/10.1103/PhysRevLett.106.046102} {\bibfield  {journal}
  {\bibinfo  {journal} {Physical Review Letters}\ }\textbf {\bibinfo {volume}
  {106}},\ \bibinfo {pages} {046102} (\bibinfo {year} {2011})}\BibitemShut
  {NoStop}%
\bibitem [{\citenamefont {Wu}(2022)}]{Wu2022}%
  \BibitemOpen
  \bibfield  {author} {\bibinfo {author} {\bibfnamefont {J.}~\bibnamefont
  {Wu}},\ }\bibfield  {title} {\bibinfo {title} {Understanding the {{Electric
  Double-Layer Structure}}, {{Capacitance}}, and {{Charging Dynamics}}},\
  }\href {https://doi.org/10.1021/acs.chemrev.2c00097} {\bibfield  {journal}
  {\bibinfo  {journal} {Chemical Reviews}\ }\textbf {\bibinfo {volume} {122}},\
  \bibinfo {pages} {10821} (\bibinfo {year} {2022})}\BibitemShut {NoStop}%
\bibitem [{\citenamefont {Janssen}(2019)}]{Janssen2019b}%
  \BibitemOpen
  \bibfield  {author} {\bibinfo {author} {\bibfnamefont {M.}~\bibnamefont
  {Janssen}},\ }\bibfield  {title} {\bibinfo {title} {Curvature affects
  electrolyte relaxation: {{Studies}} of spherical and cylindrical
  electrodes},\ }\href {https://doi.org/10.1103/PhysRevE.100.042602} {\bibfield
   {journal} {\bibinfo  {journal} {Physical Review E}\ }\textbf {\bibinfo
  {volume} {100}},\ \bibinfo {pages} {042602} (\bibinfo {year}
  {2019})}\BibitemShut {NoStop}%
\bibitem [{\citenamefont {Asta}\ \emph {et~al.}(2019)\citenamefont {Asta},
  \citenamefont {Palaia}, \citenamefont {Trizac}, \citenamefont {Levesque},\
  and\ \citenamefont {Rotenberg}}]{Asta2019JCP}%
  \BibitemOpen
  \bibfield  {author} {\bibinfo {author} {\bibfnamefont {A.~J.}\ \bibnamefont
  {Asta}}, \bibinfo {author} {\bibfnamefont {I.}~\bibnamefont {Palaia}},
  \bibinfo {author} {\bibfnamefont {E.}~\bibnamefont {Trizac}}, \bibinfo
  {author} {\bibfnamefont {M.}~\bibnamefont {Levesque}},\ and\ \bibinfo
  {author} {\bibfnamefont {B.}~\bibnamefont {Rotenberg}},\ }\bibfield  {title}
  {\bibinfo {title} {{Lattice Boltzmann electrokinetics simulation of
  nanocapacitors}},\ }\href {https://doi.org/10.1063/1.5119341} {\bibfield
  {journal} {\bibinfo  {journal} {The Journal of Chemical Physics}\ }\textbf
  {\bibinfo {volume} {151}},\ \bibinfo {pages} {114104} (\bibinfo {year}
  {2019})}\BibitemShut {NoStop}%
\bibitem [{\citenamefont {Lian}\ \emph {et~al.}(2020)\citenamefont {Lian},
  \citenamefont {Janssen}, \citenamefont {Liu},\ and\ \citenamefont {van
  Roij}}]{Lian2020}%
  \BibitemOpen
  \bibfield  {author} {\bibinfo {author} {\bibfnamefont {C.}~\bibnamefont
  {Lian}}, \bibinfo {author} {\bibfnamefont {M.}~\bibnamefont {Janssen}},
  \bibinfo {author} {\bibfnamefont {H.}~\bibnamefont {Liu}},\ and\ \bibinfo
  {author} {\bibfnamefont {R.}~\bibnamefont {van Roij}},\ }\bibfield  {title}
  {\bibinfo {title} {{Blessing and Curse: How a Supercapacitor's Large
  Capacitance Causes its Slow Charging}},\ }\href
  {https://doi.org/10.1103/PhysRevLett.124.076001} {\bibfield  {journal}
  {\bibinfo  {journal} {Physical Review Letters}\ }\textbf {\bibinfo {volume}
  {124}},\ \bibinfo {pages} {076001} (\bibinfo {year} {2020})}\BibitemShut
  {NoStop}%
\bibitem [{\citenamefont {Ma}\ \emph {et~al.}(2022)\citenamefont {Ma},
  \citenamefont {Janssen}, \citenamefont {Lian},\ and\ \citenamefont {{van
  Roij}}}]{Ma2022}%
  \BibitemOpen
  \bibfield  {author} {\bibinfo {author} {\bibfnamefont {K.}~\bibnamefont
  {Ma}}, \bibinfo {author} {\bibfnamefont {M.}~\bibnamefont {Janssen}},
  \bibinfo {author} {\bibfnamefont {C.}~\bibnamefont {Lian}},\ and\ \bibinfo
  {author} {\bibfnamefont {R.}~\bibnamefont {{van Roij}}},\ }\bibfield  {title}
  {\bibinfo {title} {Dynamic density functional theory for the charging of
  electric double layer capacitors},\ }\href
  {https://doi.org/10.1063/5.0081827} {\bibfield  {journal} {\bibinfo
  {journal} {The Journal of Chemical Physics}\ }\textbf {\bibinfo {volume}
  {156}},\ \bibinfo {pages} {084101} (\bibinfo {year} {2022})}\BibitemShut
  {NoStop}%
\bibitem [{\citenamefont {Yang}\ \emph {et~al.}(2022)\citenamefont {Yang},
  \citenamefont {Janssen}, \citenamefont {Lian},\ and\ \citenamefont {{van
  Roij}}}]{Yang2022}%
  \BibitemOpen
  \bibfield  {author} {\bibinfo {author} {\bibfnamefont {J.}~\bibnamefont
  {Yang}}, \bibinfo {author} {\bibfnamefont {M.}~\bibnamefont {Janssen}},
  \bibinfo {author} {\bibfnamefont {C.}~\bibnamefont {Lian}},\ and\ \bibinfo
  {author} {\bibfnamefont {R.}~\bibnamefont {{van Roij}}},\ }\bibfield  {title}
  {\bibinfo {title} {Simulating the charging of cylindrical electrolyte-filled
  pores with the modified {{Poisson}}--{{Nernst}}--{{Planck}} equations},\
  }\href {https://doi.org/10.1063/5.0094553} {\bibfield  {journal} {\bibinfo
  {journal} {The Journal of Chemical Physics}\ }\textbf {\bibinfo {volume}
  {156}},\ \bibinfo {pages} {214105} (\bibinfo {year} {2022})}\BibitemShut
  {NoStop}%
\bibitem [{\citenamefont {Varela}\ \emph {et~al.}(2021)\citenamefont {Varela},
  \citenamefont {Andraus}, \citenamefont {Trizac},\ and\ \citenamefont
  {T{\'e}llez}}]{Varela2021}%
  \BibitemOpen
  \bibfield  {author} {\bibinfo {author} {\bibfnamefont {L.}~\bibnamefont
  {Varela}}, \bibinfo {author} {\bibfnamefont {S.}~\bibnamefont {Andraus}},
  \bibinfo {author} {\bibfnamefont {E.}~\bibnamefont {Trizac}},\ and\ \bibinfo
  {author} {\bibfnamefont {G.}~\bibnamefont {T{\'e}llez}},\ }\bibfield  {title}
  {\bibinfo {title} {Relaxation dynamics of two interacting electrical
  double-layers in a {{1D Coulomb}} system},\ }\href
  {https://doi.org/10.1088/1361-648X/ac1237} {\bibfield  {journal} {\bibinfo
  {journal} {Journal of Physics: Condensed Matter}\ }\textbf {\bibinfo {volume}
  {33}},\ \bibinfo {pages} {394001} (\bibinfo {year} {2021})}\BibitemShut
  {NoStop}%
\bibitem [{\citenamefont {Breitsprecher}\ \emph {et~al.}(2018)\citenamefont
  {Breitsprecher}, \citenamefont {Holm},\ and\ \citenamefont
  {Kondrat}}]{Breitsprecher2018}%
  \BibitemOpen
  \bibfield  {author} {\bibinfo {author} {\bibfnamefont {K.}~\bibnamefont
  {Breitsprecher}}, \bibinfo {author} {\bibfnamefont {C.}~\bibnamefont
  {Holm}},\ and\ \bibinfo {author} {\bibfnamefont {S.}~\bibnamefont
  {Kondrat}},\ }\bibfield  {title} {\bibinfo {title} {Charge me slowly, {I} am
  in a hurry: Optimizing charge-discharge cycles in nanoporous
  supercapacitors},\ }\href {https://doi.org/10.1021/acsnano.8b04785}
  {\bibfield  {journal} {\bibinfo  {journal} {ACS Nano}\ }\textbf {\bibinfo
  {volume} {12}},\ \bibinfo {pages} {9733} (\bibinfo {year}
  {2018})}\BibitemShut {NoStop}%
\bibitem [{\citenamefont {Palaia}(2019)}]{phd}%
  \BibitemOpen
  \bibfield  {author} {\bibinfo {author} {\bibfnamefont {I.}~\bibnamefont
  {Palaia}},\ }\emph {\bibinfo {title} {{Charged systems in, out of, and driven
  to equilibrium: from nanocapacitors to cement}}},\ \href
  {http://www.theses.fr/2019SACLS398} {Ph.D. thesis},\ \bibinfo  {school}
  {Paris-Saclay University} (\bibinfo {year} {2019})\BibitemShut {NoStop}%
\bibitem [{\citenamefont {Breitsprecher}\ \emph {et~al.}(2020)\citenamefont
  {Breitsprecher}, \citenamefont {Janssen}, \citenamefont {Srimuk},
  \citenamefont {Mehdi}, \citenamefont {Presser}, \citenamefont {Holm},\ and\
  \citenamefont {Kondrat}}]{Breitsprecher2020}%
  \BibitemOpen
  \bibfield  {author} {\bibinfo {author} {\bibfnamefont {K.}~\bibnamefont
  {Breitsprecher}}, \bibinfo {author} {\bibfnamefont {M.}~\bibnamefont
  {Janssen}}, \bibinfo {author} {\bibfnamefont {P.}~\bibnamefont {Srimuk}},
  \bibinfo {author} {\bibfnamefont {B.~L.}\ \bibnamefont {Mehdi}}, \bibinfo
  {author} {\bibfnamefont {V.}~\bibnamefont {Presser}}, \bibinfo {author}
  {\bibfnamefont {C.}~\bibnamefont {Holm}},\ and\ \bibinfo {author}
  {\bibfnamefont {S.}~\bibnamefont {Kondrat}},\ }\bibfield  {title} {\bibinfo
  {title} {How to speed up ion transport in nanopores},\ }\href
  {https://doi.org/10.1038/s41467-020-19903-6} {\bibfield  {journal} {\bibinfo
  {journal} {Nature Communications}\ }\textbf {\bibinfo {volume} {11}},\
  \bibinfo {pages} {6085} (\bibinfo {year} {2020})}\BibitemShut {NoStop}%
\bibitem [{\citenamefont {Bazant}\ \emph {et~al.}(2004)\citenamefont {Bazant},
  \citenamefont {Thornton},\ and\ \citenamefont {Ajdari}}]{Bazant2004}%
  \BibitemOpen
  \bibfield  {author} {\bibinfo {author} {\bibfnamefont {M.~Z.}\ \bibnamefont
  {Bazant}}, \bibinfo {author} {\bibfnamefont {K.}~\bibnamefont {Thornton}},\
  and\ \bibinfo {author} {\bibfnamefont {A.}~\bibnamefont {Ajdari}},\
  }\bibfield  {title} {\bibinfo {title} {{Diffuse-charge dynamics in
  electrochemical systems}},\ }\href
  {https://doi.org/10.1103/PhysRevE.70.021506} {\bibfield  {journal} {\bibinfo
  {journal} {Physical Review E}\ }\textbf {\bibinfo {volume} {70}},\ \bibinfo
  {pages} {021506} (\bibinfo {year} {2004})}\BibitemShut {NoStop}%
\bibitem [{\citenamefont {Janssen}\ and\ \citenamefont
  {Bier}(2018)}]{Janssen2018}%
  \BibitemOpen
  \bibfield  {author} {\bibinfo {author} {\bibfnamefont {M.}~\bibnamefont
  {Janssen}}\ and\ \bibinfo {author} {\bibfnamefont {M.}~\bibnamefont {Bier}},\
  }\bibfield  {title} {\bibinfo {title} {{Transient dynamics of electric
  double-layer capacitors: Exact expressions within the Debye-Falkenhagen
  approximation}},\ }\href {https://doi.org/10.1103/PhysRevE.97.052616}
  {\bibfield  {journal} {\bibinfo  {journal} {Physical Review E}\ }\textbf
  {\bibinfo {volume} {97}},\ \bibinfo {pages} {052616} (\bibinfo {year}
  {2018})}\BibitemShut {NoStop}%
\bibitem [{\citenamefont {Palaia}\ \emph {et~al.}(2025)\citenamefont {Palaia},
  \citenamefont {Asta}, \citenamefont {Dutta}, \citenamefont {Warren},
  \citenamefont {Rotenberg},\ and\ \citenamefont {Trizac}}]{mainpaper}%
  \BibitemOpen
  \bibfield  {author} {\bibinfo {author} {\bibfnamefont {I.}~\bibnamefont
  {Palaia}}, \bibinfo {author} {\bibfnamefont {A.~J.}\ \bibnamefont {Asta}},
  \bibinfo {author} {\bibfnamefont {M.}~\bibnamefont {Dutta}}, \bibinfo
  {author} {\bibfnamefont {P.~B.}\ \bibnamefont {Warren}}, \bibinfo {author}
  {\bibfnamefont {B.}~\bibnamefont {Rotenberg}},\ and\ \bibinfo {author}
  {\bibfnamefont {E.}~\bibnamefont {Trizac}},\ }\bibfield  {title} {\bibinfo
  {title} {Charging dynamics of electric double layer nanocapacitors in
  mean-field},\ }\bibfield  {journal} {\bibinfo  {journal} {Physical Review
  Letters}\ }\href {https://doi.org/10.1103/72b9-c8cq} {10.1103/72b9-c8cq}
  (\bibinfo {year} {2025}),\ \bibinfo {note} {{}}\BibitemShut {NoStop}%
\bibitem [{\citenamefont {Hunter}(2001)}]{Hunter}%
  \BibitemOpen
  \bibfield  {author} {\bibinfo {author} {\bibfnamefont {R.~J.}\ \bibnamefont
  {Hunter}},\ }\href {https://doi.org/10.1016/s0927-7757(02)00170-x} {\emph
  {\bibinfo {title} {{Foundations of Colloid Science}}}},\ \bibinfo {edition}
  {2nd}\ ed.\ (\bibinfo  {publisher} {Oxford University Press},\ \bibinfo
  {year} {2001})\BibitemShut {NoStop}%
\bibitem [{\citenamefont {Andelman}(2006)}]{Andelman2010}%
  \BibitemOpen
  \bibfield  {author} {\bibinfo {author} {\bibfnamefont {D.}~\bibnamefont
  {Andelman}},\ }\bibfield  {title} {\bibinfo {title} {Introduction to
  electrostatics in soft and biological matter},\ }in\ \href
  {https://doi.org/10.1201/9781420003338.ch6} {\emph {\bibinfo {booktitle}
  {Soft {{Condensed Matter Physics}} in {{Molecular}} and {{Cell Biology}}}}},\
  \bibinfo {editor} {edited by\ \bibinfo {editor} {\bibfnamefont {W.~C.~K.}\
  \bibnamefont {Poon}}\ and\ \bibinfo {editor} {\bibfnamefont {D.}~\bibnamefont
  {Andelman}}}\ (\bibinfo  {publisher} {{CRC Press}},\ \bibinfo {year} {2006})\
  pp.\ \bibinfo {pages} {97--122}\BibitemShut {NoStop}%
\bibitem [{\citenamefont {Barbero}\ and\ \citenamefont
  {Alexe‐Ionescu}(2005)}]{Barbero2005b}%
  \BibitemOpen
  \bibfield  {author} {\bibinfo {author} {\bibfnamefont {G.}~\bibnamefont
  {Barbero}}\ and\ \bibinfo {author} {\bibfnamefont {A.~L.}\ \bibnamefont
  {Alexe‐Ionescu}},\ }\bibfield  {title} {\bibinfo {title} {Role of the
  diffuse layer of the ionic charge on the impedance spectroscopy of a cell of
  liquid},\ }\href {https://doi.org/10.1080/02678290500228105} {\bibfield
  {journal} {\bibinfo  {journal} {Liquid Crystals}\ }\textbf {\bibinfo {volume}
  {32}},\ \bibinfo {pages} {943} (\bibinfo {year} {2005})}\BibitemShut
  {NoStop}%
\bibitem [{\citenamefont {Barbero}\ \emph {et~al.}(2008)\citenamefont
  {Barbero}, \citenamefont {Batalioto},\ and\ \citenamefont
  {Figueiredo~Neto}}]{Barbero2008}%
  \BibitemOpen
  \bibfield  {author} {\bibinfo {author} {\bibfnamefont {G.}~\bibnamefont
  {Barbero}}, \bibinfo {author} {\bibfnamefont {F.}~\bibnamefont {Batalioto}},\
  and\ \bibinfo {author} {\bibfnamefont {A.~M.}\ \bibnamefont
  {Figueiredo~Neto}},\ }\bibfield  {title} {\bibinfo {title} {Theory of
  small-signal ac response of a dielectric liquid containing two groups of
  ions},\ }\href {https://doi.org/10.1063/1.2908044} {\bibfield  {journal}
  {\bibinfo  {journal} {Applied Physics Letters}\ }\textbf {\bibinfo {volume}
  {92}},\ \bibinfo {pages} {172908} (\bibinfo {year} {2008})}\BibitemShut
  {NoStop}%
\bibitem [{\citenamefont {MaCdonald}(1970)}]{MaCdonald1970}%
  \BibitemOpen
  \bibfield  {author} {\bibinfo {author} {\bibfnamefont {J.~R.}\ \bibnamefont
  {MaCdonald}},\ }\bibfield  {title} {\bibinfo {title} {Double layer
  capacitance and relaxation in electrolytes and solids},\ }\href
  {https://doi.org/10.1039/tf9706600943} {\bibfield  {journal} {\bibinfo
  {journal} {Transactions of the Faraday Society}\ }\textbf {\bibinfo {volume}
  {66}},\ \bibinfo {pages} {943} (\bibinfo {year} {1970})}\BibitemShut
  {NoStop}%
\bibitem [{\citenamefont {Kornyshev}\ and\ \citenamefont
  {Vorotyntsev}(1977)}]{Kornyshev1977}%
  \BibitemOpen
  \bibfield  {author} {\bibinfo {author} {\bibfnamefont {A.~A.}\ \bibnamefont
  {Kornyshev}}\ and\ \bibinfo {author} {\bibfnamefont {M.~A.}\ \bibnamefont
  {Vorotyntsev}},\ }\bibfield  {title} {\bibinfo {title} {Electric current
  across the metal-solid electrolyte interface {{II}}. low-amplitude
  alternating current},\ }\href {https://doi.org/10.1002/pssa.2210390225}
  {\bibfield  {journal} {\bibinfo  {journal} {physica status solidi (a)}\
  }\textbf {\bibinfo {volume} {39}},\ \bibinfo {pages} {573} (\bibinfo {year}
  {1977})}\BibitemShut {NoStop}%
\bibitem [{\citenamefont {Kornyshev}\ and\ \citenamefont
  {Vorotyntsev}(1981)}]{Kornyshev1981}%
  \BibitemOpen
  \bibfield  {author} {\bibinfo {author} {\bibfnamefont {A.}~\bibnamefont
  {Kornyshev}}\ and\ \bibinfo {author} {\bibfnamefont {M.}~\bibnamefont
  {Vorotyntsev}},\ }\bibfield  {title} {\bibinfo {title} {Conductivity and
  space charge phenomena in solid electrolytes with one mobile charge carrier
  species, a review with original material},\ }\href
  {https://doi.org/10.1016/0013-4686(81)85017-7} {\bibfield  {journal}
  {\bibinfo  {journal} {Electrochimica Acta}\ }\textbf {\bibinfo {volume}
  {26}},\ \bibinfo {pages} {303} (\bibinfo {year} {1981})}\BibitemShut
  {NoStop}%
\bibitem [{\citenamefont {Bocquet}\ and\ \citenamefont
  {Charlaix}(2010)}]{BocquetRev2010}%
  \BibitemOpen
  \bibfield  {author} {\bibinfo {author} {\bibfnamefont {L.}~\bibnamefont
  {Bocquet}}\ and\ \bibinfo {author} {\bibfnamefont {E.}~\bibnamefont
  {Charlaix}},\ }\bibfield  {title} {\bibinfo {title} {{Nanofluidics, from bulk
  to interfaces}},\ }\href {https://doi.org/10.1039/B909366B} {\bibfield
  {journal} {\bibinfo  {journal} {Chemical Society Reviews}\ }\textbf {\bibinfo
  {volume} {39}},\ \bibinfo {pages} {1073} (\bibinfo {year}
  {2010})}\BibitemShut {NoStop}%
\bibitem [{Lyk(1995)}]{Lyklema1995}%
  \BibitemOpen
  \bibfield  {title} {\bibinfo {title} {{{Electric Double Layers}}},\ }in\
  \href {https://doi.org/10.1016/S1874-5679(06)80006-1} {\emph {\bibinfo
  {booktitle} {Fundamentals of {{Interface}} and {{Colloid Science}}}}},\
  Vol.~\bibinfo {volume} {2},\ \bibinfo {editor} {edited by\ \bibinfo {editor}
  {\bibfnamefont {J.}~\bibnamefont {Lyklema}}}\ (\bibinfo  {publisher}
  {{Academic Press}},\ \bibinfo {year} {1995})\BibitemShut {NoStop}%
\bibitem [{\citenamefont {Macdonald}(1954)}]{Macdonald1954}%
  \BibitemOpen
  \bibfield  {author} {\bibinfo {author} {\bibfnamefont {J.~R.}\ \bibnamefont
  {Macdonald}},\ }\bibfield  {title} {\bibinfo {title} {Theory of the
  {{Differential Capacitance}} of the {{Double Layer}} in {{Unadsorbed
  Electrolytes}}},\ }\href {https://doi.org/10.1063/1.1739933} {\bibfield
  {journal} {\bibinfo  {journal} {The Journal of Chemical Physics}\ }\textbf
  {\bibinfo {volume} {22}},\ \bibinfo {pages} {1857} (\bibinfo {year}
  {1954})}\BibitemShut {NoStop}%
\bibitem [{\citenamefont {Beunis}\ \emph {et~al.}(2008)\citenamefont {Beunis},
  \citenamefont {Strubbe}, \citenamefont {Marescaux}, \citenamefont {Beeckman},
  \citenamefont {Neyts},\ and\ \citenamefont {Verschueren}}]{Beunis2008}%
  \BibitemOpen
  \bibfield  {author} {\bibinfo {author} {\bibfnamefont {F.}~\bibnamefont
  {Beunis}}, \bibinfo {author} {\bibfnamefont {F.}~\bibnamefont {Strubbe}},
  \bibinfo {author} {\bibfnamefont {M.}~\bibnamefont {Marescaux}}, \bibinfo
  {author} {\bibfnamefont {J.}~\bibnamefont {Beeckman}}, \bibinfo {author}
  {\bibfnamefont {K.}~\bibnamefont {Neyts}},\ and\ \bibinfo {author}
  {\bibfnamefont {A.~R.~M.}\ \bibnamefont {Verschueren}},\ }\bibfield  {title}
  {\bibinfo {title} {Dynamics of charge transport in planar devices},\ }\href
  {https://doi.org/10.1103/PhysRevE.78.011502} {\bibfield  {journal} {\bibinfo
  {journal} {Physical Review E}\ }\textbf {\bibinfo {volume} {78}},\ \bibinfo
  {pages} {011502} (\bibinfo {year} {2008})}\BibitemShut {NoStop}%
\bibitem [{\citenamefont {Beunis}\ \emph {et~al.}(2007)\citenamefont {Beunis},
  \citenamefont {Strubbe}, \citenamefont {Marescaux}, \citenamefont {Neyts},\
  and\ \citenamefont {Verschueren}}]{Beunis2007}%
  \BibitemOpen
  \bibfield  {author} {\bibinfo {author} {\bibfnamefont {F.}~\bibnamefont
  {Beunis}}, \bibinfo {author} {\bibfnamefont {F.}~\bibnamefont {Strubbe}},
  \bibinfo {author} {\bibfnamefont {M.}~\bibnamefont {Marescaux}}, \bibinfo
  {author} {\bibfnamefont {K.}~\bibnamefont {Neyts}},\ and\ \bibinfo {author}
  {\bibfnamefont {A.~R.~M.}\ \bibnamefont {Verschueren}},\ }\bibfield  {title}
  {\bibinfo {title} {Diffuse double layer charging in nonpolar liquids},\
  }\href {https://doi.org/10.1063/1.2805229} {\bibfield  {journal} {\bibinfo
  {journal} {Applied Physics Letters}\ }\textbf {\bibinfo {volume} {91}},\
  \bibinfo {pages} {182911} (\bibinfo {year} {2007})}\BibitemShut {NoStop}%
\bibitem [{\citenamefont {Grahame}(1953)}]{Grahame1953}%
  \BibitemOpen
  \bibfield  {author} {\bibinfo {author} {\bibfnamefont {D.~C.}\ \bibnamefont
  {Grahame}},\ }\bibfield  {title} {\bibinfo {title} {{Diffuse double layer
  theory for electrolytes of unsymmetrical valence types}},\ }\href
  {https://doi.org/10.1063/1.1699109} {\bibfield  {journal} {\bibinfo
  {journal} {The Journal of Chemical Physics}\ }\textbf {\bibinfo {volume}
  {21}},\ \bibinfo {pages} {1054} (\bibinfo {year} {1953})}\BibitemShut
  {NoStop}%
\bibitem [{\citenamefont {Lelidis}\ and\ \citenamefont
  {Barbero}(2005)}]{Lelidis2005}%
  \BibitemOpen
  \bibfield  {author} {\bibinfo {author} {\bibfnamefont {I.}~\bibnamefont
  {Lelidis}}\ and\ \bibinfo {author} {\bibfnamefont {G.}~\bibnamefont
  {Barbero}},\ }\bibfield  {title} {\bibinfo {title} {Effect of different
  anionic and cationic mobilities on the impedance spectroscopy measurements},\
  }\href {https://doi.org/10.1016/j.physleta.2005.06.038} {\bibfield  {journal}
  {\bibinfo  {journal} {Physics Letters A}\ }\textbf {\bibinfo {volume}
  {343}},\ \bibinfo {pages} {440} (\bibinfo {year} {2005})}\BibitemShut
  {NoStop}%
\bibitem [{\citenamefont {Barbero}\ and\ \citenamefont
  {Lelidis}(2007)}]{Barbero2007}%
  \BibitemOpen
  \bibfield  {author} {\bibinfo {author} {\bibfnamefont {G.}~\bibnamefont
  {Barbero}}\ and\ \bibinfo {author} {\bibfnamefont {I.}~\bibnamefont
  {Lelidis}},\ }\bibfield  {title} {\bibinfo {title} {{Evidence of the
  ambipolar diffusion in the impedance spectroscopy of an electrolytic cell}},\
  }\href {https://doi.org/10.1103/PhysRevE.76.051501} {\bibfield  {journal}
  {\bibinfo  {journal} {Physical Review E}\ }\textbf {\bibinfo {volume} {76}},\
  \bibinfo {pages} {051501} (\bibinfo {year} {2007})}\BibitemShut {NoStop}%
\bibitem [{\citenamefont {Chassagne}\ \emph {et~al.}(2016)\citenamefont
  {Chassagne}, \citenamefont {Dubois}, \citenamefont {Jim{\'{e}}nez},
  \citenamefont {van~der Ploeg},\ and\ \citenamefont {van
  Turnhout}}]{Chassagne2016}%
  \BibitemOpen
  \bibfield  {author} {\bibinfo {author} {\bibfnamefont {C.}~\bibnamefont
  {Chassagne}}, \bibinfo {author} {\bibfnamefont {E.}~\bibnamefont {Dubois}},
  \bibinfo {author} {\bibfnamefont {M.~L.}\ \bibnamefont {Jim{\'{e}}nez}},
  \bibinfo {author} {\bibfnamefont {J.~P.~M.}\ \bibnamefont {van~der Ploeg}},\
  and\ \bibinfo {author} {\bibfnamefont {J.}~\bibnamefont {van Turnhout}},\
  }\bibfield  {title} {\bibinfo {title} {{Compensating for Electrode
  Polarization in Dielectric Spectroscopy Studies of Colloidal Suspensions:
  Theoretical Assessment of Existing Methods}},\ }\href
  {https://doi.org/10.3389/fchem.2016.00030} {\bibfield  {journal} {\bibinfo
  {journal} {Frontiers in Chemistry}\ }\textbf {\bibinfo {volume} {4}},\
  \bibinfo {pages} {1} (\bibinfo {year} {2016})}\BibitemShut {NoStop}%
\bibitem [{\citenamefont {Antonova}\ \emph {et~al.}(2020)\citenamefont
  {Antonova}, \citenamefont {Barbero}, \citenamefont {Evangelista},\ and\
  \citenamefont {Tilli}}]{Antonova2020}%
  \BibitemOpen
  \bibfield  {author} {\bibinfo {author} {\bibfnamefont {A.}~\bibnamefont
  {Antonova}}, \bibinfo {author} {\bibfnamefont {G.}~\bibnamefont {Barbero}},
  \bibinfo {author} {\bibfnamefont {L.~R.}\ \bibnamefont {Evangelista}},\ and\
  \bibinfo {author} {\bibfnamefont {P.}~\bibnamefont {Tilli}},\ }\bibfield
  {title} {\bibinfo {title} {Ambipolar diffusion in the low frequency impedance
  response of electrolytic cells},\ }\href
  {https://doi.org/10.1088/1742-5468/ab7a23} {\bibfield  {journal} {\bibinfo
  {journal} {Journal of Statistical Mechanics: Theory and Experiment}\ }\textbf
  {\bibinfo {volume} {2020}},\ \bibinfo {pages} {043202} (\bibinfo {year}
  {2020})}\BibitemShut {NoStop}%
\bibitem [{\citenamefont {Robinson}\ and\ \citenamefont
  {Stokes}(2002)}]{Robinson1959}%
  \BibitemOpen
  \bibfield  {author} {\bibinfo {author} {\bibfnamefont {R.~A.}\ \bibnamefont
  {Robinson}}\ and\ \bibinfo {author} {\bibfnamefont {R.~H.}\ \bibnamefont
  {Stokes}},\ }\href@noop {} {\emph {\bibinfo {title} {{Electrolyte solutions:
  Second Revised Edition}}}}\ (\bibinfo  {publisher} {Dover Publications},\
  \bibinfo {year} {2002})\ pp.\ \bibinfo {pages} {286--292}\BibitemShut
  {NoStop}%
\bibitem [{\citenamefont {Zhao}(2011)}]{Zhao2011}%
  \BibitemOpen
  \bibfield  {author} {\bibinfo {author} {\bibfnamefont {H.}~\bibnamefont
  {Zhao}},\ }\bibfield  {title} {\bibinfo {title} {Diffuse-charge dynamics of
  ionic liquids in electrochemical systems},\ }\href
  {https://doi.org/10.1103/PhysRevE.84.051504} {\bibfield  {journal} {\bibinfo
  {journal} {Physical Review E}\ }\textbf {\bibinfo {volume} {84}},\ \bibinfo
  {pages} {051504} (\bibinfo {year} {2011})}\BibitemShut {NoStop}%
\bibitem [{\citenamefont {Fertig}\ and\ \citenamefont
  {Janssen}(2025)}]{Fertig2025}%
  \BibitemOpen
  \bibfield  {author} {\bibinfo {author} {\bibfnamefont {D.}~\bibnamefont
  {Fertig}}\ and\ \bibinfo {author} {\bibfnamefont {M.}~\bibnamefont
  {Janssen}},\ }\href {https://doi.org/10.48550/arXiv.2503.15171} {\bibinfo
  {title} {Charging dynamics of electric double layer capacitors including
  beyond-mean-field electrostatic correlations}} (\bibinfo {year} {2025}),\
  \Eprint {https://arxiv.org/abs/2503.15171} {arXiv:2503.15171 [physics]}
  \BibitemShut {NoStop}%
\bibitem [{\citenamefont {Netz}\ and\ \citenamefont
  {Orland}(2000)}]{Netz2000EPJE}%
  \BibitemOpen
  \bibfield  {author} {\bibinfo {author} {\bibfnamefont {R.~R.}\ \bibnamefont
  {Netz}}\ and\ \bibinfo {author} {\bibfnamefont {H.}~\bibnamefont {Orland}},\
  }\bibfield  {title} {\bibinfo {title} {{Beyond Poisson-Boltzmann: Fluctuation
  effects and correlation functions}},\ }\href
  {https://doi.org/10.1007/s101890050023} {\bibfield  {journal} {\bibinfo
  {journal} {The European Physical Journal E}\ }\textbf {\bibinfo {volume}
  {1}},\ \bibinfo {pages} {203} (\bibinfo {year} {2000})}\BibitemShut {NoStop}%
\bibitem [{\citenamefont {Naji}\ \emph {et~al.}(2013)\citenamefont {Naji},
  \citenamefont {Kandu{\v{c}}}, \citenamefont {Forsman},\ and\ \citenamefont
  {Podgornik}}]{Naji2013}%
  \BibitemOpen
  \bibfield  {author} {\bibinfo {author} {\bibfnamefont {A.}~\bibnamefont
  {Naji}}, \bibinfo {author} {\bibfnamefont {M.}~\bibnamefont {Kandu{\v{c}}}},
  \bibinfo {author} {\bibfnamefont {J.}~\bibnamefont {Forsman}},\ and\ \bibinfo
  {author} {\bibfnamefont {R.}~\bibnamefont {Podgornik}},\ }\bibfield  {title}
  {\bibinfo {title} {{Perspective: Coulomb fluids—Weak coupling, strong
  coupling, in between and beyond}},\ }\href
  {https://doi.org/10.1063/1.4824681} {\bibfield  {journal} {\bibinfo
  {journal} {The Journal of Chemical Physics}\ }\textbf {\bibinfo {volume}
  {139}},\ \bibinfo {pages} {150901} (\bibinfo {year} {2013})}\BibitemShut
  {NoStop}%
\bibitem [{\citenamefont {Trizac}\ and\ \citenamefont {Samaj}(2012)}]{Varenna}%
  \BibitemOpen
  \bibfield  {author} {\bibinfo {author} {\bibfnamefont {E.}~\bibnamefont
  {Trizac}}\ and\ \bibinfo {author} {\bibfnamefont {L.}~\bibnamefont {Samaj}},\
  }\bibfield  {title} {\bibinfo {title} {{Like-charge colloidal attraction: a
  simple argument}},\ }in\ \href {https://doi.org/10.3254/978-1-61499-278-3-61}
  {\emph {\bibinfo {booktitle} {Proceedings of the International School of
  Physics Enrico Fermi}}},\ Vol.\ \bibinfo {volume} {184},\ \bibinfo {editor}
  {edited by\ \bibinfo {editor} {\bibfnamefont {C.}~\bibnamefont {Bechinger}},
  \bibinfo {editor} {\bibfnamefont {F.}~\bibnamefont {Sciortino}},\ and\
  \bibinfo {editor} {\bibfnamefont {P.}~\bibnamefont {Ziherl}}}\ (\bibinfo
  {year} {2012})\ pp.\ \bibinfo {pages} {61--73},\ \Eprint
  {https://arxiv.org/abs/1210.5843} {arXiv:1210.5843} \BibitemShut {NoStop}%
\bibitem [{\citenamefont {Palaia}\ \emph {et~al.}(2022)\citenamefont {Palaia},
  \citenamefont {Goyal}, \citenamefont {Del~Gado}, \citenamefont {{\v S}amaj},\
  and\ \citenamefont {Trizac}}]{Palaia2022a}%
  \BibitemOpen
  \bibfield  {author} {\bibinfo {author} {\bibfnamefont {I.}~\bibnamefont
  {Palaia}}, \bibinfo {author} {\bibfnamefont {A.}~\bibnamefont {Goyal}},
  \bibinfo {author} {\bibfnamefont {E.}~\bibnamefont {Del~Gado}}, \bibinfo
  {author} {\bibfnamefont {L.}~\bibnamefont {{\v S}amaj}},\ and\ \bibinfo
  {author} {\bibfnamefont {E.}~\bibnamefont {Trizac}},\ }\bibfield  {title}
  {\bibinfo {title} {Like-{{Charge Attraction}} at the {{Nanoscale}}:
  {{Ground-State Correlations}} and {{Water Destructuring}}},\ }\href
  {https://doi.org/10.1021/acs.jpcb.2c00028} {\bibfield  {journal} {\bibinfo
  {journal} {The Journal of Physical Chemistry B}\ }\textbf {\bibinfo {volume}
  {126}},\ \bibinfo {pages} {3143} (\bibinfo {year} {2022})}\BibitemShut
  {NoStop}%
\bibitem [{\citenamefont {Moreira}\ and\ \citenamefont
  {Netz}(2000)}]{Moreira2000}%
  \BibitemOpen
  \bibfield  {author} {\bibinfo {author} {\bibfnamefont {A.~G.}\ \bibnamefont
  {Moreira}}\ and\ \bibinfo {author} {\bibfnamefont {R.~R.}\ \bibnamefont
  {Netz}},\ }\bibfield  {title} {\bibinfo {title} {{Strong-coupling theory for
  counter-ion distributions}},\ }\href
  {https://doi.org/10.1209/epl/i2000-00495-1} {\bibfield  {journal} {\bibinfo
  {journal} {Europhysics Letters (EPL)}\ }\textbf {\bibinfo {volume} {52}},\
  \bibinfo {pages} {705} (\bibinfo {year} {2000})}\BibitemShut {NoStop}%
\bibitem [{\citenamefont {Santangelo}(2006)}]{Santangelo2006}%
  \BibitemOpen
  \bibfield  {author} {\bibinfo {author} {\bibfnamefont {C.~D.}\ \bibnamefont
  {Santangelo}},\ }\bibfield  {title} {\bibinfo {title} {{Computing counterion
  densities at intermediate coupling}},\ }\href
  {https://doi.org/10.1103/PhysRevE.73.041512} {\bibfield  {journal} {\bibinfo
  {journal} {Physical Review E}\ }\textbf {\bibinfo {volume} {73}},\ \bibinfo
  {pages} {041512} (\bibinfo {year} {2006})}\BibitemShut {NoStop}%
\bibitem [{\citenamefont {{\v{S}}amaj}\ \emph {et~al.}(2018)\citenamefont
  {{\v{S}}amaj}, \citenamefont {Trulsson},\ and\ \citenamefont
  {Trizac}}]{Samaj2018a}%
  \BibitemOpen
  \bibfield  {author} {\bibinfo {author} {\bibfnamefont {L.}~\bibnamefont
  {{\v{S}}amaj}}, \bibinfo {author} {\bibfnamefont {M.}~\bibnamefont
  {Trulsson}},\ and\ \bibinfo {author} {\bibfnamefont {E.}~\bibnamefont
  {Trizac}},\ }\bibfield  {title} {\bibinfo {title} {{Strong-coupling theory of
  counterions between symmetrically charged walls: From crystal to fluid
  phases}},\ }\href {https://doi.org/10.1039/c8sm00571k} {\bibfield  {journal}
  {\bibinfo  {journal} {Soft Matter}\ }\textbf {\bibinfo {volume} {14}},\
  \bibinfo {pages} {4040} (\bibinfo {year} {2018})}\BibitemShut {NoStop}%
\bibitem [{\citenamefont {Zhao}\ \emph {et~al.}(2024)\citenamefont {Zhao},
  \citenamefont {Yang}, \citenamefont {Jin},\ and\ \citenamefont
  {Wu}}]{Zhao2024}%
  \BibitemOpen
  \bibfield  {author} {\bibinfo {author} {\bibfnamefont {C.}~\bibnamefont
  {Zhao}}, \bibinfo {author} {\bibfnamefont {T.}~\bibnamefont {Yang}}, \bibinfo
  {author} {\bibfnamefont {S.}~\bibnamefont {Jin}},\ and\ \bibinfo {author}
  {\bibfnamefont {B.}~\bibnamefont {Wu}},\ }\bibfield  {title} {\bibinfo
  {title} {Measurement of {{Electric Double Layer Charging Dynamics}} on
  {{Platinum Electrodes}} in {{Aqueous Solutions}} of {{Alkali Sulfates}} and
  {{Nitrates}}},\ }\href {https://doi.org/10.1021/acs.jpcc.3c08105} {\bibfield
  {journal} {\bibinfo  {journal} {The Journal of Physical Chemistry C}\
  }\textbf {\bibinfo {volume} {128}},\ \bibinfo {pages} {5964} (\bibinfo {year}
  {2024})}\BibitemShut {NoStop}%
\bibitem [{\citenamefont {Lobaskin}\ and\ \citenamefont
  {Netz}(2016)}]{Lobaskin2016}%
  \BibitemOpen
  \bibfield  {author} {\bibinfo {author} {\bibfnamefont {V.}~\bibnamefont
  {Lobaskin}}\ and\ \bibinfo {author} {\bibfnamefont {R.~R.}\ \bibnamefont
  {Netz}},\ }\bibfield  {title} {\bibinfo {title} {Diffusive-convective
  transition in the non-equilibrium charging of an electric double layer},\
  }\href {https://doi.org/10.1209/0295-5075/116/58001} {\bibfield  {journal}
  {\bibinfo  {journal} {Europhysics Letters}\ }\textbf {\bibinfo {volume}
  {116}},\ \bibinfo {pages} {58001} (\bibinfo {year} {2016})}\BibitemShut
  {NoStop}%
\bibitem [{\citenamefont {Balu}\ and\ \citenamefont {Khair}(2018)}]{Balu2018}%
  \BibitemOpen
  \bibfield  {author} {\bibinfo {author} {\bibfnamefont {B.}~\bibnamefont
  {Balu}}\ and\ \bibinfo {author} {\bibfnamefont {A.~S.}\ \bibnamefont
  {Khair}},\ }\bibfield  {title} {\bibinfo {title} {Role of stefan–maxwell
  fluxes in the dynamics of concentrated electrolytes},\ }\href
  {https://doi.org/10.1039/C8SM01222A} {\bibfield  {journal} {\bibinfo
  {journal} {Soft Matter}\ }\textbf {\bibinfo {volume} {14}},\ \bibinfo {pages}
  {8267} (\bibinfo {year} {2018})}\BibitemShut {NoStop}%
\bibitem [{\citenamefont {Armand}\ \emph {et~al.}(2009)\citenamefont {Armand},
  \citenamefont {Endres}, \citenamefont {MacFarlane}, \citenamefont {Ohno},\
  and\ \citenamefont {Scrosati}}]{Armand2009}%
  \BibitemOpen
  \bibfield  {author} {\bibinfo {author} {\bibfnamefont {M.}~\bibnamefont
  {Armand}}, \bibinfo {author} {\bibfnamefont {F.}~\bibnamefont {Endres}},
  \bibinfo {author} {\bibfnamefont {D.~R.}\ \bibnamefont {MacFarlane}},
  \bibinfo {author} {\bibfnamefont {H.}~\bibnamefont {Ohno}},\ and\ \bibinfo
  {author} {\bibfnamefont {B.}~\bibnamefont {Scrosati}},\ }\bibfield  {title}
  {\bibinfo {title} {Ionic-liquid materials for the electrochemical challenges
  of the future},\ }\href {https://doi.org/10.1038/nmat2448} {\bibfield
  {journal} {\bibinfo  {journal} {Nature Materials}\ }\textbf {\bibinfo
  {volume} {8}},\ \bibinfo {pages} {621} (\bibinfo {year} {2009})}\BibitemShut
  {NoStop}%
\bibitem [{\citenamefont {Storey}\ and\ \citenamefont
  {Bazant}(2012)}]{Storey2012}%
  \BibitemOpen
  \bibfield  {author} {\bibinfo {author} {\bibfnamefont {B.~D.}\ \bibnamefont
  {Storey}}\ and\ \bibinfo {author} {\bibfnamefont {M.~Z.}\ \bibnamefont
  {Bazant}},\ }\bibfield  {title} {\bibinfo {title} {{Effects of electrostatic
  correlations on electrokinetic phenomena}},\ }\href
  {https://doi.org/10.1103/PhysRevE.86.056303} {\bibfield  {journal} {\bibinfo
  {journal} {Physical Review E}\ }\textbf {\bibinfo {volume} {86}},\ \bibinfo
  {pages} {056303} (\bibinfo {year} {2012})}\BibitemShut {NoStop}%
\bibitem [{\citenamefont {Kandu{\v{c}}}\ \emph {et~al.}(2010)\citenamefont
  {Kandu{\v{c}}}, \citenamefont {Naji}, \citenamefont {Forsman},\ and\
  \citenamefont {Podgornik}}]{Kanduc2010}%
  \BibitemOpen
  \bibfield  {author} {\bibinfo {author} {\bibfnamefont {M.}~\bibnamefont
  {Kandu{\v{c}}}}, \bibinfo {author} {\bibfnamefont {A.}~\bibnamefont {Naji}},
  \bibinfo {author} {\bibfnamefont {J.}~\bibnamefont {Forsman}},\ and\ \bibinfo
  {author} {\bibfnamefont {R.}~\bibnamefont {Podgornik}},\ }\bibfield  {title}
  {\bibinfo {title} {{Dressed counterions: Strong electrostatic coupling in the
  presence of salt}},\ }\href {https://doi.org/10.1063/1.3361672} {\bibfield
  {journal} {\bibinfo  {journal} {The Journal of Chemical Physics}\ }\textbf
  {\bibinfo {volume} {132}},\ \bibinfo {pages} {124701} (\bibinfo {year}
  {2010})}\BibitemShut {NoStop}%
\bibitem [{\citenamefont {Bonneau}\ \emph {et~al.}(2023)\citenamefont
  {Bonneau}, \citenamefont {D{\'e}mery},\ and\ \citenamefont
  {Rapha{\"e}l}}]{Bonneau2023}%
  \BibitemOpen
  \bibfield  {author} {\bibinfo {author} {\bibfnamefont {H.}~\bibnamefont
  {Bonneau}}, \bibinfo {author} {\bibfnamefont {V.}~\bibnamefont
  {D{\'e}mery}},\ and\ \bibinfo {author} {\bibfnamefont {E.}~\bibnamefont
  {Rapha{\"e}l}},\ }\bibfield  {title} {\bibinfo {title} {Temporal response of
  the conductivity of electrolytes},\ }\href
  {https://doi.org/10.1088/1742-5468/acdced} {\bibfield  {journal} {\bibinfo
  {journal} {Journal of Statistical Mechanics: Theory and Experiment}\ }\textbf
  {\bibinfo {volume} {2023}},\ \bibinfo {pages} {073205} (\bibinfo {year}
  {2023})}\BibitemShut {NoStop}%
\bibitem [{\citenamefont {Adar}\ \emph {et~al.}(2017)\citenamefont {Adar},
  \citenamefont {Markovich},\ and\ \citenamefont {Andelman}}]{Adar2017}%
  \BibitemOpen
  \bibfield  {author} {\bibinfo {author} {\bibfnamefont {R.~M.}\ \bibnamefont
  {Adar}}, \bibinfo {author} {\bibfnamefont {T.}~\bibnamefont {Markovich}},\
  and\ \bibinfo {author} {\bibfnamefont {D.}~\bibnamefont {Andelman}},\
  }\bibfield  {title} {\bibinfo {title} {Bjerrum pairs in ionic solutions: {{A
  Poisson-Boltzmann}} approach},\ }\href {https://doi.org/10.1063/1.4982885}
  {\bibfield  {journal} {\bibinfo  {journal} {The Journal of Chemical Physics}\
  }\textbf {\bibinfo {volume} {146}},\ \bibinfo {pages} {194904} (\bibinfo
  {year} {2017})}\BibitemShut {NoStop}%
\bibitem [{\citenamefont {Borukhov}\ \emph {et~al.}(1997)\citenamefont
  {Borukhov}, \citenamefont {Andelman},\ and\ \citenamefont
  {Orland}}]{Borukhov1997}%
  \BibitemOpen
  \bibfield  {author} {\bibinfo {author} {\bibfnamefont {I.}~\bibnamefont
  {Borukhov}}, \bibinfo {author} {\bibfnamefont {D.}~\bibnamefont {Andelman}},\
  and\ \bibinfo {author} {\bibfnamefont {H.}~\bibnamefont {Orland}},\
  }\bibfield  {title} {\bibinfo {title} {{Steric effects in electrolytes: A
  modified Poisson-Boltzmann equation}},\ }\href
  {https://doi.org/10.1103/PhysRevLett.79.435} {\bibfield  {journal} {\bibinfo
  {journal} {Physical Review Letters}\ }\textbf {\bibinfo {volume} {79}},\
  \bibinfo {pages} {435} (\bibinfo {year} {1997})}\BibitemShut {NoStop}%
\bibitem [{\citenamefont {Ben-Yaakov}\ \emph {et~al.}(2009)\citenamefont
  {Ben-Yaakov}, \citenamefont {Andelman}, \citenamefont {Harries},\ and\
  \citenamefont {Podgornik}}]{Ben-Yaakov2009}%
  \BibitemOpen
  \bibfield  {author} {\bibinfo {author} {\bibfnamefont {D.}~\bibnamefont
  {Ben-Yaakov}}, \bibinfo {author} {\bibfnamefont {D.}~\bibnamefont
  {Andelman}}, \bibinfo {author} {\bibfnamefont {D.}~\bibnamefont {Harries}},\
  and\ \bibinfo {author} {\bibfnamefont {R.}~\bibnamefont {Podgornik}},\
  }\bibfield  {title} {\bibinfo {title} {{Beyond standard Poisson-Boltzmann
  theory: Ion-specific interactions in aqueous solutions}},\ }\href
  {https://doi.org/10.1088/0953-8984/21/42/424106} {\bibfield  {journal}
  {\bibinfo  {journal} {Journal of Physics Condensed Matter}\ }\textbf
  {\bibinfo {volume} {21}},\ \bibinfo {pages} {424106} (\bibinfo {year}
  {2009})}\BibitemShut {NoStop}%
\bibitem [{\citenamefont {Jiang}\ \emph {et~al.}(2014)\citenamefont {Jiang},
  \citenamefont {Cao}, \citenamefont {Jiang},\ and\ \citenamefont
  {Wu}}]{Jiang2014}%
  \BibitemOpen
  \bibfield  {author} {\bibinfo {author} {\bibfnamefont {J.}~\bibnamefont
  {Jiang}}, \bibinfo {author} {\bibfnamefont {D.}~\bibnamefont {Cao}}, \bibinfo
  {author} {\bibfnamefont {D.~E.}\ \bibnamefont {Jiang}},\ and\ \bibinfo
  {author} {\bibfnamefont {J.}~\bibnamefont {Wu}},\ }\bibfield  {title}
  {\bibinfo {title} {{Time-dependent density functional theory for ion
  diffusion in electrochemical systems}},\ }\href
  {https://doi.org/10.1088/0953-8984/26/28/284102} {\bibfield  {journal}
  {\bibinfo  {journal} {Journal of Physics Condensed Matter}\ }\textbf
  {\bibinfo {volume} {26}},\ \bibinfo {pages} {284102} (\bibinfo {year}
  {2014})}\BibitemShut {NoStop}%
\bibitem [{\citenamefont {H{\"a}rtel}(2017)}]{Hartel2017}%
  \BibitemOpen
  \bibfield  {author} {\bibinfo {author} {\bibfnamefont {A.}~\bibnamefont
  {H{\"a}rtel}},\ }\bibfield  {title} {\bibinfo {title} {Structure of electric
  double layers in capacitive systems and to what extent (classical) density
  functional theory describes it},\ }\href
  {https://doi.org/10.1088/1361-648X/aa8342} {\bibfield  {journal} {\bibinfo
  {journal} {Journal of Physics: Condensed Matter}\ }\textbf {\bibinfo {volume}
  {29}},\ \bibinfo {pages} {423002} (\bibinfo {year} {2017})}\BibitemShut
  {NoStop}%
\bibitem [{\citenamefont {Gupta}\ \emph {et~al.}(2020)\citenamefont {Gupta},
  \citenamefont {Govind~Rajan}, \citenamefont {Carter},\ and\ \citenamefont
  {Stone}}]{Gupta2020}%
  \BibitemOpen
  \bibfield  {author} {\bibinfo {author} {\bibfnamefont {A.}~\bibnamefont
  {Gupta}}, \bibinfo {author} {\bibfnamefont {A.}~\bibnamefont {Govind~Rajan}},
  \bibinfo {author} {\bibfnamefont {E.~A.}\ \bibnamefont {Carter}},\ and\
  \bibinfo {author} {\bibfnamefont {H.~A.}\ \bibnamefont {Stone}},\ }\bibfield
  {title} {\bibinfo {title} {Ionic {{Layering}} and {{Overcharging}} in
  {{Electrical Double Layers}} in a {{Poisson-Boltzmann Model}}},\ }\href
  {https://doi.org/10.1103/PhysRevLett.125.188004} {\bibfield  {journal}
  {\bibinfo  {journal} {Physical Review Letters}\ }\textbf {\bibinfo {volume}
  {125}},\ \bibinfo {pages} {188004} (\bibinfo {year} {2020})}\BibitemShut
  {NoStop}%
\bibitem [{\citenamefont {B{\"u}ltmann}\ and\ \citenamefont
  {H{\"a}rtel}(2022)}]{Bultmann2022}%
  \BibitemOpen
  \bibfield  {author} {\bibinfo {author} {\bibfnamefont {M.}~\bibnamefont
  {B{\"u}ltmann}}\ and\ \bibinfo {author} {\bibfnamefont {A.}~\bibnamefont
  {H{\"a}rtel}},\ }\bibfield  {title} {\bibinfo {title} {The primitive model in
  classical density functional theory: Beyond the standard mean-field
  approximation},\ }\href {https://doi.org/10.1088/1361-648X/ac5e7a} {\bibfield
   {journal} {\bibinfo  {journal} {Journal of Physics: Condensed Matter}\
  }\textbf {\bibinfo {volume} {34}},\ \bibinfo {pages} {235101} (\bibinfo
  {year} {2022})}\BibitemShut {NoStop}%
\bibitem [{\citenamefont {Schlaich}\ \emph {et~al.}(2019)\citenamefont
  {Schlaich}, \citenamefont {{Dos Santos}},\ and\ \citenamefont
  {Netz}}]{Schlaich2019}%
  \BibitemOpen
  \bibfield  {author} {\bibinfo {author} {\bibfnamefont {A.}~\bibnamefont
  {Schlaich}}, \bibinfo {author} {\bibfnamefont {A.~P.}\ \bibnamefont {{Dos
  Santos}}},\ and\ \bibinfo {author} {\bibfnamefont {R.~R.}\ \bibnamefont
  {Netz}},\ }\bibfield  {title} {\bibinfo {title} {{Simulations of
  Nanoseparated Charged Surfaces Reveal Charge-Induced Water Reorientation and
  Nonadditivity of Hydration and Mean-Field Electrostatic Repulsion}},\ }\href
  {https://doi.org/10.1021/acs.langmuir.8b03474} {\bibfield  {journal}
  {\bibinfo  {journal} {Langmuir}\ }\textbf {\bibinfo {volume} {35}},\ \bibinfo
  {pages} {551} (\bibinfo {year} {2019})}\BibitemShut {NoStop}%
\bibitem [{\citenamefont {{Mendez-Morales}}\ \emph {et~al.}(2018)\citenamefont
  {{Mendez-Morales}}, \citenamefont {Burbano}, \citenamefont {Haefele},
  \citenamefont {Rotenberg},\ and\ \citenamefont
  {Salanne}}]{Mendez-Morales2018}%
  \BibitemOpen
  \bibfield  {author} {\bibinfo {author} {\bibfnamefont {T.}~\bibnamefont
  {{Mendez-Morales}}}, \bibinfo {author} {\bibfnamefont {M.}~\bibnamefont
  {Burbano}}, \bibinfo {author} {\bibfnamefont {M.}~\bibnamefont {Haefele}},
  \bibinfo {author} {\bibfnamefont {B.}~\bibnamefont {Rotenberg}},\ and\
  \bibinfo {author} {\bibfnamefont {M.}~\bibnamefont {Salanne}},\ }\bibfield
  {title} {\bibinfo {title} {Ion-ion correlations across and between
  electrified graphene layers},\ }\href {https://doi.org/10.1063/1.5012761}
  {\bibfield  {journal} {\bibinfo  {journal} {The Journal of Chemical Physics}\
  }\textbf {\bibinfo {volume} {148}},\ \bibinfo {pages} {193812} (\bibinfo
  {year} {2018})}\BibitemShut {NoStop}%
\bibitem [{\citenamefont {Werkhoven}\ \emph {et~al.}(2018)\citenamefont
  {Werkhoven}, \citenamefont {Everts}, \citenamefont {Samin},\ and\
  \citenamefont {{van Roij}}}]{Werkhoven2018}%
  \BibitemOpen
  \bibfield  {author} {\bibinfo {author} {\bibfnamefont {B.~L.}\ \bibnamefont
  {Werkhoven}}, \bibinfo {author} {\bibfnamefont {J.~C.}\ \bibnamefont
  {Everts}}, \bibinfo {author} {\bibfnamefont {S.}~\bibnamefont {Samin}},\ and\
  \bibinfo {author} {\bibfnamefont {R.}~\bibnamefont {{van Roij}}},\ }\bibfield
   {title} {\bibinfo {title} {{Flow-Induced Surface Charge Heterogeneity in
  Electrokinetics due to Stern-Layer Conductance Coupled to Reaction
  Kinetics}},\ }\href {https://doi.org/10.1103/PhysRevLett.120.264502}
  {\bibfield  {journal} {\bibinfo  {journal} {Physical Review Letters}\
  }\textbf {\bibinfo {volume} {120}},\ \bibinfo {pages} {264502} (\bibinfo
  {year} {2018})}\BibitemShut {NoStop}%
\bibitem [{\citenamefont {Janssen}\ and\ \citenamefont
  {Bier}(2019)}]{Janssen2019}%
  \BibitemOpen
  \bibfield  {author} {\bibinfo {author} {\bibfnamefont {M.}~\bibnamefont
  {Janssen}}\ and\ \bibinfo {author} {\bibfnamefont {M.}~\bibnamefont {Bier}},\
  }\bibfield  {title} {\bibinfo {title} {{Transient response of an electrolyte
  to a thermal quench}},\ }\href {https://doi.org/10.1103/PhysRevE.99.042136}
  {\bibfield  {journal} {\bibinfo  {journal} {Physical Review E}\ }\textbf
  {\bibinfo {volume} {99}},\ \bibinfo {pages} {042136} (\bibinfo {year}
  {2019})}\BibitemShut {NoStop}%
\bibitem [{\citenamefont {Press}\ \emph {et~al.}(2007)\citenamefont {Press},
  \citenamefont {Teukolsky}, \citenamefont {Vetterling},\ and\ \citenamefont
  {Flannery}}]{NumericalRecipes}%
  \BibitemOpen
  \bibfield  {author} {\bibinfo {author} {\bibfnamefont {W.~H.}\ \bibnamefont
  {Press}}, \bibinfo {author} {\bibfnamefont {S.~A.}\ \bibnamefont
  {Teukolsky}}, \bibinfo {author} {\bibfnamefont {W.~T.}\ \bibnamefont
  {Vetterling}},\ and\ \bibinfo {author} {\bibfnamefont {B.~P.}\ \bibnamefont
  {Flannery}},\ }\href@noop {} {\emph {\bibinfo {title} {{Numerical Recipes 3rd
  Edition: The Art of Scientific Computing}}}},\ \bibinfo {edition} {3rd}\ ed.\
  (\bibinfo  {publisher} {Cambridge University Press},\ \bibinfo {address} {New
  York, NY, USA},\ \bibinfo {year} {2007})\BibitemShut {NoStop}%
\end{thebibliography}


%

\end{document}